\documentclass[fleqn,usenatbib]{mnras}

\usepackage{newtxtext,newtxmath}

\usepackage[T1]{fontenc}

\DeclareRobustCommand{\VAN}[3]{#2}
\let\VANthebibliography\thebibliography
\def\thebibliography{\DeclareRobustCommand{\VAN}[3]{##3}\VANthebibliography}

\usepackage{graphicx}	
\usepackage{amsmath}	
\usepackage{booktabs}
\defcitealias{EliasChavez2021}{Paper-I}

\title[AGN X-ray spectral analysis in the XMM-UNDF survey]{XMM-Newton Ultra Narrow Deep Field survey II: X-ray spectral analysis of the brightest AGN population}

\author[Elías-Chávez et al.]{M. Elías-Chávez,$^{1}$\thanks{E-mail: melias@astro.unam.mx}
A. L. Longinotti,$^{2}$
Y. Krongold,$^{2}$
D. Rosa-González,$^{3}$
C. Vignali,$^{4,5}$
\newauthor
S. Mathur,$^{6,7}$
T. Miyaji,$^{1}$
Y. D. Mayya,$^{3}$
F. Nicastro$^{8,9}$
\\
$^{1}$ Universidad Nacional Autónoma de México. Instituto de Astronomía. Apdo. Postal 106, 22800. Ensenada, B.C., México\\
$^{2}$ Universidad Nacional Autónoma de México. Instituto de Astronomía. Apdo. Postal 70-264, 04510. Ciudad de México, México\\
$^{3}$ Instituto Nacional de Astrofísica, Óptica y Electrónica, Luis E. Erro 1, Tonantzintla, Puebla, Apdo. Postal 72840, Mexico\\
$^{4}$ Dipartimento di Fisica e Astronomia, Universit\`a di Bologna, viale Berti Pichat 6/2, 40127 Bologna, Italy\\
$^{5}$ INAF - Osservatorio di Astrofisica e Scienza dello Spazio di Bologna, Via Gobetti 93/3, I-40129 Bologna, Italy\\
$^{6}$ Department of Astronomy, The Ohio State University, 140 West 18th Avenue, Columbus, OH 43210, USA\\
$^{7}$ Center for Cosmology and Astroparticle Physics, 191 West Woodruff Avenue, Columbus, OH 43210, USA\\
$^{8}$ Osservatorio Astronomico di Roma-INAF, Via di Frascati 33, 1-00040 Monte Porzio Catone, RM, Italy\\
$^{9}$ Department of Astronomy, Xiamen University, Xiamen, Fujian 361005, China\\
}

\date{Accepted XXX. Received YYY; in original form ZZZ}

\pubyear{2022}

\begin{document}
\label{firstpage}
\pagerange{\pageref{firstpage}--\pageref{lastpage}}
\maketitle

\begin{abstract}
In this work, we present the results of a detailed X-ray spectral analysis of the brightest AGNs detected in the XMM-Newton 1.75 Ms Ultra Narrow Deep Field. We analyzed 23 AGNs that have a luminosity range of  $\sim 10^{42} - 10^{46}\, \rm{erg}\, \rm{s}^{-1}$ in the $2 - 10\, \rm{keV}$ energy band, redshifts up to 2.66, and $\sim 10,000$ X-ray photon counts in the $0.3 - 10\, \rm{keV}$ energy band. Our analysis confirms the “Iwasawa-Taniguchi effect”, an anti-correlation between the X-ray luminosity ($L_x$) and the Fe-k$\alpha$ Equivalent Width ($EW_{Fe}$) possibly associated with the decreasing of the torus covering factor as the AGN luminosity increases. We investigated the relationship among black hole mass ($M_{BH}$), $L_x$, and X-ray variability, quantified by the Normalized Excess Variance ($\sigma^2_{rms}$). Our analysis suggest an anti-correlation in both $M_{BH} - \sigma^2_{rms}$ and $L_x- \sigma^2_{rms}$ relations. The first is described as $\sigma^2_{rms} \propto M^{-0.26 \pm 0.05}_{BH}$, while the second presents a similar trend with $\sigma^2_{rms} \propto L_{x}^{-0.31 \pm 0.04}$. These results support the idea that the luminosity-variability anti-correlation is a byproduct of an intrinsic relationship between the BH mass and the X-ray variability, through the size of the emitting region. Finally, we found a strong correlation among the Eddington ratio ($\lambda_{Edd}$), the hard X-ray photon index ($\Gamma$), and the illumination factor $\log(A)$, which is related to the ratio between the number of Compton scattered photons and the number of seed photons. The $\log(\lambda_{Edd})-\Gamma-\log(A)$ plane could arise naturally from the connection between the accretion flow and the hot corona.

\end{abstract}




\section{Introduction}

Active Galactic Nuclei (AGNs) are among the most luminous objects in the Universe. They reside in galaxies that harbor a Super Massive Black Hole (SMBH) at their centers, with a mass range of $M_{BH} \sim 10^{5-10} M_{\sun}$, powered by an accreting disc of gas \citep{Kormendy1995}. The huge amount of energy generated in their nuclear region can reach X-ray luminosities higher than $10^{46}\, \rm{erg}\, \rm{s}^{-1}$ with bolometric luminosities  $L_{bol} \approx 10^{42-48} \, \rm{erg}\, \rm{s}^{-1}$ \citep{Hickox2018}. The X-ray radiation observed in AGNs is thought to be produced primarily by the process of comptonization. In this process, optical/UV disk photons are scattered by a corona of hot electrons located above the accretion disk  \citep{Haardt1991,George1991,Matt1997}. 

X-ray surveys conducted by satellites such as XMM-Newton, Chandra, and more recently eROSITA, serve as highly effective methods for AGN identification \citep[e.g.][]{Luo2017,Chen2018,Liu2022}. For instance, X-ray emission resulting from the accretion process onto SMBHs can penetrate through high hydrogen column densities ($N_H \approx 10^{21}$ - $10^{24.5}\, \rm{cm}^{-2}$), and experience minimal dilution by starlight from the host galaxy \citep{Brandt2015}. As a result, X-ray surveys allow to census large samples of both obscured  ($N_H \ge 10^{22}\, \rm{cm}^{-2}$) and unobscured  ($N_H < 10^{22}\, \rm{cm}^{-2}$) AGNs and their host galaxies across different redshift ranges \citep{Hickox2018}. This feature facilitates studies such as the connection between SMBHs and galaxy formation, the contribution of these sources to the Cosmic X-ray Background \citep[XRB,][]{Gilli2007}, and to test models of quasar formation and AGN evolution  \citep{Scoville2007,Kellermann2008,Rosen2016,Brandt2017}.

X-ray spectral analysis is a powerful diagnostic tool to investigate the physical properties of AGNs. It enables the estimation of parameters such as intrinsic absorption $N_H$, black hole mass $M_{BH}$, X-ray luminosity $L_x$, and to characterize the nuclear region surrounding the SMBH \citep{Brandt2017}. One of the main spectral features commonly observed in most AGNs is the Fe-K$\alpha$ emission line at 6.4 $\rm{keV}$, which is generated as a consequence of the X-ray fluorescence process. The analysis of the Fe-K$\alpha$ is fundamental to studying the structure of the torus and inner regions of AGNs, such as the accretion disk properties \citep{Fragile2005}. For instance \citet{Iwasawa1993} and \citet{Bianchi2007}  have demonstrated the presence of an anti-correlation between the Fe-K$\alpha$ equivalent width ($EW_{Fe}$) and the X-ray luminosity in the $2-10\, \rm{keV}$ energy range. This relationship, described as $EW_{Fe} \propto L_x^{-0.2}$, has been referred to as the “Iwasawa-Taniguchi effect” or the “X-Ray Baldwin Effect”. Some studies have proposed that it arises from the decreasing of the opening angle of the torus as a function of the increasing of the AGN luminosity \citep{Bianchi2007, Ricci2014}. Another possibility is the decreasing amount of low-ionization material available to generate the fluorescence Fe-K$\alpha$ line, due to the increase of the X-ray luminosity that ionizes the neutral iron in the torus \citep{Shu2010}. During our analysis, we will investigate the "Iwasawa-Taniguchi effect" using a collection of new high-quality AGN spectra.

Most AGNs exhibit short-term and long-term X-ray variability, which is thought to be generated due to changes in the accretion flow \citep{Yuan2014}, instabilities in the disk corona \citep{McHardy2004}, or variable-density absorptions, among other factors influenced by the accretion dynamics and the surrounding environment \citep{Beuchert2015}. Taking advantage of this feature, certain X-ray studies have contributed with novel techniques to identify and characterize AGNs samples, infer their general properties, and investigate the contribution of $M_{BH}$ to the variability-luminosity relation. For example, \citet{Nikolajuk2004} and \citet{Ponti2012} employed the normalized excess variance ($\sigma^2_{rms}$), a parameter that quantifies the X-ray flux variation, to estimate the BH mass in local unobscured Radio-Quiet AGNs. \citet{Lanzuisi2014} analyzed the long-term variability of the brightest AGNs detected in the XMM Cosmic Evolution Survey (XMM-COSMOS). They reported a significant anti-correlation between the X-ray luminosity and the X-ray flux variability, suggesting the possibility that the observed luminosity-variability relationship is a consequence of an intrinsic $M_{BH}$-variability relation. 

The accretion process in AGNs plays a crucial role in determining the cosmic evolution of SMBHs. A meaningful physical parameter that offers valuable insights into the BH growth is the Eddington ratio $\lambda_{Edd} = L_{bol}/L_{Edd}$ \citep{Trump2011,Georgakakis2017,Laurenti2022},  which is defined as the ratio between the bolometric luminosity $L_{bol} = \eta\dot{M}_{BH}c^2$ where $\eta$ describes the efficiency of the accretion process, and the Eddington luminosity $L_{Edd}$. $L_{bol}$ can be estimated from the X-ray luminosity \citep{Netzer2013}. $L_{Edd}$ represents the maximum luminosity allowed by a steady-state accretion, at which the radiation pressure balances the gravitational force. For instance, \citet{Trump2011} demonstrated the utility of accretion rate to distinguish between unobscured broad-line, narrow-line, and lineless AGNs, after analyzing a sample of 153 AGNs. The research revealed that high accretion rates ($\lambda_{Edd} > 0.01$) were predominantly associated with broad-line AGNs and some potentially obscured narrow-line AGNs. In contrast, Narrow-line and lineless AGNs exhibited lower specific accretion rates $(\lambda_{Edd} < 0.01)$ and higher radio-to-optical/UV emission ratios. In the highest accretion regime, \citet{Laurenti2022} reported significant dispersions in key spectral parameters (e.g. $\Gamma = 1.3 - 2.5$) for a small group of highly accreting AGNs $(\lambda_{Edd}>1)$, with approximately 30\% classified as X-ray weak quasars.

In this paper, we analyzed the main X-ray spectral properties of the brightest AGNs detected in one of the deepest surveys observed by XMM-Newton. We searched for any relationship among the accretion process and the X-ray variability represented by $\lambda_{Edd}$ and $\sigma^2_{rms}$, respectively, and other physical parameters of our AGN sample, including X-ray luminosity, BH mass, K$\alpha$ iron emission line, photon index, and intrinsic column density. This paper is organized as follows: in Section \ref{AGNsample}, we provided a description of the AGN sample and the XMM-UNDF survey. We outlined the XMM-Newton observations, the available multi-wavelength data, and the main properties of the AGNs. The X-ray spectral analysis, including the modeling and its results, is presented in Section \ref{X-ray_spectral_analysis}. Section \ref{analysis} focuses on the study of the primary spectral parameters. The X-ray variability, the BH mass estimations, and a linear regression Monte Carlo simulation are presented in Section \ref{vary}. Then, in Section \ref{Eddington} we present an analysis of the accretion rate distribution of our AGNs and in Section \ref{conclusions} we summarize the key outcomes of our analysis.

Throughout this work, we adopted the cosmological parameters $H_0 = 70\, \rm{km}\, \rm{s}^{-1}\, \rm{Mpc}^{-1}$, $\Omega_m = 0.3$ and $\Omega_\Lambda = 0.7$.

\section{X-ray survey and sample selection}
\label{AGNsample}

\subsection{The XMM-UNDF Survey}
\label{XMM-UNDF}

\begin{figure}
\centering
\includegraphics[scale=0.51]{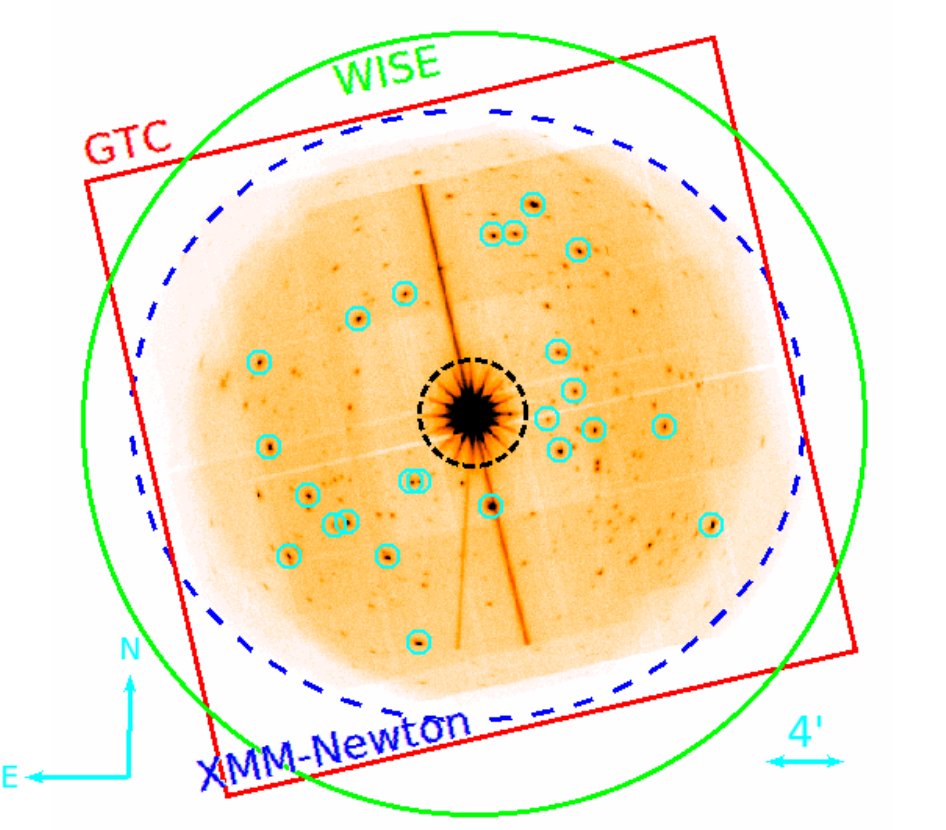} 
\caption{Mosaic image of the X-ray survey at $0.3 - 10\, \rm{keV}$ band. Our AGN sample is represented with cyan circles. The full optical GTC-Field is marked by a red square of $33'$ sides, the blue ellipse refers to the XMM-UNDF of about $30' \times 30'$, and the green circle of $20'$ radius is for the infrared WISE-field.} 
\label{xmm1}
\end{figure}

The XMM-Newton Ultra Narrow Deep Field survey, detailed in our previous study of \citet{EliasChavez2021} (\citetalias{EliasChavez2021} hereafter) and renamed as XMM-UNDF, consists of 13 observations taken over 2 years with a total exposure time of 1.75 million seconds (Ms) in a field of  900 $\rm{arcmin}^2$ \citep{Nicastro2018} around the high luminous blazar 1ES 1553+113 ($F_{0.3-10\, \rm{keV}}  \approx 2 \times 10^{-11} \rm{erg}\, \rm{s}^{-1}\, \rm{cm}^{-2}$). With a flux limit of $4.03 \times 10^{-16}\, \rm{erg}\, \rm{s}^{-1}\, \rm{cm}^{-2}$ in the $0.2 - 2.0\, \rm{keV}$ energy band. The XMM-UNDF is the deepest survey observed by XMM-Newton centered around a bright source and the third with the highest sensitivity. It was complemented with deep optical broadband images with the Sloan Digital Sky Survey (SDSS) filters $u'$, $g'$, $r'$, $i'$, and $z'$ obtained with the OSIRIS instrument mounted on the Gran Telescopio Canarias (GTC), down to magnitude $r \sim 24.5$. Additionally, this field presents IR coverage with WISE and 2MASS observatories in the Mid ($W_1$, $W_2$, $W_3$, and $W_4$) and Near ($J$, $H$, and $K_s$) infrared bands, respectively, with detections at signal-to-noise of SNR > 5 \citep{Cutri2014}. Figure \ref{xmm1} presents the mosaic of X-ray images of the field with optical (GTC) and infrared (WISE) observational coverage. The cyan circles highlight the AGN sample of our analysis. 

\subsection{X-ray Data Products}

The X-ray data processing was performed with the XMM-Newton Science Analysis Software version 18 \citep[SAS,][]{Gabriel2004}. The tasks \texttt{epproc} and \texttt{emproc} provided by the \texttt{epicproc} package were utilized to prepare the observations and generate data products such as light-curves, spectra, images from the EPIC instruments (PN, MOS1, MOS2 cameras). Afterward, source detection was performed using the task \texttt{edetec\_stack}, specifically developed for multi-epoch XMM-Newton observations \citep{Traulsen2019,Traulsen2020}. For more details on the AGN identification process, X-ray source detection, and data reduction, refer to  \citetalias{EliasChavez2021}.

\subsection{The AGN sample}
In \citetalias{EliasChavez2021}, we reported an X-ray-Optical-IR catalog consisting of 301 sources detected at a significance level of $3\sigma$ in the XMM-UNDF survey\footnote{The full X-ray catalog is available online at \url{https://doi.org/10.26093/cds/vizier.19190018}}. The majority (244; 81\%) of the objects in the catalog possess at least one optical or infrared counterpart association. Among these sources, 204 were classified as AGNs based on criteria involving X-ray luminosity, X-ray/optical, and X-ray/IR flux ratios, as carried out in other X-ray surveys \citep{Xue2011,Luo2017,Chen2018}. 

\begin{figure}
\centering
\includegraphics[scale=0.21]{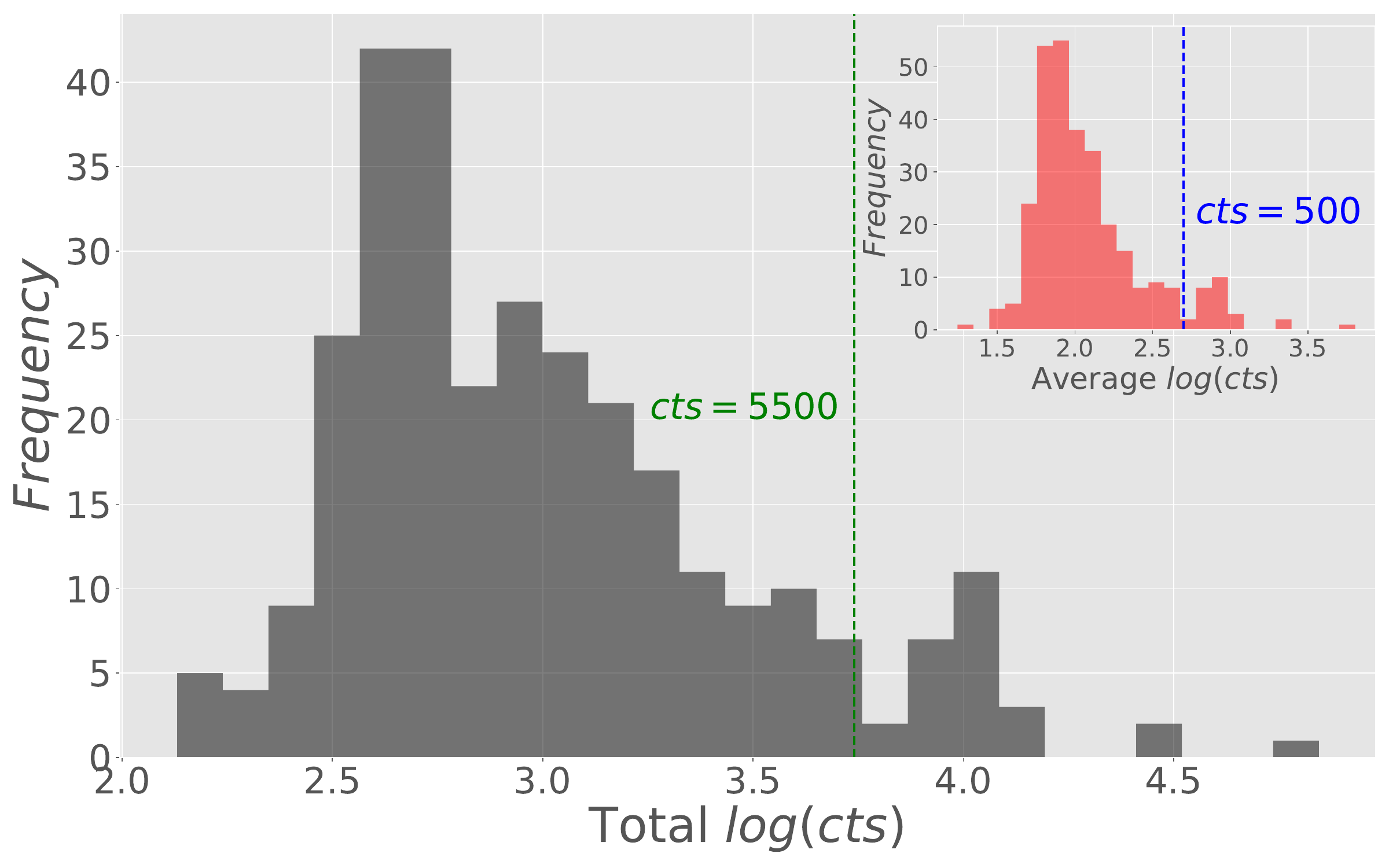} 
\caption{Histograms of the stacked observations (black) and average-per-observation (red) X-ray counts of our AGN population in the XMM-UNDF survey in the $0.3-10\, \rm{keV}$ band, the green and blue dotted lines limit the regions for those objects that were detected with 5500 counts in total and 500 in average-per-observation with the PN camera, respectively.}
\label{cts_hist}
\end{figure}

In the present analysis, we selected a subsample of the brightest AGNs that were detected with at least 500 photon counts (cts) on average per observation in the $0.3-10\, \rm{keV}$ energy band with the PN camera. For instance, a source detected in 11 observations will have at least 5500 cts in total with the PN camera to satisfy this criterion. This threshold ensured a sufficient level of statistical quality in each individual X-ray spectrum. As a result, we reduced the list to 23 AGNs that met this requirement with a median X-ray count of $\sim 10,000$ cts and X-ray flux range from $3 \times 10^{-13}$ to $2 \times 10^{-14}\, \rm{erg}\, \rm{s}^{-1}\, \rm{cm}^{-2}$ in the $0.3-10\, \rm{keV}$ band, all with optical and infrared counterparts, except for three sources for which we did not detect WISE infrared emission. The total and average photon count distributions of our AGN sample are presented in Figure \ref{cts_hist} with the black and red histograms, respectively. Table \ref{photometric} reports their main multi-wavelength properties.

\begin{table*}
\centering
\caption{Multi-wavelength parameters of our bright AGN sample detected in the XMM-UNDF survey. From XMM-Newton ($cts$, $F_{0.3-10\, \rm{keV}}$ and $L_{0.3-10\, \rm{keV}}$) and GTC ($u'g'r'i'$ and $z'$) observations, and WISE ($W_1$, $W_3$, $W_3$, and $W_4$) public data. We maintained the same X-ray identification names (XID) from \citetalias{EliasChavez2021}.}
\label{photometric}
\begin{tabular}{ccccccccccccccc}
\toprule 
XID & $z$ & $z_f$$^a$ & cts & $F_{0.3-10\, \rm{keV}}$ & $L_{0.3-10\, \rm{keV}}$ & $u'$ & $g'$ & $r'$ & $i'$ & $z'$ & $W_1$ & $W_2$ & $W_3$ & $W_4$ \\
 & &  &  & $10^{-14}\, \rm{erg}\, \rm{s}^{-1}\, \rm{cm}^{-2}$ & $10^{44}\, \rm{erg}\, \rm{s}^{-1}$ & mag & mag & mag & mag & mag & mag & mag & mag & mag \\ \midrule 
1 & 2.66 & s & 70,046  & 30.44 & 260.77 & 20.2 & 19.4 & 19.3 & 19.3 & 19 & 16 & 15.2 & 11.5 & 8.42 \\
2 & 0.134 & s & 27,437  & 14.38 & 0.007 & 20 & 18.8 & 18 & 17.5 & 17.1 & 14.9 & 14.7 & 12.2 & 9.09 \\
3 & 0.757 & s & 9,698  & 11.79 & 3.67 & 19.7 & 19.4 & 19.5 & 19.4 & 19 & 14.7 & 13.5 & 10.6 & 8.09 \\
4 & 1.13 & s & 25,947  & 8.55 & 7.53 & 20.9 & 20.7 & 20.4 & 20.4 & 20.2 & 16.2 & 15.2 & 12.4 & 8.97 \\
5 & 0.948 & s & 15,617  & 8.41 & 4.7 & 22.7 & 21.7 & 21.3 & 20.9 & 20.2 & - & - & - & - \\
6 & 1.15 & p & 11,314  & 6.42 & 6.01 & 21.3 & 21.3 & 20.9 & 20.7 & 20.6 & 16.9 & 15.7 & 12.3 & 8.96 \\
7 & 1.04 & p & 12,989  & 7.08 & 5.07 & 22.1 & 21.5 & 21.3 & 21.1 & 20.7 & 16 & 15 & 12 & 8.8 \\
8 & 0.621 & p & 12,094  & 5.5 & 1.03 & 23.6 & 22.4 & 22.2 & 21.5 & 21.2 & 17.7 & 16.8 & 12.2 & 8.85 \\
9 & 0.998 & p & 12,625  & 7.78 & 4.98 & 21.1 & 20.7 & 20.5 & 20.4 & 20.1 & 15.9 & 15.7 & 12.4 & 8.99 \\
10 & 0.879 & s & 10,959  & 6.18 & 2.84 & 21.6 & 21.2 & 20.9 & 20.5 & 20.1 & 15.9 & 15.3 & 12 & 9.02 \\
11 & 0.126 & p & 9,640  & 4.92 & 0.002 & 20.1 & 19.7 & 19.5 & 19 & 19 & 16.4 & 15.2 & 12.1 & 8.64 \\
12 & 0.749 & p & 10,243  & 4.97 & 1.51 & 21.9 & 21.3 & 21.3 & 20.8 & 20.9 & 18 & 16.2 & 12.1 & 8.25 \\
13 & 0.842 & p & 10,105  & 4.81 & 1.97 & 22.6 & 22.5 & 21.9 & 21.7 & 21.3 & 16.8 & 15.6 & 12 & 9.05 \\
14 & 1.43 & s & 9,953  & 5.03 & 8.29 & 21.2 & 21 & 20.7 & 20.5 & 20.4 & 16.9 & 15.7 & 12 & 8.95 \\
15 & 0.344 & p & 8,393  & 5.97 & 0.26 & 22.4 & 21.4 & 20.6 & 20.2 & 19.9 & 16.2 & 14.9 & 11.4 & 8.5 \\
16 & 0.386 & p & 9,664  & 4.5 & 0.26 & 19.9 & 19.5 & 19.4 & 19.2 & 19.2 & 16.2 & 15.1 & 12.5 & 8.7 \\
17 & 0.434 & s & 8901  & 3.27 & 0.25 & 23 & 22 & 20.5 & 19.8 & 19.3 & 15.9 & 15.5 & 12.7 & 9.09 \\
18 & 0.61 & p & 7,914  & 3.6 & 0.65 & 20.7 & 20.2 & 20.1 & 19.9 & 19.8 & 16.2 & 15 & 11.9 & 8.8 \\
19 & 0.427 & p & 8,749  & 4.31 & 0.32 & 23.2 & 22.1 & 21.8 & 21.3 & 21 & 16.6 & 15.4 & 11.9 & 8.55 \\
20 & 0.589 & p & 10,418  & 4.21 & 0.69 & 23.8 & 23 & 22.7 & 22.1 & 22 & - & - & - & - \\
8 & 0.722 & s & 9,164  & 3.35 & 0.92 & 22.7 & 22 & 21.6 & 20.8 & 20.4 & 16.1 & 15.7 & 12.2 & 8.66 \\
33 & 0.458 & p & 6,265  & 2.05 & 0.18 & 23 & 22.4 & 21.6 & 21.3 & 21.1 & - & - & - & - \\
36 & 0.949 & p & 6,415  & 1.91 & 1.07 & 19.8 & 19.7 & 19.7 & 19.4 & 19.1 & 16.4 & 15.3 & 12.2 & 8.29  \\ \bottomrule
\end{tabular}
\begin{flushleft}
\footnotesize 
$^a$ Redshift flag, p and s correspond to photometric and spectroscopic redshift, respectively. Photometric redshifts present a normalized standard deviation of $\sigma_{norm} = 0.064$.
\end{flushleft}
\end{table*}

\section{X-ray spectral analysis}
\label{X-ray_spectral_analysis}
\subsection{Stacking Multiple Spectra}
Given that the XMM-UNDF survey consists of multiple observations centered in the same field, we employed a spectral stacking approach to enhance the signal-to-noise ratio and minimize statistical uncertainties. We extracted and combined the individual spectra of each AGN from our sample using 11 out of 13 observations that were conducted in a PN full window mode, i.e. the 2 PN small window observations (0761100701 and 0790381001) were not considered in this analysis\footnote{In \citetalias{EliasChavez2021}, the central region of the observations around the bright blazar (marked with the black circle in Figure \ref{xmm1}) was avoided during the source detection process. This region covers most of the small window field of view.}. To combine the spectra, we followed the XMM-Newton data analysis thread "Combining the spectra of the 3 epic cameras"\footnote{\url{https://www.cosmos.esa.int/web/xmm-newton/sas-thread-epic-merging}}. We used the task \texttt{epicspeccombine} to merge the spectra of the three EPIC cameras (PN, MOS1, and MOS2) whenever they were available from the 11 observations. This resulted in a single spectrum with its corresponding calibration matrices (rmf, arf) and background (bkg) files. The procedure we followed is outlined below:

\begin{enumerate}
\item  We begin by using the task \texttt{evselect} to extract all source and background spectra from manually selected regions in the 3 cameras, we used circular areas of $15\arcsec$ and $30\arcsec$, respectively. These regions correspond to about 75\% of the encircled energy fraction. An example is presented in Figure \ref{XID7} with the source XID-7.

\begin{figure}
\centering
\includegraphics[scale=0.28]{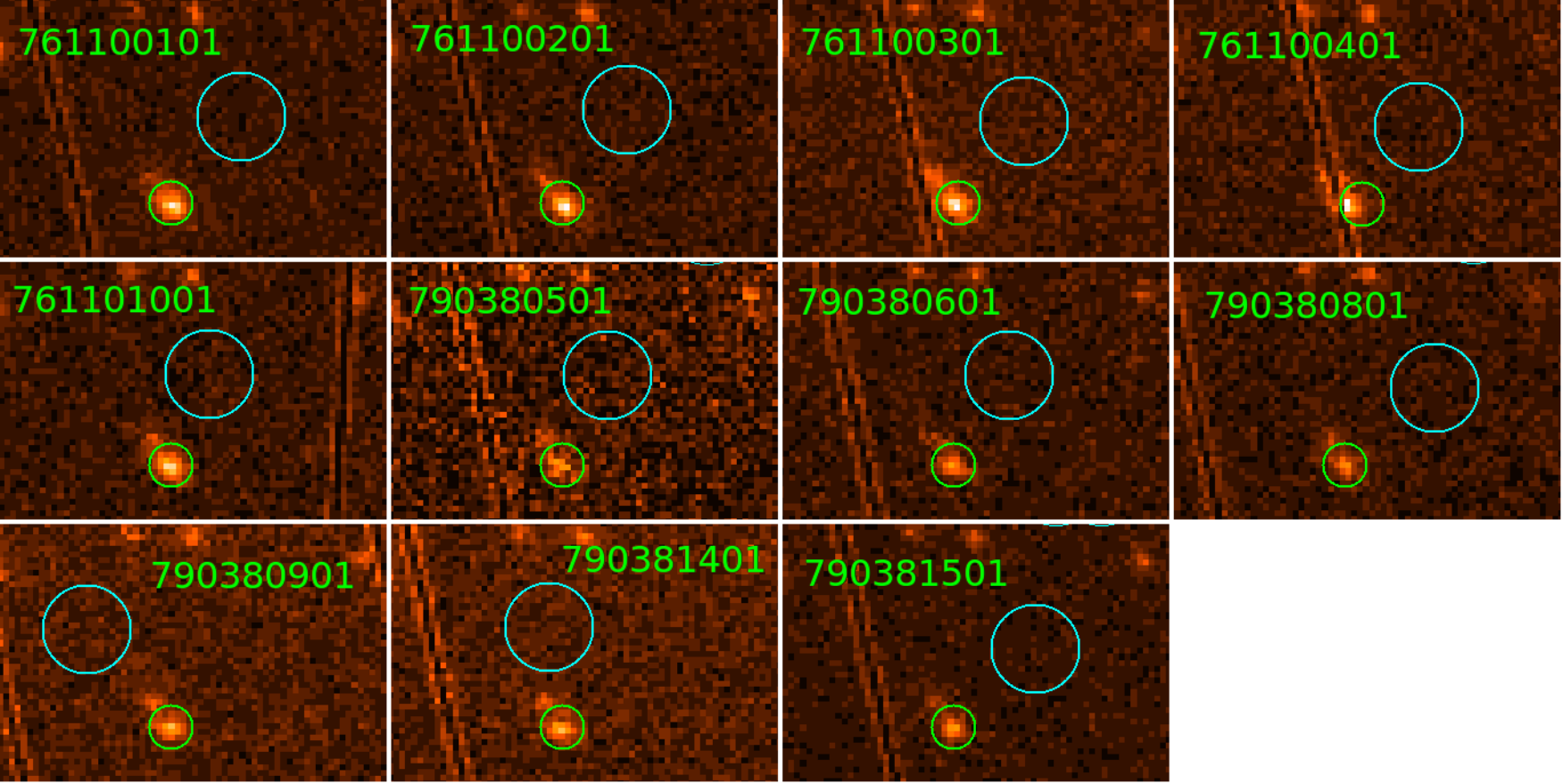} 
\caption{PN images in the $0.3 - 10\, \rm{keV}$ band of the 11 observations. The green and cyan circles with a radius of $15\arcsec$ and $30\arcsec$, respectively, mark the position of the  XID-7 source and the adopted used background region.}
\label{XID7}
\end{figure}

\item Then, the \texttt{backscale} task was used to calculate the areas of the source and background regions.

\item For each extracted spectrum, the redistribution matrix file (rmf) and auxiliary response file (arf) were generated using the \texttt{rmfgen} and \texttt{arfgen} tasks.

\item To ensure adequate statistical quality and prevent oversampling of the energy resolution by more than a factor of 3, the \texttt{specgroup} task was used to rebin the spectrum and link the associated files to have at least 25 counts for each background-subtracted spectral channel.

\item Finally, all individual spectra from the three cameras and the 11 observations were combined into one single spectrum using the \texttt{epicspeccombine} task.

\end{enumerate}

\subsection{Spectral Fitting and Modelling}
\label{fitting}

For our study, we used the software XSPEC version 12.10.0 to perform the X-ray spectral fitting. We employed a set of simple absorbed power-law models to obtain the best description of the spectral shape. The models used are: 

\begin{enumerate}
    \item \texttt{tbabs*powerlaw}
    \item \texttt{tbabs*powerlaw*zphabs}
    \item \texttt{tbabs*(powerlaw*zphabs + zgauss)} 
    \item \texttt{tbabs*((powerlaw + zbb)*zphabs + zgauss)} 
\end{enumerate}

We used an approach similar to that presented in previous X-ray analyses of bright AGNs \citep[e.g.][]{Corral2011,Iwasawa2020} with a high number of counts (>200). Our objective is to derive crucial parameters ($\Gamma$, $N_H$, and $EW_{Fe}$) essential for our study, avoiding the introduction of higher complexity in the form of additional components (e.g. ionized absorptions, reflected component). 

The first base model consists of a simple power-law (\texttt{powerlaw}) with a Galactic absorption \citep[\texttt{tbabs},][]{Wilms2000}. The second model incorporates a neutral intrinsic absorber associated with the AGN or its host galaxy (\texttt{zphabs}). The third model increases the complexity of the second model by including a Fe-K$\alpha$ emission line. For sources that were not well fit by our previous three models, we included a black body component (\texttt{zbb}) at temperature $kT$ to account for the soft-excess emission.

In our analysis, we adopted a Galactic absorption of $N_H = 3.56 \times 10^{20}\, \rm{cm}^{-2}$ in the line of sight of the X-ray field (as reported in \citetalias{EliasChavez2021}) and we let the power-law and intrinsic absorption parameters free-to-vary. To ensure physically reliable estimations, we fixed the neutral emission line energy at $E = 6.4\, \rm{keV}$ with a narrow line width at $\sigma = 0.01\, \rm{keV}$. Error bars for our spectral analysis were estimated with a $90\%$ confidence level. We employed a $\chi^2$-statistics. 

To assess whether there is an improvement in the accuracy of our results when we increase the complexity of the model, we employed the Akaike’s Information Criterion \citep[AIC,][]{Akaike1974} defined in Equation \ref{AIC}. It is a statistical test for nested models that estimates the relative quality of our models \citep{Hebbar2019,Krongold2021}.

\begin{equation} \label{AIC}
    AIC = 2k + \chi^{2}_{stat}
\end{equation}

where $\chi^{2}_{stat}$ is the $\chi^{2}$-statistic value and $k$ is the number of parameters. Then, we estimated the relative likelihood ($l_{AIC}$) of our models using Equation \ref{LAIC} to quantify if we are losing information due to excluding a new spectral component $x_i$. We used the inverse of $l_{AIC}$ as the factor $\rho_{x_i} = l_{AIC}^{-1}$ by which a more complex model with extra spectral components is preferred over a simpler model. A threshold of $\rho_{x_i}>5$ to confirm the detection of a new component was implemented.  
 
\begin{equation}\label{LAIC}
    l_{AIC} = exp \left( \frac{AIC(x_0,x_i) - AIC(x_0)}{2} \right)
\end{equation}

We employed a 95\% of confidence for our statistical test. This threshold roughly corresponds to the $2\sigma$ level, supporting the use of the second model over the first to describe our data.

\begin{figure}
\centering
\includegraphics[scale=0.31]{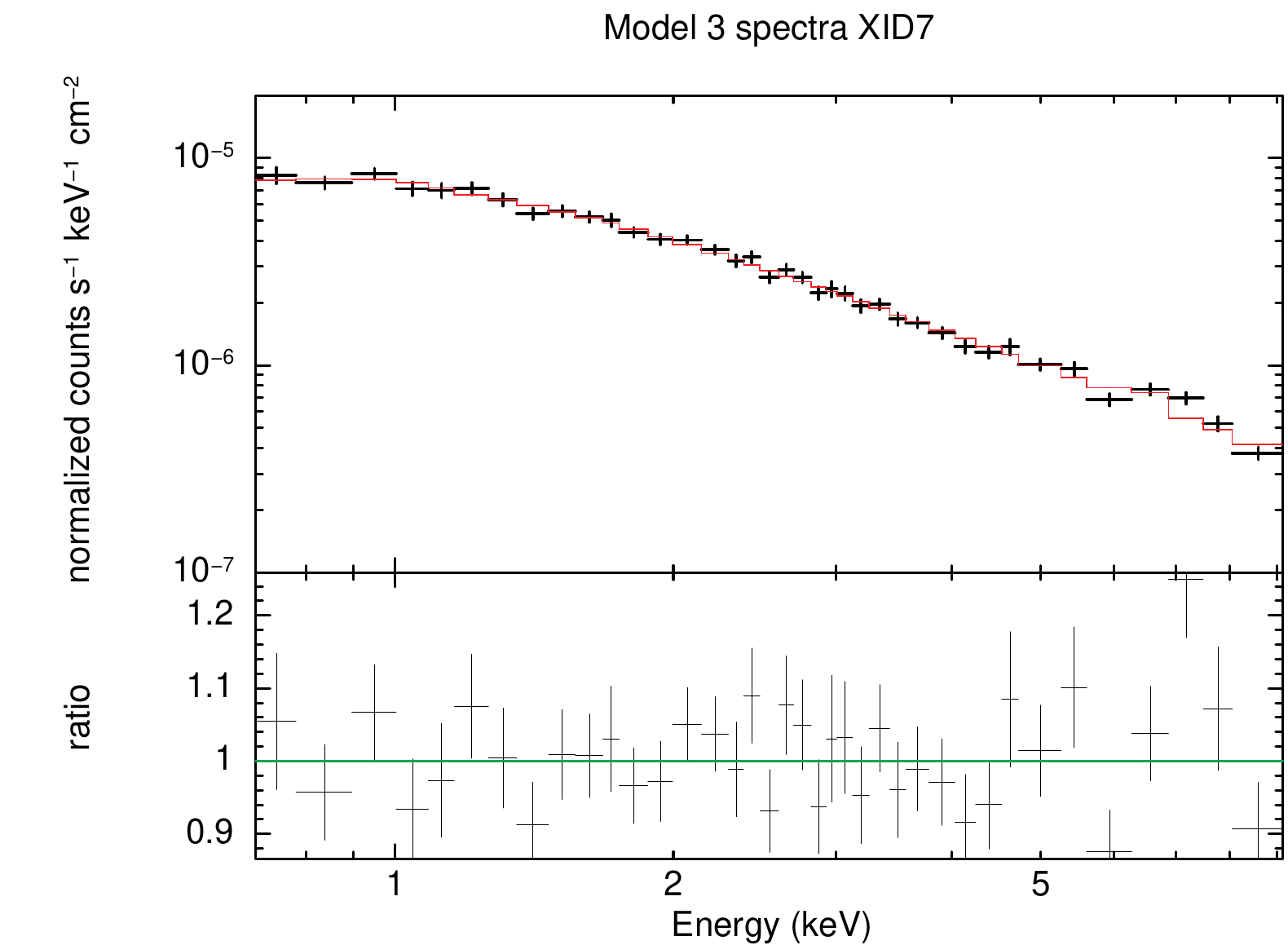}
\caption{Upper: combined rest-frame spectrum (black points) of the AGN XID-7.  The red line represents the best-fit model, i.e. \texttt{tbabs*(powerlaw*zphabs + zgauss)}. Lower: residuals of the fit (data divided by the model).}
\label{mod3_XID7}
\end{figure}

\begin{figure}
\centering
\includegraphics[scale=0.3]{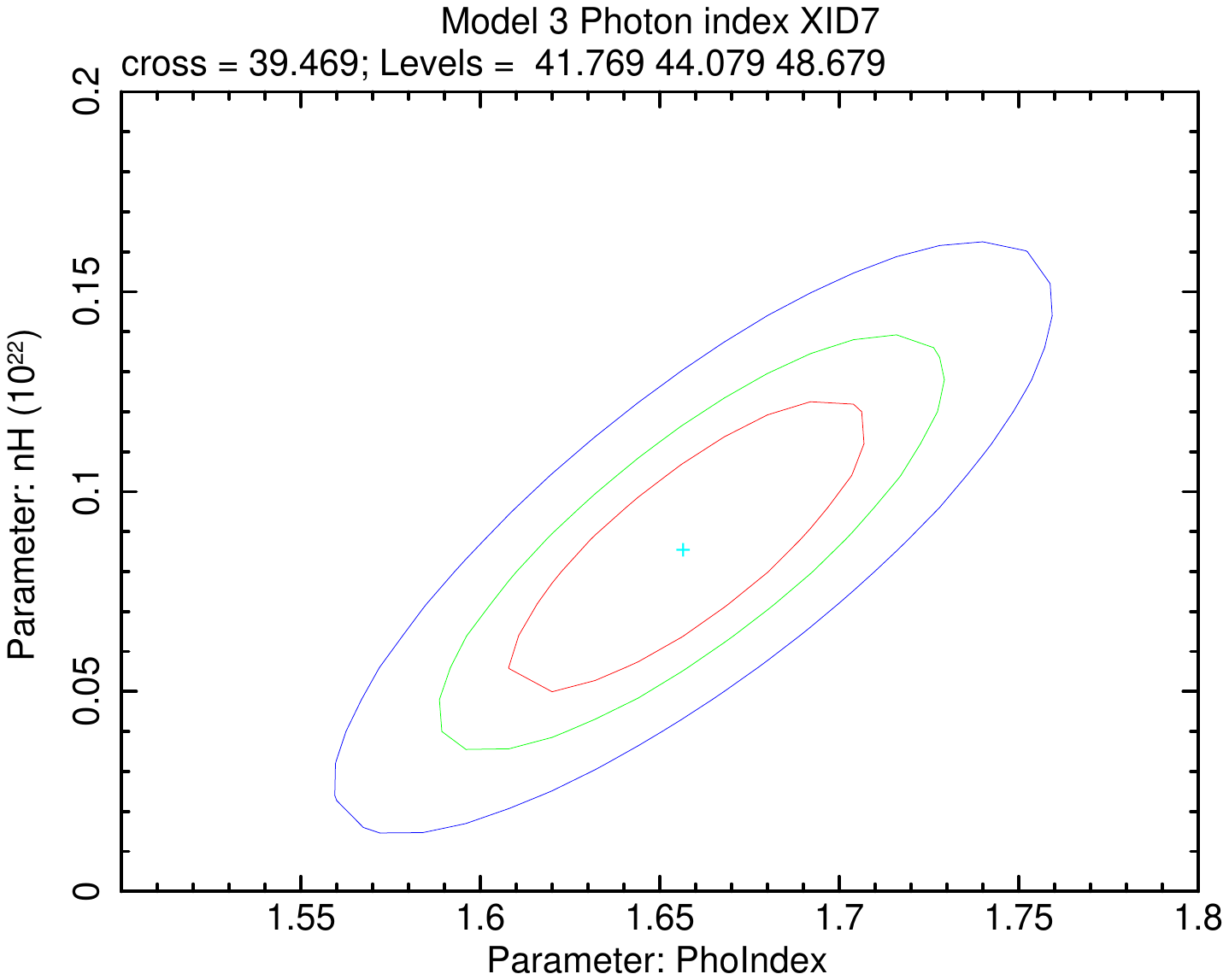} 
\caption{Confidence contours for the photon index and column density parameters derived by the best-fit model applied to source XID-7. The red, green and blue countors refer to 1$\sigma$, 2$\sigma$, and 3$\sigma$ of confidence.} 
\label{contornos}
\end{figure}

Figure \ref{mod3_XID7} presents an example of our analysis. In the upper panel, we display the combined $0.3 - 10\, \rm{keV}$ rest-frame spectrum (black dots) of the source XID-7, fitted with model 3 (red line). In the lower panel, we present the residuals (data divided by the folded model). We detected a residual feature at $\sim 7\, \rm{keV}$, which could be generated as a result of the FeXXVI line at 6.97 $\rm{keV}$. This component was not considered during our analysis. Then, in Figure \ref{contornos} we present the confidence contour plot of the photon index as a function of $N_H$ for the source XID-7. The high counting statistics criterion allow us to reduce the degeneracy between the spectral parameters, $\Gamma$ and $N_H$ \citep{Mateos2008}. 

A summary of the best-fit parameters obtained with the spectral analysis of the 23 sources is presented in Table \ref{statistic1}. The table includes values of $\chi^2$ and degrees of freedom $(dof)$ for the best-fit model. Additionally, the results of the AIC criterion are provided in the $\rho_{N_H}$, and $\rho_{Fe}$ columns, respectively. Measurements of some properties such as $\Gamma$, $N_H$, $kT$, $EW_{Fe}$, and luminosities in the $2-10\, \rm{keV}$ band are also included. 

Out of the 23 AGNs analyzed, we found that 9 of them exhibit statistically significant intrinsic absorption component $N_H$, with a mean and standard deviation of $\log(N_H) = 20.92 \pm 0.18\, \rm{cm}^{-2}$. For the remaining 14 AGNs, we provide upper-limits at $90\%$ of confidence level. We found statistically reliable detection of the Fe-K$\alpha$ line for 12 AGNs, with a mean and standard deviation of their equivalent width of $ EW_{Fe} = 0.14 \pm 0.11\, \rm{keV}$. For the remaining 11, we have $EW_{Fe}$ upper-limits. The combined spectrum of each source fitted with its best model, according to the last column of table \ref{statistic1}, can be found in appendix \ref{Spectra}. We obtained a mean photon index of $\Gamma = 1.88 \pm 0.16$ for the whole sample, estimated from the best-fit model for each spectrum, which is a typical value for type-1 or unabsorbed AGNs \citep{Mateos2010,Corral2011}. Finally, we found that 7 sources show an improvement with $90\%$ confidence when including a black body component (model 4).

\begin{table*}
\centering
\caption{Summary table of the spectral parameters with the power-law model of our bright AGN sample detected in the XMM-UNDF survey. We display the values obtained with the best-fit model, which is reported in the last column with the digits 1, 2, 3, and 4. The results of the statistical test are presented in the columns $\rho_{N_H}$ for the intrinsic absorption, and  $\rho_{N_H}$ for the Fe-K$\alpha$ emission line. Only sources fitted with model 4 present $kT$ measurements}
\label{statistic1}
\begin{tabular}{cccccccccc}
\toprule 
XID & $\chi^2/dof$ & $\rho_{NH}$ & $\rho_{Fe}$ & $\Gamma$ & $\log(N_H)$ & $EW_{Fe}$ & $kT$ & $\log(L_{2-10\, keV})$ & Best-fit \\
 &   &   &  &  & $\rm{cm}^{-2}$ & $\rm{keV}$ & $\rm{keV}$ & $\rm{erg}\, s^{-1}$ &  \\ \midrule 
1  & 47/49  & 33.28 & 2.65 & $1.79_{-0.03}^{+0.03}$ & $20.95_{-0.51}^{+0.26}$ & $<0.01$ & $-$ & $45.34_{-0.01}^{+0.01}$ & 2 \\
2  & 49/25  & 2.72 & 8062.67 & $2.03_{-0.07}^{+0.05}$ & $<20$ & $0.32_{-0.18}^{+0.2}$ & $0.15_{-0.04}^{+0.03}$ & $42.48_{-0.02}^{+0.02}$ &4 \\
3  & 11/21  & 2.72 & 4.36E+7 & $1.91_{-0.07}^{+0.09}$ & $<21$ & $0.31_{-0.13}^{+0.15}$ & $< 0.2$ & $44.32_{-0.01}^{+0.01}$ & 4 \\
4 &  56/37  & 2.76 & 62.5 & $2.06_{-0.04}^{+0.05}$ & $<20.8$ & $0.06_{-0.04}^{+0.04}$ & $-$ & $44.45_{-0.01}^{+0.01}$ & 3 \\
5  &  56/38  & 1.79E+19  & 333.33 & $1.49_{-0.04}^{+0.04}$ & $21.30_{-0.09}^{+0.08}$ & $0.07_{-0.04}^{+0.04}$ & $-$ & $44.12_{-0.01}^{+0.01}$ & 3 \\
6  &  12/20  & 2.72 & 2.73 & $1.86_{-0.06}^{+0.06}$ & $<21.2$ & $<0.05$ & $0.17_{-0.12}^{+0.09}$ & $44.35_{-0.02}^{+0.01}$ & 4 \\
7  &  40/34  & 439.22 & 27.03 & $1.65_{-0.05}^{+0.05}$ & $20.93_{-0.27}^{+0.17}$ & $0.06_{-0.05}^{+0.05}$ & $-$ & $44.11_{-0.01}^{+0.01}$ & 3 \\
8  & 46/46 &  2.72 & 3.46 & $1.67_{-0.03}^{+0.03}$ & $<20.29$ & $<0.09$ & $-$ & $43.58_{-0.01}^{+0.01}$ & 1 \\
9  & 32/24 &  2.72 & 66.67 & $1.92_{-0.08}^{+0.09}$ & $<21.2$ &$0.09_{-0.06}^{+0.07}$ & $0.14_{-0.07}^{+0.11}$ & $44.22_{-0.02}^{+0.01}$ & 4 \\
10  & 30/22 &  2.19E+4 & 5.89 & $1.87_{-0.07}^{+0.07}$ & $20.97_{-0.20}^{+0.14}$ & $0.06_{-0.06}^{+0.09}$ & $-$ & $43.94_{-0.02}^{+0.01}$ & 2 \\
11  & 45/25 &  2.8 & 7.87 & $2.07_{-0.04}^{+0.08}$ & $<20.3$ & $0.31_{-0.30}^{+0.38}$ & $-$ & $41.94_{-0.02}^{+0.03}$ & 3 \\
12  & 28/25 &  1502.67 & 2.72 & $2.01_{-0.06}^{+0.06}$ & $20.74_{-0.23}^{+0.20}$ & $<0.07$ & $-$ & $43.71_{-0.02}^{+0.02}$ & 2 \\
13  & 34/29 &  2.72 & 3.7 & $1.81_{-0.04}^{+0.04}$ & $<20.3$ & $<0.09$ & $-$ & $43.81_{-0.02}^{+0.01}$ & 1 \\
14  & 19/21 &  2.76 & 10.75 & $1.94_{-0.05}^{+0.08}$ & $<20.9$ & $0.06_{-0.04}^{+0.05}$ & $-$ & $44.31_{-0.02}^{+0.01}$ & 3 \\
15  & 39/31 & 2.82 & 5.71 & $1.92_{-0.03}^{+0.07}$ & $<20.3$ & $0.10_{-0.09}^{+0.25}$ & $-$ & $43.05_{-0.02}^{+0.02}$ & 1 \\
16  & 17/18 &  2.72 & 5.41 & $1.93_{-0.08}^{+0.13}$ & $<20.1$ & $<0.23$ & $0.14_{-0.04}^{+0.03}$ & $43.09_{-0.02}^{+0.02}$ & 4 \\
17  & 20/22  & 314.19 & 2.72 & $1.73_{-0.07}^{+0.07}$ & $20.67_{-0.24}^{+0.21}$ & $<0.12$ & $-$ & $43.05_{-0.02}^{+0.02}$ &2 \\
18  & 44/21 &  2.72 & 2.73 & $2.04_{-0.05}^{+0.07}$ & $<20.65$ & $<0.09$ & $< 0.2$ & $43.42_{-0.02}^{+0.02}$ & 4 \\
19  & 28/27 &  2.02E+10 & 2.25 & $1.86_{-0.06}^{+0.07}$ & $20.95_{-0.11}^{+0.12}$ & $<0.12$ & $-$ & $43.19_{-0.02}^{+0.03}$ & 2 \\
20  & 43/33 &  3.14E+4 & 2.72 & $1.67_{-0.06}^{+0.06}$ & $20.90_{-0.17}^{+0.18}$ & $<0.16$ & $-$ & $43.43_{-0.02}^{+0.02}$ &  2 \\
28  & 14/22 &  2.72 & 250 & $1.88_{-0.12}^{+0.12}$ & $<21.16$ & $0.16_{-0.14}^{+0.15}$ & $0.13_{-0.05}^{+0.05}$ & $43.42_{-0.03}^{+0.02}$ & 4 \\
33  & 26/22 &  3944.19 & 2.72 & $2.02_{-0.11}^{+0.11}$ & $20.87_{-0.12}^{+0.25}$ & $<0.27$ & $-$ & $42.82_{-0.03}^{+0.04}$ & 2 \\
36  & 20/24 &  2.72 & 5.02 & $2.22_{-0.08}^{+0.08}$ & $<20.85$ & $0.08_{-0.07}^{+0.26}$ & $-$ & $43.48_{-0.03}^{+0.03}$ & 1 \\ \bottomrule 
\end{tabular}
\end{table*}

\section{Correlation analysis of the spectral parameters}
\label{analysis}
To investigate the existence of physical correlations in our AGN sample, we performed a correlation analysis on their main spectral properties. In order to incorporate sources with upper-limits in our ultra-narrow pencil beam survey, we employed a Monte-Carlo (MC) approach based on the linear regression algorithm proposed by  \citet{Bianchi2007}. The steps involved in the analysis are as follows:

\begin{enumerate}

    \item For each source, we generated a set of 1000 random values for the y-axis variable (e.g. $N_H$ or Fe-K$\alpha$).  For sources with upper-limits, we used a uniform distribution ranging from 0 to the upper-limit, while for the remaining sources, we used a normal distribution with their errors as the standard deviation. 
    
    \item  We computed a least-square linear regression fit for each set of the simulated data, considering one of the physical parameters of our AGN sample as the x-axis variable (e.g. $z$ or $L_x$). The best fit and its statistical uncertainty will be determined as the mean and standard deviation, respectively, of the 1000 linear regressions.   

    \item To assess the strength of the correlations, we calculated the Spearman Rank Coefficient $S$ and the $p_{value}$ at the 95\% of confidence level to determine the statistical significance of any observed correlation \citep{Zwillinger2000}. 

\end{enumerate}

The distributions obtained through the MC procedure reveal interesting relationships among the spectral properties of our sample, which will be presented next.

\subsection{Iwasawa-Taniguchi effect}
\label{Iwa-Ta}
\begin{figure}
\centering
\includegraphics[scale=0.27]{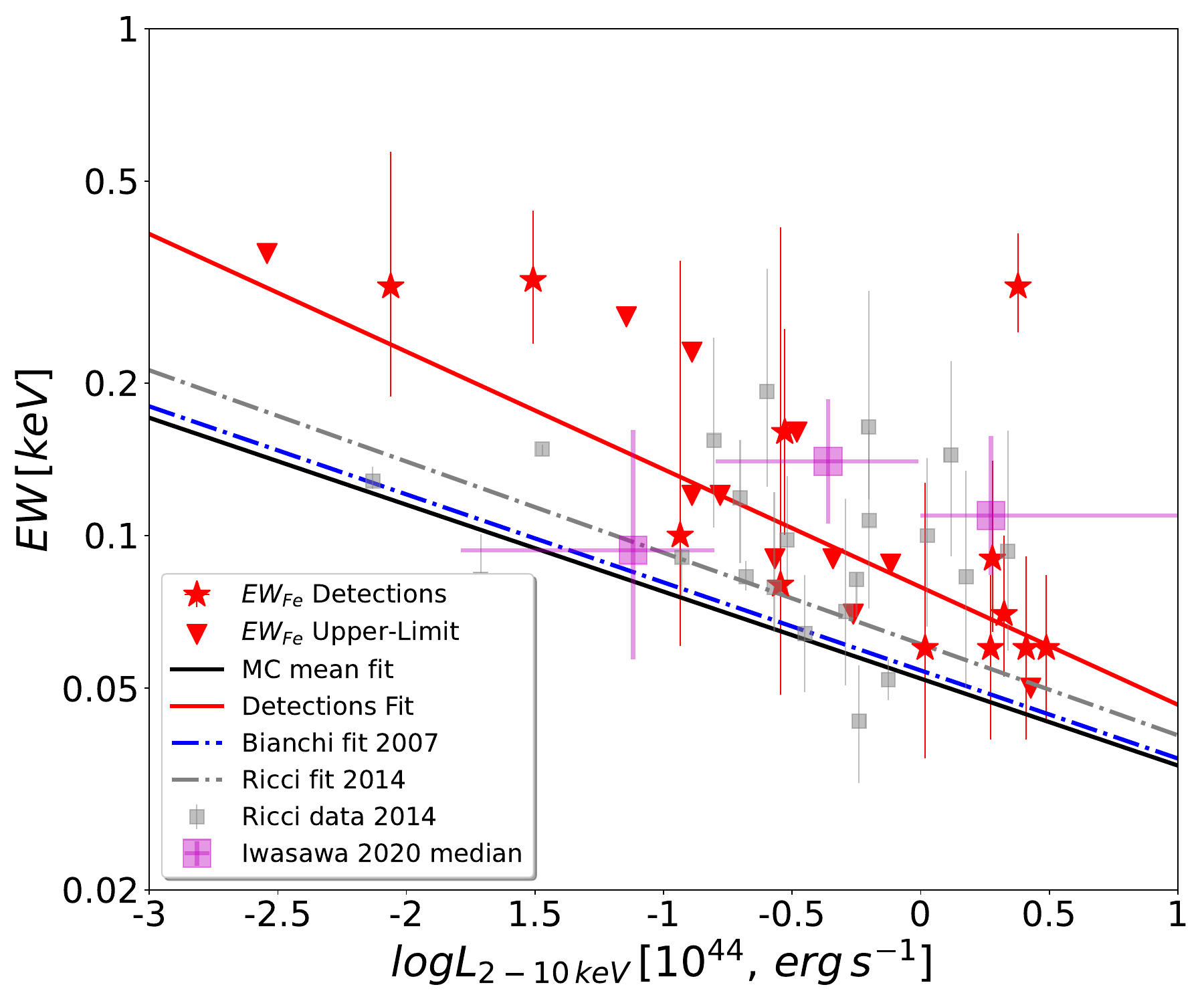}
\caption{Fe-K$\alpha$ Equivalent Width vs. the X-ray luminosity at $2-10\, \rm{keV}$ energy band of our AGN sample. The black and red lines represent the best MC linear fit for the whole sample and the Fe-K$\alpha$ detections (red star), respectively. We included the data and results of previous works of \citet{Bianchi2007} (blue line), \citet{Ricci2014} (gray line and points), and \citet{Iwasawa2020} (pink squares).}
\label{Lx_vs_EWb}
\end{figure}

In Figure \ref{Lx_vs_EWb} we present the distribution of the rest-frame X-ray luminosity at $2 - 10\, \rm{keV}$ energy band in units of $10^{44}\, \rm{erg}\, \rm{s}^{-1}$ vs. $EW_{Fe}$ of our sample in $\rm{keV}$. We found an anti-correlation with a flat slope of $m = -0.17 \pm 0.08$ and Spearman rank of  $S = - 0.21$ with low significance $p_{value} > 0.05$. However, when excluding sources with $EW_{Fe}$ upper-limits (red triangles), the Spearman rank increases to $S = - 0.6$ with a high significance of $p_{value} < 0.05$. The best fit for the whole sample (black line) and only $EW_{Fe}$ detections (red line) are expressed in equation \ref{EWall} and \ref{EWdet}, respectively: 

\begin{align}
    \log \left(\frac{EW_{Fe}}{keV} \right )_{all} =  (-1.28 \pm 0.08) + (-0.17 \pm 0.08)  \log \left(\frac{Lx}{10^{44} \rm{erg}\, s^{-1}}\right) \label{EWall} \\
    \log \left(\frac{EW_{Fe}}{keV} \right)_{det} = (-1.10\pm 0.06) + (-0.23 \pm 0.07)  \log \left(\frac{Lx}{10^{44} \rm{erg}\, s^{-1}}\right) \label{EWdet}
\end{align} 

Our results agree with the observed $L_x-EW_{Fe}$ anti-correlation reported in previous studies. For instance, \citet{Bianchi2007} found a similar trend for nearby AGNs (blue dotted line), while \citet{Ricci2014} observed this effect in two samples of Seyfert-2 and Seyfert-1 galaxies (grey points and dotted line) with a slope of $m = 0.18 \pm 0.06$. They also included upper-limits with their best fits in agreement with our results. 

The underlying physical mechanism responsible for the  “Iwasawa-Taniguchi effect” \citep{Iwasawa1993}, remains unknown. One potential explanation is that brighter AGNs may induce higher degrees of ionization in the surrounding material, resulting in the fading of the fluorescence Fe-K$\alpha$ line emitted by low-ionization matter \citep{Shu2010}. Furthermore, the iron emission originating from cooler gas near the supermassive black hole could undergo scattering due to radiation pressure or thermal dissipation \citep[e.g.][]{Fabian2008}. Alternatively, it is plausible that the decrease in the torus covering factor and/or column density of the cold gas responsible for the iron emission is correlated with an increase in AGN luminosity. This suggests that as the luminosity of the AGN rises, the covering factor and/or column density of the cold gas in the torus declines \citep{Bianchi2007, Ricci2014}.  

\subsection{Intrinsic Column Density versus redshift}

Observational results and synthesis models have shown that obscured AGNs represent a significant fraction of the entire AGN population, which increases with redshift \citep{Amato2020}. For instance, \citet{Gilli2007} and \citet{Burlon2011} estimated that the fraction of Compton-thick AGNs $(N_H \ge 1.5 \times 10^{24}\, cm^{-2})$ in the local universe comprises about 20\% to 30\% of the total AGN population. 

\begin{figure}
\centering
\includegraphics[scale=0.27]{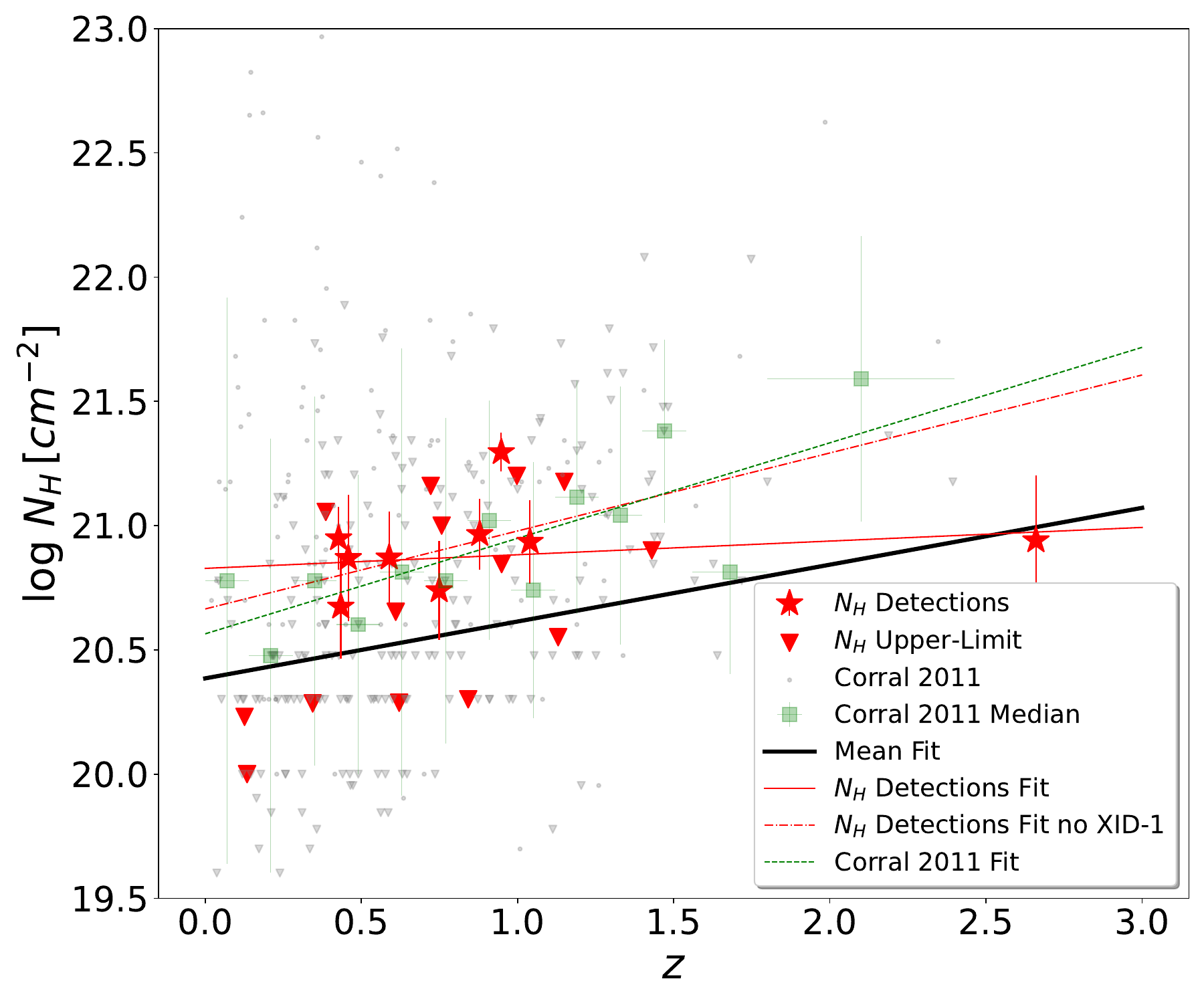} 

\caption{Redshift vs. intrinsic column density absorption of our bright AGN sample. The black and red solid lines correspond to the best fit for the whole sample and only the $N_H$ detections (red star), respectively. The gray points and green squares refer to the XBS AGN sample reported by  \citet{Corral2011}  with its corresponding best fit (green dark dashed line). The red dotted line corresponds to the best fit of the $N_H$ detections after removing the XID-1 source at $z = 2.66$.}
\label{z_vs_Nh}
\end{figure}

In our analysis presented in Figure \ref{z_vs_Nh}, we did not find a significant correlation ($p_{value}>0.05$) with Spearman rank of $S \approx 0.23$ between the intrinsic absorption $N_H$ of our AGN sample as a function of the redshift. This result could be due to the limited redshift range covered for our data. For instance, our AGN sample includes only one high-redshift source, the XID-1 at $z = 2.66$. Due to the lack of more sources to sample this high-redshift regime, the best fit for the hole sample (black solid line) and the $N_H$ detection (red solid line) might be dominated by this AGN, resulting in misleading results. In the case of removing the highest redshift source, we found a moderate ($S = 0.45$) but still low significant ($p_{value}>0.05$) correlation where the linear fit (red-dotted line) increases from a flat $m \approx 0.05$ to a steeper slope $m = 0.31$, expressed by the equation:

\begin{equation}
    \label{z_NH}
        \log(N_H) =  (20.66 \pm 0.16) + (0.31 \pm 0.23) z 
\end{equation}

We compared our results with a set of AGNs detected in the XMM-Newton Bright Serendipitous Survey \citep[XBS,][]{Corral2011}. The XBS is composed of 305 AGNs (grey dots) detected in a sky coverage of $28.1\, \deg^2$ and flux limit of $7 \times 10^{-14}\, \rm{erg}\, \rm{cm}^{-2}\, \rm{s}^{-1}$ \citep{Della2004}. Most of the AGNs detected in the XBS are distributed in the range of $z=0 - 1.5$ (293 out of 305) with bright X-ray sources similar to our survey. To analyze the XBS catalog, we computed the median and standard deviation in 13 redshift bins of $\Delta z =  0.14$ with at least 15 elements per bin, except for the last two bins at $z>1.5$ with  $\Delta z =  0.24$ and $0.6$, respectively, both with 6 elements each. We found a strong and statistically reliable correlation (green dotted line) with $S = 0.761$ and $p_{value} = 0.003$ that agrees with the linear fit obtained from our AGNs with intrinsic absorption, after removing the source with the highest redshift. 

The observed column densities could be associated with various factors, such as the presence of dust lanes in the host galaxy or underlying presence of BAL QSOs. For sources at higher redshift \citep[$z>3$, see][]{Gilli2022}, the underlying physical mechanism driving this behavior could be related to the increase of the reservoir of gas available towards early cosmic epochs, leading to an evolution of the fraction of obscured AGNs, as reported by \citet{Liu2017} and \citet{Iwasawa2020} with the 7Ms Chandra Deep Field-South Survey (CDFS).

We investigated some other physical relationships, such as $EW_{Fe}$ vs. $N_H$, $L_x$ vs. $N_H$, and $EW_{Fe}$ vs. redshift. We observed no significant correlation between $EW_{Fe}$ and $N_H$, as well as between  $L_x$ and $N_H$. However, we detected a significant anti-correlation between $EW_{Fe}$ and $z$, which could be explained by the $EW_{Fe}-L_x$ relation presented in Section \ref{Iwa-Ta}, i.e. at higher redshifts, we observe intrinsically more luminous sources with lower $EW_{Fe}$, following the Iwasawa-Taniguchi effect.

\section{X-ray Variability and Black Hole Mass-Luminosity Relation}
\label{vary}

\subsection{The Normal Excess Variances}

Since we are working with multi-epoch observations, we can study how the X-ray variability of our sample evolves as a function of their SMBH mass. Following \citet{Lanzuisi2014}, we used the Normal Excess Variances $\sigma^2_{rms}$  to measure the amplitude flux variation of our AGN sample, as follows: 

\begin{equation}
\label{sigma2}
\sigma^2_{rms} = \frac{1}{(N_{obs} -1)\bar{x}^2}\sum^{N_{obs}}_{i=1} (x-\bar{x})^2 - \frac{1}{N_{obs} \bar{x}^2}\sum^{N_{obs}}_{i=1} \sigma^2_{err,i}
\end{equation}

where $N_{obs}$ is the number of observations, $\bar{x}$ is the average flux of the source in the full survey, and $x_i$ is the individual flux per observation with error $\sigma_{err,i}$. The Normal Excess error $err(\sigma^2_{rms})$ is defined as follows: 

\begin{equation}
\label{sigma2error}
    err(\sigma^2_{rms}) = \sqrt{\left( \sqrt{\frac{2}{N_{obs}} \frac{\overline{\sigma^2_{err}}}{\bar{x}^2}} \right)^2 + \left( \sqrt{\frac{\overline{\sigma^2_{err}}}{N_{obs}} \frac{2F_{var}}{\bar{x}}} \right)^2 }
\end{equation}

where $F_{var} = \sqrt{\sigma^2_{rms}}$ is the fractional variability  and the component $\overline{\sigma^2_{err}} = \frac{1}{N_{obs}}\sum^{N}_{i=1} \sigma^{2}_{err,i}$ is the mean square error. For those sources that $err(\sigma^2_{rms}) > \sigma^2_{rms}$, we will use only their upper limit, which is defined as $\sigma^2_{rms,UL}= \sigma^2_{rms} + err(\sigma^2_{rms})$.

To include those sources with upper limits, we employed the same MC linear regression simulation, as described before. Our results are presented in Table \ref{spec_par}. We found an average Excess Variance for the whole bright AGN sample of $\sigma^2_{rms} = 0.091 \pm 0.028$. 

We did not detect a relation between $\sigma^2_{rms}$ and the X-ray counts. This suggests that our sample possesses a sufficient number of counts to have a statistically reliable measure of the excess variance. For example, \citet{Lanzuisi2014} found an anti-correlation between $\sigma^2_{rms}$ and the average X-ray counts. They reported that this anti-correlation arises due to a selection effect that disappears within their brightest AGN sample.

\subsection{The BH mass estimation}

Since we do not possess optical spectroscopic data for most of our sources, we do not have measurements of optical lines such as MgII $\lambda$2798 \AA\,, CIV $\lambda$1549 \AA\,, H$\beta$ or H$\alpha$. Therefore, we decided to use two indirect methods to estimate the SMBH mass of our AGN sample and subsequently compare the results. 

\subsubsection{Estimating the BH mass from X-ray luminosity, $M_{BH,L_x}$}
\label{MBHLx}

The first method involved the X-ray luminosity in the $2-10\, \rm{keV}$ energy band with Equation \ref{L5100} to estimate the optical continuum at 5100 \AA\, ($L_{5100}$). This equation was derived by \citet{Netzer2019}, from the tight correlation between the UV and X-ray luminosities $\alpha_{OX}$ and arises from theoretical calculations of optically thick, geometrically thin accretion disks, and observations of X-ray properties in type-1 AGN. According to the estimated column densities, we can classify our sources into this class.

\begin{equation}
\label{L5100}
    \log(L_{5100}) = 1.4\times \log(L_{2-10\, \rm{keV}}) - 16.8
\end{equation}

Then, we estimated the BH masses ($M_{BH,L_x}$) for the whole sample using the $M_{BH} - L_{5100}$ empirical relation described in Equation \ref{Ml}, which was originally reported by \citet{Peterson2004} based on black hole mass measurements using reverberation analysis in AGNs. We found a mean and standard deviation of  $\log(M_{BH,L_x}/M_{\sun}) = 7.59 \pm 0.59$ (see Table \ref{spec_par}).  

\begin{equation}
\label{Ml}
    \log \left (\frac{M}{10^8\, M_{\sun}} \right) = -0.12(\pm 0.07) + 0.79(\pm 0.09) \log \left (\frac{L_{5100}}{10^{44}\, \rm{erg}\, \rm{s}^{-1}} \right)
\end{equation}

\subsubsection{Estimating the BH mass from X-ray scaling, $M_{BH,X}$}
\label{MBHX}

For the second method, we used the X-ray scaling estimation technique \citep{Gliozzi2011,Gliozzi2021}. This approach scaled the spectral parameters of our sample with a set of reference sources with known mass and distance, allowing us to estimate the AGNs black hole mass. \citet{Gliozzi2011} tested this method using a set of AGN black hole masses estimated from reverberation mapping. They reported a strong agreement between their estimation and those from reverberation mapping.

During our analysis, we have already fitted our spectra with a combination of a simple power law, a Galactic absorption, a free-to-vary intrinsic absorption, and a narrow gaussian at fixed rest-frame energy of 6.4 $\rm{keV}$ for an iron line. However, to estimate $M_{BH}$ using the X-ray scaling method, we need to refit our spectra while including the Bulk Motion Comptonization model (BMC). The BMC model is designed for modeling X-ray spectra of accreting black holes and consists of the convolution of thermal seed photons producing a power law. This model comprises four parameters, the BMC normalization $N_{BMC}$, the photon temperature $kT$, the spectral index $\alpha  = \Gamma_{BMC} - 1$ where $\Gamma_{BMC}$ is the photon index, and the parameter $\log(A)$ is the so-called “illumination factor” related to the fraction of scattered seed photons $f$, i.e. $f$ is the ratio between the number of Compton scattered photons and the number of seed photons, described by the equation $f = A/(1+A)$ \citep{Shrader2003,Farinelli2008,Shaposhnikov2009,Williams2023}. 

The considerations that support this method for accreting compact objects, presented by \cite{Shaposhnikov2009} and \cite{Gliozzi2011}, can be summarized as follows: 1) The break frequency X-ray variability of the power spectrum is inversely proportional to the black hole mass. 2) The BMC normalization is proportional to the distance and luminosity, i.e. $N_{BMC} \propto L/d^2$. 3) The luminosity of an accreting BH can be expressed as $L \propto \eta M_{BH}\dot{m}$, where $\eta$ describe the radiative efficiency and $\dot{m}$  the accretion rate. 4) $\Gamma$ is a reliable indicator for the source’s spectral state regardless of the BH mass \citep{Shaposhnikov2009}.

Following the methodology described in \citet{Gliozzi2021}, the black hole mass is estimated in three steps. First, we have to estimate $N_{BMC}$ from our refitting process, which is performed in the 2–10 keV energy range to avoid the complexity associated with the soft X-ray band, such as the soft excess (e.g. observed in 7 sources) and the potential presence of warm absorbers. We used the best fit obtained from Section \ref{fitting}, however, we replaced the \texttt{powerlaw} component with the BMC model (e.g. \texttt{tbabs*BMC*zphabs}). 

The BMC parameters are free-to-vary ($N_{BMC}$, $\log(A)$, $\alpha$), except for $kT$, which was fixed to 0.1 $\rm{keV}$ based on the result obtained by \citet{Gliozzi2011} with their set of AGNs, and we used the value of $\log(A)$ from the first fit iteration. They reported that the parameters $kT$ and $\log(A)$ have a negligible effect on the estimation of $M_{BH}$.

The second step consists of computing the BMC normalization of the reference sources ($N_{BMC,r}$)  with Equation \ref{Nbmc}. These reference sources served as calibrations and comprised Galactic stellar-mass black holes with known masses and distances.

\begin{equation}
\label{Nbmc}
N_{BMC,r}(\Gamma_{BMC}) = N_{tr}\times \left (1-ln \left [e^{\frac{a-\Gamma_{BMC}}{B}} -1 \right ] \right )^{1/\beta}
\end{equation}

where $\Gamma_{BMC}$ is obtained from the spectral index $\alpha$, while  $a$, $B$, $N_{tr}$, and $\beta$ are the reference sources patterns reported by \citet{Gliozzi2011} and presented in their Table 2. Finally, we used the Equation \ref{Msp}, to estimate the Black Hole masses of our sample. 

\begin{equation}\label{Msp}
    M_{BH,t} = M_{BH,r}\times \left(\frac{N_{BMC,t}}{N_{BMC,r}} \right) \times \left(\frac{d_t}{d_r}\right)^2
\end{equation}

where $d$ is the distance and the $t$ and $r$ subscripts denote the target and the reference source, respectively. The best estimation of $M_{BH}$ will be the average of the masses inferred from all the available reference sources. Using the X-ray scaling method, we obtained a mean SMBH mass of $\log(M_{BH,X}/M_{\sun}) = 7.26 \pm  0.68$ for the entire AGN sample.

\begin{figure}
\centering
\includegraphics[scale=0.27]{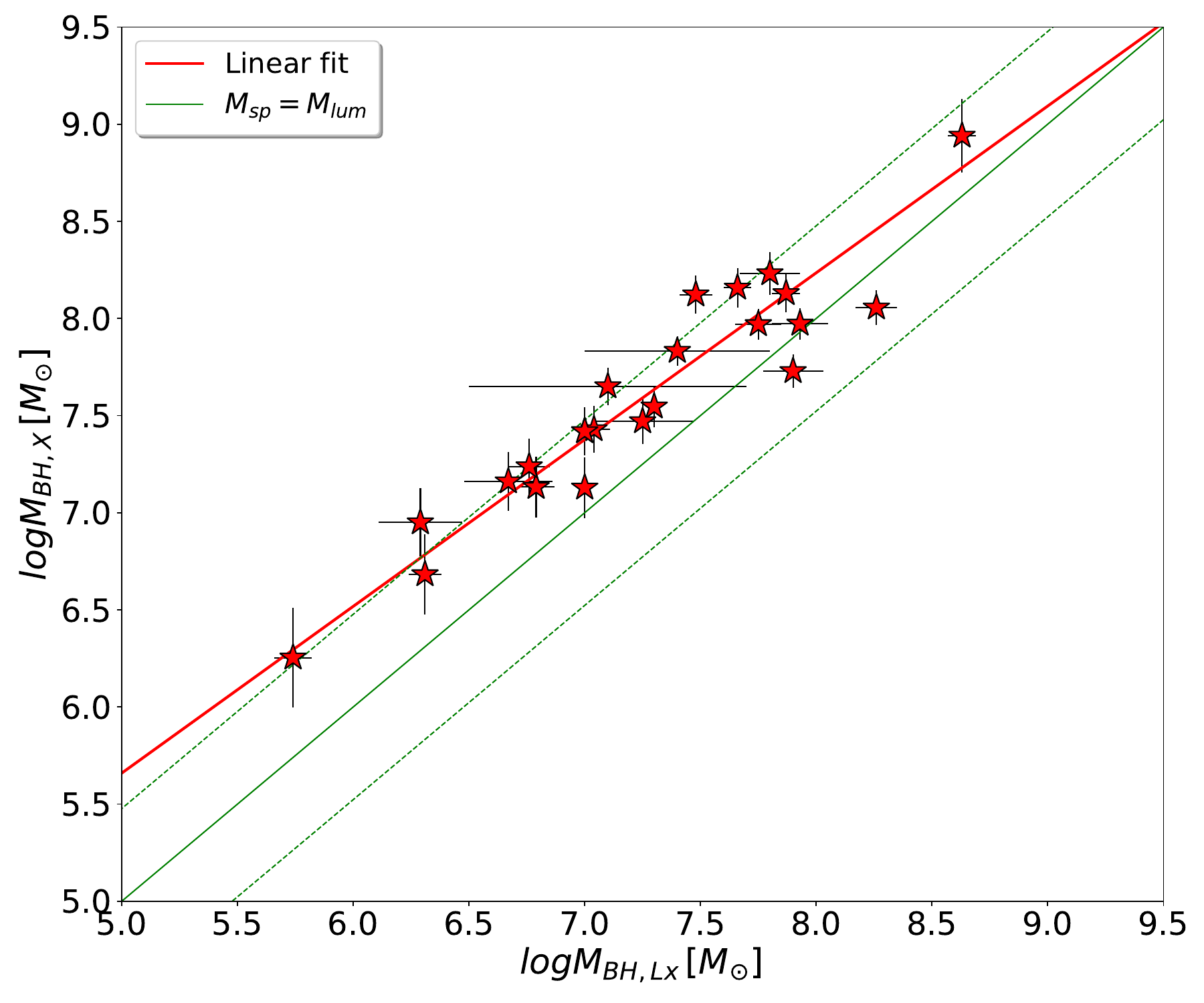}
\caption{BH masses comparison between the X-ray scaling method (Section \ref{MBHLx}) vs. the luminosity method (Section \ref{MBHX}). The red and green solid lines represent the best linear fit and a perfect one-to-one correspondence, respectively, while the dashed green lines indicate the ratios of 3 and 1/3.}
\label{Ms_Ml}
\end{figure}

Figure \ref{Ms_Ml} presents a comparison between $M_{BH,L_x}$ and $M_{BH,X}$, where we observed a trend of lower masses for the luminosity method, described as $M_{BH,L_x} \sim 0.33\times M_{BH,X}$. The red and green solid lines represent the best fit and the ideal case when $M_{BH,X} = M_{BH,L_x}$, respectively. A summary of our spectral results and the BH mass estimations are provided in Table \ref{spec_par} (see also  \citet{Mathur2001} for measuring $M_{BH,X}$). 

\begin{table*}
\centering
\caption{Summary table of the spectral parameters with the BMC model and the black hole masses estimated with both methods.}
\label{spec_par}
\begin{tabular}{ccccccccc}
\hline
XID & $\Gamma_{BMC}^{a}$ & $\log(A)$ & $N_{BMC}^{b}$ & $\chi^2/dof$ & $\log(L_{5100})^{c}$ & $\log(M_{BH,X})^{d}$ & $\log(M_{BH,L_x})^{e}$ & $\sigma^2_{rms}$\\ 
 & & & $(10^{-6})$ &   & $(\rm{erg}\, \rm{s}^{-1})$ & $(\rm{erg}\, \rm{s}^{-1})$ & $(M_{\sun})$ & $(M_{\sun})$ \\ \midrule
1 & 1.82 $\pm$ 0.04 & 5.44  & 0.96 $\pm$ 0.03 & 39/41  & 46.7 & 8.63 $\pm$ 0.06 & 8.94 $\pm$ 0.19 & 0.031 $\pm$ 0.004 \\
2 & 1.77 $\pm$ 0.07 & 1.17 & 0.47 $\pm$ 0.04 & 11/5 &  42.7 & 6.31 $\pm$ 0.07 & 6.68 $\pm$ 0.21 & 0.370 $\pm$ 0.032 \\
3 & 1.74 $\pm$ 0.04 & 0.98 & 0.68 $\pm$ 0.02 & 20/6 & 45.2 & 7.87 $\pm$ 0.06 & 8.13 $\pm$ 0.10 & 0.011 $\pm$ 0.017 \\
4 & 2.00 $\pm$ 0.05 & -1.41  & 0.48 $\pm$ 0.16 & 43/24 &  45.4 & 7.80 $\pm$ 0.13 & 8.23 $\pm$ 0.11 & 0.020 $\pm$ 0.006 \\
5 & 1.55 $\pm$ 0.06 & 1.11 & 0.31 $\pm$ 0.01 & 35/22 &  45 & 7.93 $\pm$ 0.12 & 7.97 $\pm$ 0.08 & 0.020 $\pm$ 0.007 \\
6 & 1.84 $\pm$ 0.04 & 7.04  & 0.28 $\pm$ 0.01 & 9/10 &  45.3 & 7.66 $\pm$ 0.06 & 8.16 $\pm$ 0.10 & 0.016 $\pm$ 0.015 \\
7 & 1.62 $\pm$ 0.08 & 0.87  & 0.55 $\pm$ 0.02 & 34/20 &  45 & 7.75 $\pm$ 0.10 & 7.97 $\pm$ 0.08 & 0.067 $\pm$ 0.018 \\
8 & 1.66 $\pm$ 0.06 & 0.90  & 0.20 & 25/23 & 44.2 & 7.3 & 7.55 $\pm$ 0.11 & 0.364 $\pm$ 0.059 \\
9 & 1.78 $\pm$ 0.06 & -0.56  & 1.23 $\pm$ 0.35 & 16/11 &  45.1 & 8.26 $\pm$ 0.09 & 8.06 $\pm$ 0.09 & 0.049 $\pm$ 0.022 \\
10 & 1.81 $\pm$ 0.06 & 1.12  & 0.22 & 16/7 &  44.7 & 7.4 $\pm$ 0.4 & 7.83 $\pm$ 0.08 & 0.156 $\pm$ 0.031 \\
11 & 1.83 $\pm$ 0.11 & 1.39 & 0.16 $\pm$ 0.03 & 10/5 &  41.9 & 5.74 $\pm$ 0.08 & 6.25 $\pm$ 0.26 & 0.032 $\pm$ 0.022 \\
12 & 2.03 $\pm$ 0.07 & 7.22  & 0.21 & 9/8 &  44.4 & 7.1 $\pm$ 0.6 & 7.65 $\pm$ 0.10 & 0.028 $\pm$ 0.016 \\
13 & 1.74 $\pm$ 0.06 & -0.49  & 0.63 $\pm$ 0.34 & 13/11 &  44.5 & 7.90 $\pm$ 0.13 & 7.73 $\pm$ 0.09 & 0.078 $\pm$ 0.027 \\
14 & 1.94 $\pm$ 0.04 & 7.34 & 0.18 $\pm$ 0.02 & 17/11 &  45.2 & 7.48 $\pm$ 0.07 & 8.12 $\pm$ 0.10 & 0.070 $\pm$ 0.024 \\
15 & 1.74 $\pm$ 0.09 & 0.75 & 0.23 $\pm$ 0.03 & 7/7 &  43.5 & 6.79 $\pm$ 0.08 & 7.13 $\pm$ 0.16 & 0.448 $\pm$ 0.175 \\
16 & 1.95 $\pm$ 0.10 & 1.25 & 0.23 $\pm$ 0.03 & 3/2 &  43.5 & 6.67 $\pm$ 0.19 & 7.16 $\pm$ 0.15 & 0.040 $\pm$ 0.023 \\
17 & 1.76 $\pm$ 0.08 & 0.98 & 0.13 & 8/4 &  43.5 & 7.0 & 7.13 $\pm$ 0.16 & 0.058 $\pm$ 0.024 \\
18 & 1.73 $\pm$ 0.08 & 0.99  & 0.14 & 16/6 &  44 & 7.0 & 7.42 $\pm$ 0.12 & 0.030 $\pm$ 0.023 \\
19 & 1.90 $\pm$ 0.07 & 6.62  & 0.20 $\pm$ 0.05 & 14/7 &  43.7 & 6.76 $\pm$ 0.09 & 7.24 $\pm$ 0.14 & 0.041 $\pm$ 0.019 \\
20 & 1.79 $\pm$ 0.07 & 1.04  & 3.03 $\pm$ 0.01 & 14/13 &  44 & 7.04 $\pm$ 0.07 & 7.43 $\pm$ 0.12 & 0.018 $\pm$ 0.010 \\
28 & 1.78 $\pm$ 0.10 & 0.82  & 0.10 & 10/8 &  44 & 7.0 & 7.42 $\pm$ 0.12 & 0.030 $\pm$ 0.017 \\
33 & 2.07 $\pm$ 0.15 & 4.94  & 0.09 $\pm$ 0.04 & 5/6 & 43.2 & 6.29 $\pm$ 0.18 & 6.95 $\pm$ 0.18 & 0.047 $\pm$ 0.025 \\
36 & 1.97 $\pm$ 0.15 & 1.62 & 0.20 $\pm$ 0.15 & 9/7 &  44.1 & 7.25 $\pm$ 0.22 & 7.47 $\pm$ 0.12 & 0.156 $\pm$ 0.047 \\
\bottomrule
\end{tabular}

\begin{flushleft}
\footnotesize 
$^a$ Related to the spectral index $\alpha$ with the equation  $\alpha = \Gamma_{BMC} - 1$.

$^b$ BMC normalization with uncertainties at 1 sigma of confidence.

$^c$ Inferred optical continuum at 5100 \AA, estimated from Equation \ref{L5100}.

$^d$ AGN black hole masses computed with equation \ref{Ml}  as a function of $L_{5100}$.

$^e$ AGN black hole masses computed with the spectral parameters including the BMC model with Equation \ref{Msp}.

\end{flushleft}

\end{table*}

It is important to highlight that the masses estimated in this paper are based on spectral and luminosity measurements. Therefore, caution should be taken when interpreting the outcome of the relations derived for BH mass with the other parameters in the following sections.

\subsection{Normal Excess Variance vs. BH mass and Luminosity}

In this section, we studied how and if the X-ray variability of our AGN sample correlates with the X-ray luminosity and black hole mass. For this analysis, we did not include the source XID-11 because it is an outlier due to its lower mass, i.e. $\log(M_{BH,X}/M_{\sun}) < 6$. 

\begin{figure}
\centering
\includegraphics[scale=0.27]{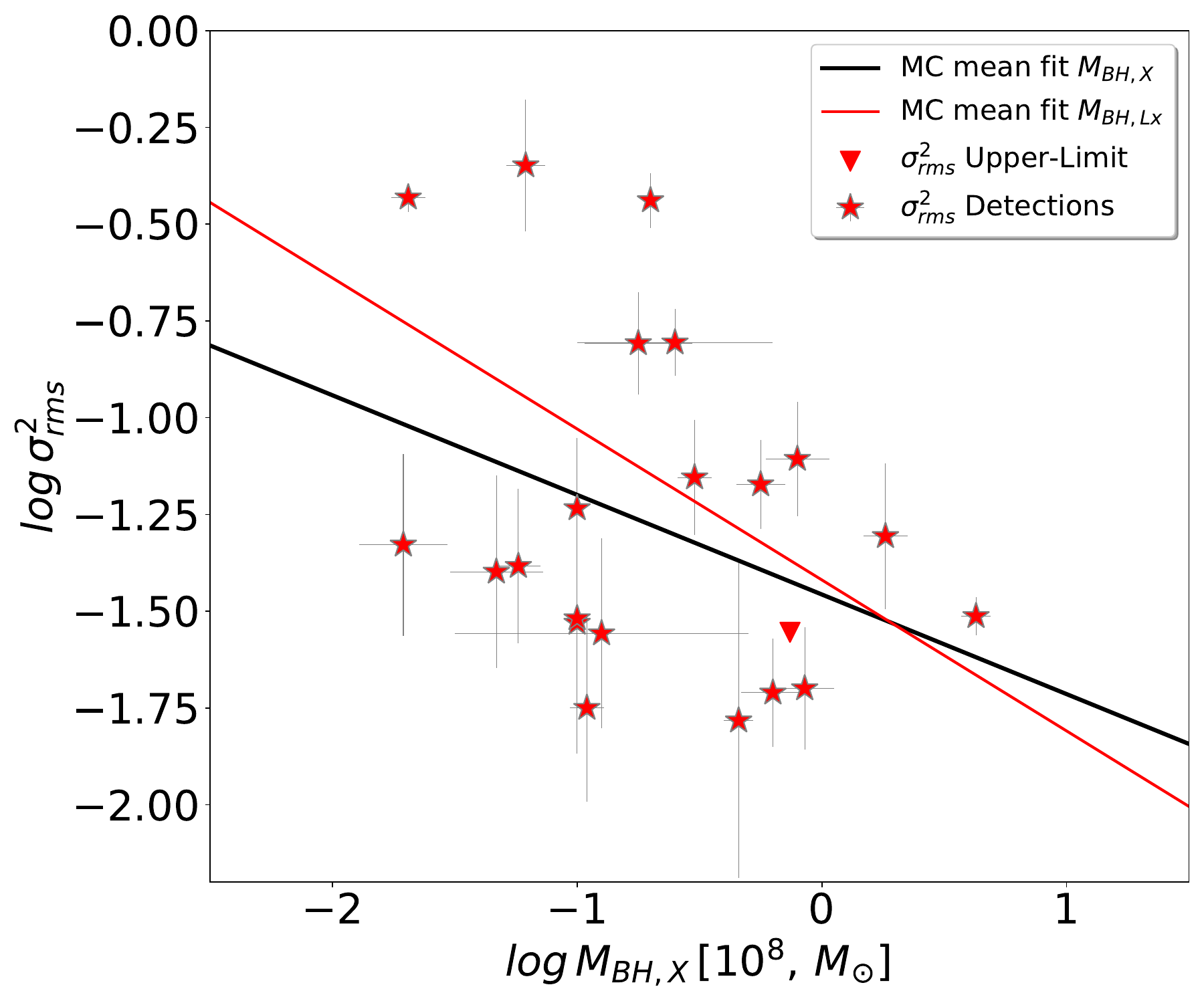}
\caption{Distribution of $\sigma^2_{rms}$ vs. $M_{BH,X}$ of our AGN sample. The best MC linear fit is presented with the black solid line, while the red line is obtained with $M_{BH,L_x}$.} 
\label{sigma_MBH}
\end{figure}

In Figure \ref{sigma_MBH}, we present the Normal Excess Variance vs. the Black Hole masses of our sample. The red and black solid lines represent the best MC linear fit using $M_{BH,L_x}$ (Equation \ref{MBH_Lx}) and $M_{BH,X}$ (Equation \ref{MBH_X}), respectively. In both cases, we found an anti-correlation with a reliable confidence level ($p_{value} \approx 0.05$). For the $\sigma^{2}_{rms} - M_{BH,L_x}$ relation we obtained a slope of $m = -0.39 \pm 0.06$ and  Spearman rank coefficient of $S =  -0.34 $, while the  $\sigma^{2}_{rms} - M_{BH,X}$ relation exhibits a flatter slope of $m = -0.26 \pm 0.05$ and $S =  -0.26$.  

\begin{align} 
\log(\sigma^{2}_{rms}) &= (-1.42 \pm 0.04) + (-0.39 \pm 0.06)\log \left ( \frac{M_{BH,L_x}}{10^8\, M_{\sun}} \right) \label{MBH_Lx} \\
\log(\sigma^{2}_{rms}) &= (-1.46 \pm 0.05) + (-0.26 \pm 0.05)\log \left (\frac{M_{BH,X}}{10^8\, M_{\sun}} \right)
\label{MBH_X}
\end{align}

Since our results are consistent with both methods and to reduce potential biases, for the remaining analysis, including the study of the Eddington ratio distribution in Section \ref{Eddington}, we will use only the masses estimated from the X-ray Scaling Method. 

\begin{figure}
\centering
\includegraphics[scale=0.27]{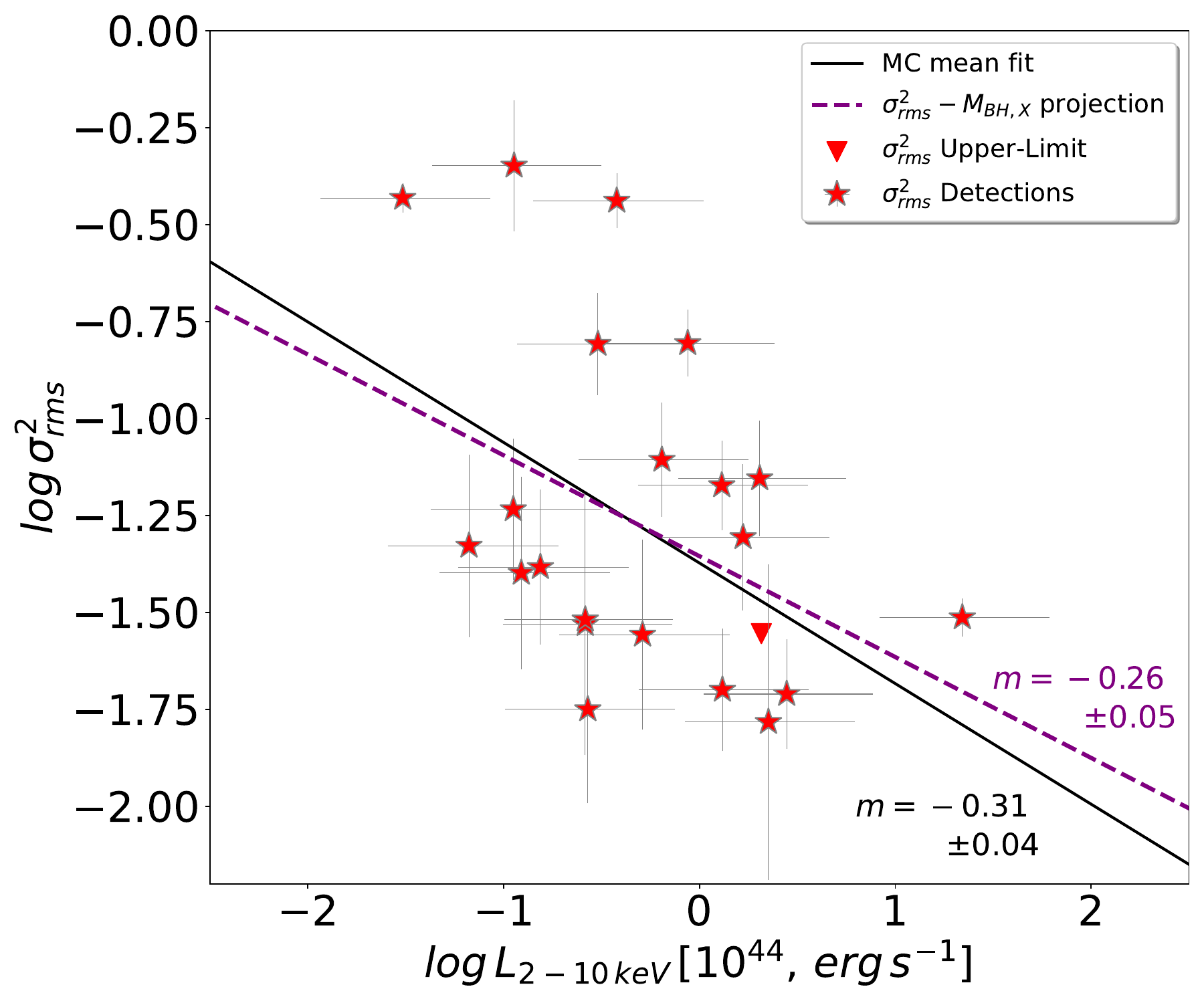}
\caption{Normal Excess Variances vs. X-ray luminosity of our AGN sample with measurements of their rest-frame X-ray luminosity at $2 - 10\, \rm{keV}$ energy range. The best MC linear fit is presented with the black line, while the purple dotted line represents the best MC linear fit (with arbitrary y-axis interceptions) obtained with the $\sigma^2_{rms}-M_{BH,X}$ relation.}  
\label{Lx_vs_sigma}
\end{figure}

In Figure \ref{Lx_vs_sigma} we present the distribution of $\sigma^2_{rms}$ as a function of the rest-frame X-ray luminosity at $2 - 10\, \rm{keV}$ energy range. A statistically significant "variability-luminosity" anti-correlation is observed, presented with the black solid line expressed in Equation \ref{sigma_lx}, with Spearman rank of $S \approx -0.4$. Additionally, the obtained linear fit closely aligns with the purple dotted line, which corresponds to the best MC linear fit (with arbitrary y-axis interceptions) obtained with $\sigma^2_{rms}-M_{BH,X}$ relation. For instance, the slope of the bright sample is $m = -0.31 \pm 0.04$, which is similar (considering the errors) to the fit obtained with the $\sigma^2_{rms}$-$M_{BH,X}$ relation, i.e. a slope of $m = -0.26 \pm 0.05$. 

\begin{equation}
\label{sigma_lx}
      \log(\sigma^{2}_{rms}) = (-1.37 \pm 0.04) + (-0.31 \pm 0.04)\log \left(\frac{L_{2-10\, \rm{keV}}}{10^{44}\, erg\, s^{-1} } \right)
\end{equation} 

Furthermore, we tested this relation with the full AGN population detected in the XMM-UNDF (i.e. 160 AGNs detected in at least three observations). We found that this anti-correlation did not change when considering a larger sample, maintaining the same trend with a slope of $m = -0.3 \pm 0.03$. Our results agree with the reported by previous deep X-ray surveys \citep[e.g.][]{Papadakis2008,Zheng2017}, and for local \citep[e.g.][]{Ponti2012} and distance  \citep[e.g.][]{Yang2016} AGNs samples.

\subsubsection{An underlying $M_{BH}$ dependence}

We obtained consistent slopes for both relations, with $m = -0.26 \pm 0.05$ for $\sigma_{rms}^{2} - M_{BH}$ and $m = -0.31 \pm 0.04$ for $\sigma_{rms}^{2} - L_x$, respectively. These results support the idea that the “Luminosity - X-ray variability” anti-correlation is generated as a byproduct of an intrinsic “BH mass - X-ray variability” relation. Furthermore, our results agree with those reported from previous surveys as \citet{Papadakis2008}, \citet{Ponti2012}, and \citet{Lanzuisi2014}. For example, \citeauthor{Papadakis2008} studied the variability - luminosity relation as a function of the redshift with a set of AGNs detected in the Lockman Hole, they found that this relationship with a steeper slope of $m = -0.66 \pm 0.12$ increases with redshift up to $z \sim 1$ and then stays roughly constant. Similarly, \citeauthor{Lanzuisi2014} with the XMM-COSMOS survey with a larger sample made by 638 AGNs, reported roughly the same reliable anti-correlations with a flatter slope of  $m = -0.23 \pm 0.03$ and Spearman coefficient of $S = -0.38$ for both $L_x - \sigma^{2}_{rms}$ and $M_{BH} - \sigma^{2}_{rms}$.

A feasible scenario that could cause these relations can be explained as follows: during the accretion process when a gas particle with mass $m_i$  interacts with the black hole, the released energy $E_{accretion}$ is proportional to the black hole mass and therefore for the size of the emitting region, given by $\Delta E_{accretion} = GM_{BH}m_i/R_s$ (where $R_s$ is the Schwarzschild radius). Most of this energy is emitted as optical and UV photons. These seed photons are then reprocessed by the hot corona generating the observed  AGN X-ray luminosity. Since the time scales associated with the accretion disk (e.g. viscous time and Sound-crossing time\footnote{The viscous time represents the time it takes for accretion rate variations to propagate across the disk, while the Sound-crossing time represents the time it takes for mechanical instabilities to cross the disk as acoustic waves, traveling at the sound speed \citep{Peterson2004}.}) are proportional to the BH mass \citep{Peterson2004}, variations in the accretion rate will primarily drive the AGN flux variability \citep{Uttley2014,Ricci2022}. Therefore, low-mass BHs (corresponding to faint AGNs) will have shorter time scales, leading to higher variations in the accretion rate and, as a result, affect the general flux emission of the AGN producing high variability, and vice versa high-mass BHs (corresponding to bright AGNs) will display lower variability due to longer time scales.

\section{The Eddington ratio distribution}
\label{Eddington}

Another important parameter that we analyzed in this paper is the Eddington ratio ($\lambda_{Edd}$), which is defined as the ratio between the bolometric luminosity ($L_{bol}$) and the Eddington luminosity ($L_{Edd}$). $\lambda_{Edd}$ represents the accretion rate relative to the Eddington limit and is a measure of how efficiently material is converted into radiation.  We used $L_{2-10\, \rm{keV}}$  to infer $L_{bol}$ from the X-ray luminosities. Following \citet{Netzer2013,Netzer2019}, we used $L_{bol}= K_{bol} \times L_{2-10\, \rm{keV}}$, where $K_{bol}$ is the bolometric correction factor defined as $K_{bol} = 69.8 - 1.4 \log(L_{2-10\, \rm{keV}}/\rm{erg}\, s^{-1})$.

\begin{figure}
\centering
\includegraphics[scale=0.27]{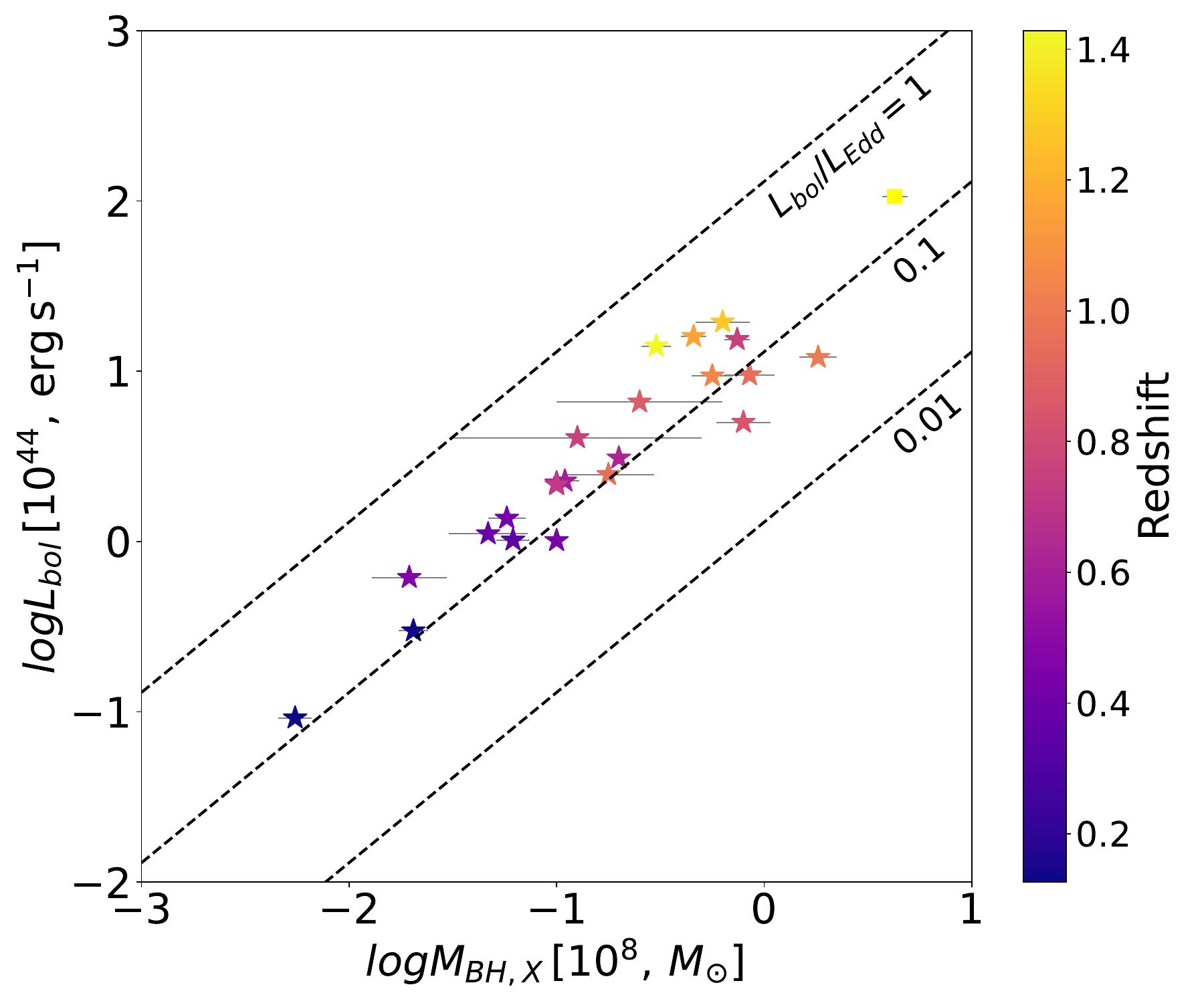}
\caption{$L_{bol}$ vs. $M_{BH,X}$ distribution of our AGN sample weighted by their redshift (color-bar). The yellow square represents the AGN XID-1 at z = 2.66.}
\label{L_M}
\end{figure}

Figure \ref{L_M} presents the distribution of $L_{bol}$ and $M_{BH,X}$ of our AGN sample. We observed a broad range of bolometric luminosities of  $42.5 < \log(L_{bol}) < 46.1$ with low dispersion of the BH masses at different redshift bins. For a better visualization of the redshift distribution, the color bar did not consider the AGN with the highest redshift $(z \sim 2.6)$ marked with the yellow square. The multiple linear regression equation of $L_{bol}$ as a function of $M_{BH}$ and $z$ is expressed as follows:

\begin{multline}
\label{LMz}
\log \left(\frac{L_{bol}}{10^{44}\, erg\, s^{-1}} \right) = (0.39 \pm 0.19) + (0.75 \pm 0.17)\log \left( \frac{M_{BH}}{10^8\, M_{\odot}} \right) \\
+ (0.57\pm 0.10) z \\
\end{multline}

\begin{figure}
\centering
\includegraphics[scale=0.27]{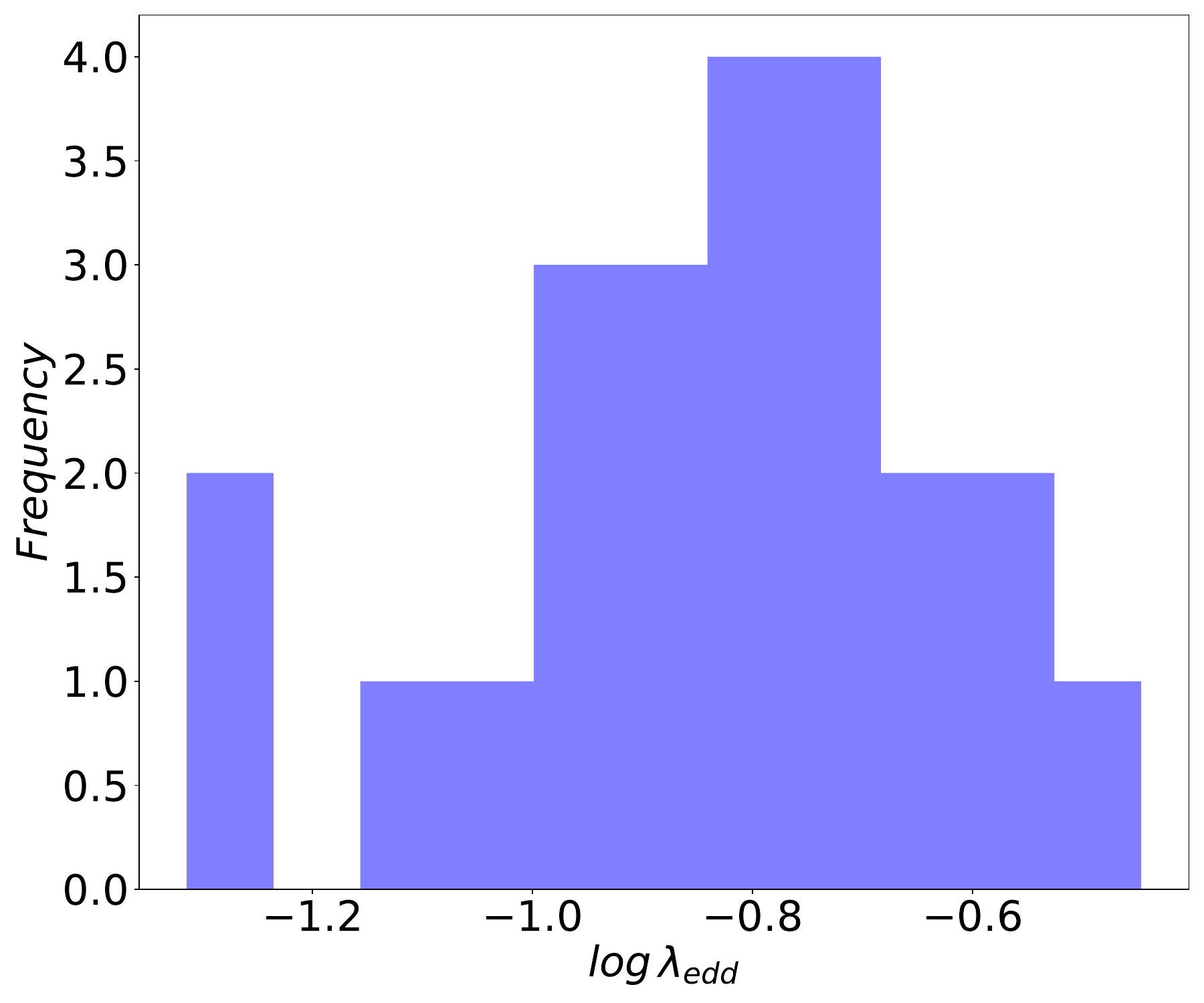}
\caption{Eddington ratio histogram of our AGN sample.}
\label{Edd_Hist}
\end{figure}

For an in-falling plasma composed mostly of ionized hydrogen, the Eddington luminosity can be calculated as $L_{Edd} \approx 1.3 \times 10^{38} (M_{BH}/M_{\sun})\, \rm{erg}\, \rm{s}^{-1}$ \citep{Rees1978}. Figure \ref{Edd_Hist} presents the Eddington ratio distribution of our sample, revealing a log-normal shape with a tail in the lowest values. We found that most of our AGNs are relatively low accretion rate systems with $\lambda_{Edd}<0.3$, having a mean of 0.16 and a dispersion of 0.07 dex. In Table \ref{Edd_table}, we summarize the estimation of $L_{bol}, L_{Edd}$, and $\lambda_{Edd}$. Our result is not unexpected since Seyfert galaxies tend to have relatively low $(\sim 0.1)$ Eddington ratios \citep[e.g.][]{Nobuta2012,Caccianiga2013}. For instance, \citet{Caccianiga2013} reported that most of their flux-limited sample, composed of 154 type-1 AGNs at redshift from 0.02 to 2 detected in the XBS Survey, peaks at an Eddington ratio of 0.1, ranging from 0.001 to 0.5. Similarly, \citet{Nobuta2012} analyzed 215 broad-line AGNs detected in the Subaru XMM-Newton field, with a mean redshift of $z \sim 1.4$. Their sample presents a $\lambda_{Edd}$ log-normal distribution with a mean and standard deviation of $0.14 \pm 0.2$. Additionally, our bolometric correction factors are consistent with the $K_{bol}$ confidence region reported by \citet{Duras2020}.

\begin{table}
\caption{Summary table of the $L_{bol}, L_{Edd}$, and $\lambda_{Edd}$ parameters, estimated from our AGN sample.}
\label{Edd_table} 
\begin{tabular}{ccccc} \toprule
XID & $\lambda_{Edd}$ & $L_{edd}$ & $L_{bol}$ & $K_{bol}$  \\
 & & $10^{44}\, \rm{erg}\, \rm{s}^{-1}$ & $10^{44}\, \rm{erg}\, \rm{s}^{-1}$ & \\  \midrule 
1 & 0.191 $\pm$ 0.027 & 554.55 $\pm$ 76.63 & 105.98 $\pm$ 1.48 & 4.82\\
2 & 0.113 $\pm$ 0.018 & 2.65 $\pm$ 0.43 & 0.30 $\pm$ 0.01 & 9.8\\
3 & 0.159 $\pm$ 0.022 & 96.37 $\pm$ 13.32 & 15.34 $\pm$ 0.14 & 7.41\\
4 & 0.237 $\pm$ 0.071 & 82.02 $\pm$ 24.56 & 19.45 $\pm$ 0.17 & 6.98\\
5 & 0.086 $\pm$ 0.024 & 110.65 $\pm$ 30.58 & 9.51 $\pm$ 0.09 & 7.26\\
6 & 0.269 $\pm$ 0.037 & 59.42 $\pm$ 8.21 & 16.00 $\pm$ 0.17 & 7.11\\
7 & 0.128 $\pm$ 0.030 & 73.10 $\pm$ 16.84 & 9.38 $\pm$ 0.09 & 7.21\\
8 & 0.119 $\pm$ 0.002 & 25.94 & 3.09 $\pm$ 0.05 & 8.18\\
9 & 0.051 $\pm$ 0.011 & 236.56 $\pm$ 49.03 & 12.06 $\pm$ 0.13 & 7.22\\
10 & 0.202 $\pm$ 0.186 & 32.65 $\pm$ 30.08 & 6.59 $\pm$ 0.09 & 7.56\\
11 & 0.129 $\pm$ 0.024 & 0.71 $\pm$ 0.13 & 0.09 $\pm$ 0.01 & 10.54\\
12 & 0.249 $\pm$ 0.344 & 16.37 $\pm$ 22.61 & 4.07 $\pm$ 0.06 & 7.95\\
13 & 0.049 $\pm$ 0.015 & 103.26 $\pm$ 30.92 & 5.01 $\pm$ 0.06 & 7.79\\
14 & 0.358 $\pm$ 0.058 & 39.26 $\pm$ 6.33 & 14.04 $\pm$ 0.17 & 6.91\\
15 & 0.127 $\pm$ 0.024 & 8.02 $\pm$ 1.48 & 1.02 $\pm$ 0.02 & 9.02\\
16 & 0.183 $\pm$ 0.080 & 6.08 $\pm$ 2.66 & 1.11 $\pm$ 0.03 & 9.03\\
17 & 0.078 $\pm$ 0.001 & 13 & 1.01 $\pm$ 0.02 & 9.05\\
18 & 0.170 $\pm$ 0.003 & 13 & 2.21 $\pm$ 0.04 & 8.47\\
19 & 0.183 $\pm$ 0.038 & 7.48 $\pm$ 1.55 & 1.37 $\pm$ 0.03 & 8.9\\
20 & 0.159 $\pm$ 0.026 & 14.25 $\pm$ 2.30 & 2.27 $\pm$ 0.04 & 8.42\\
28 & 0.166 $\pm$ 0.003 & 13 & 2.15 $\pm$ 0.04 & 8.25\\
33 & 0.243 $\pm$ 0.101 & 2.53 $\pm$ 1.05 & 0.61 $\pm$ 0.02 & 9.24\\
36 & 0.107 $\pm$ 0.054 & 23.12 $\pm$ 11.71 & 2.47 $\pm$ 0.07 & 8.15\\ \bottomrule
\end{tabular}
\end{table}

\subsection{$\lambda_{Edd}$ vs other parameters}

\begin{figure*}
\centering
\includegraphics[scale=0.32]{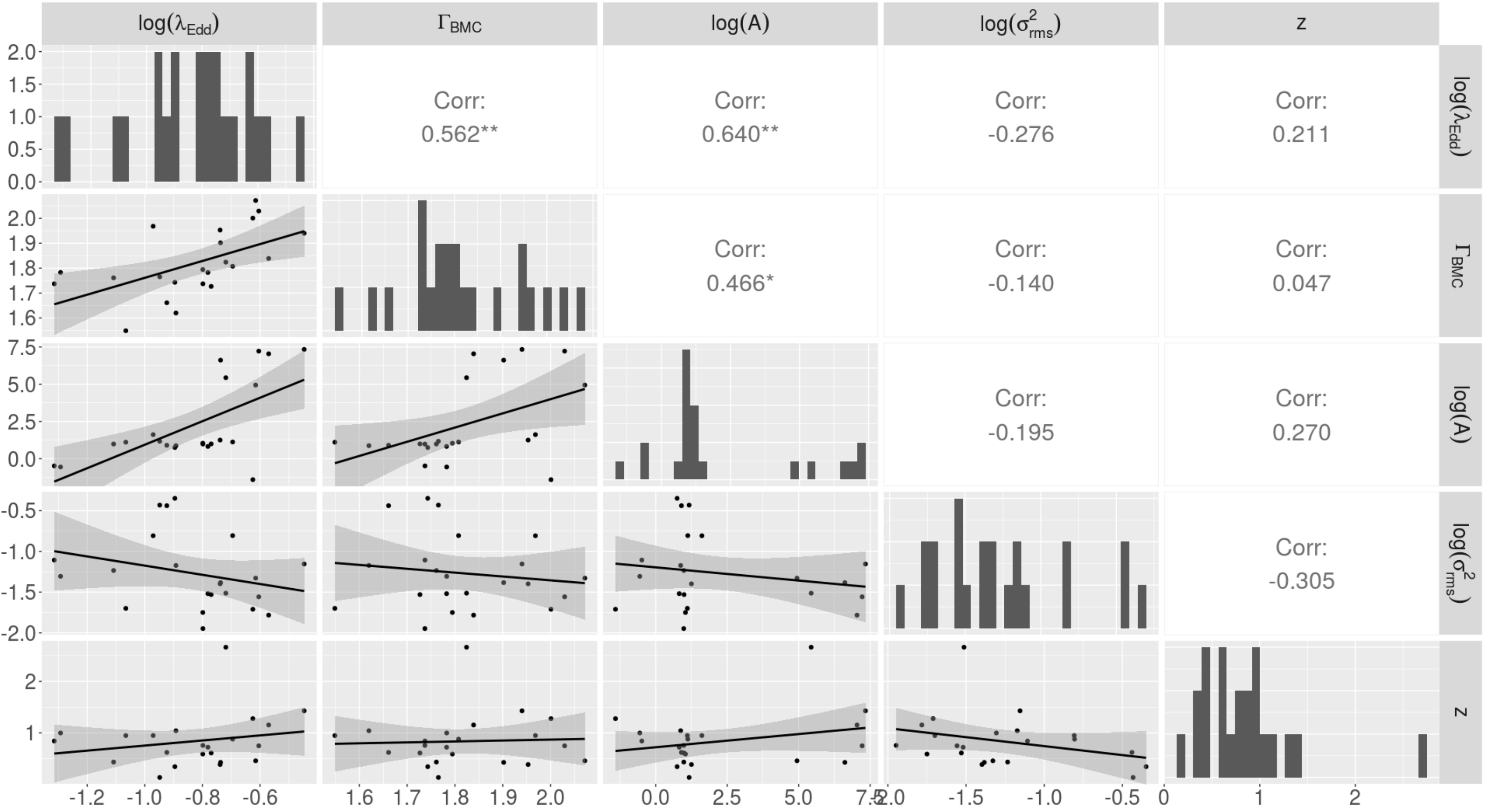}
\caption{Correlogram of  the main parameters of our AGNs including $\lambda_{Edd}$. Upper-panels: Spearman rank coefficient between each pair of variables. Diagonal: individual histograms. Lower-panels: scatter plots with regression lines and confidence intervals. The “*” and “**” symbols represent a moderate and strong correlation, respectively.}
\label{Correlogram}
\end{figure*}

We study the relationship among the Eddington ratio and other physical parameters, including $\sigma^2_{rms}$, $z$, the hard X-ray photon index $\Gamma_{BMC}$, and the illumination factor $\log(A)$. We performed a multilinear regression analysis among those parameters, resulting in the correlation matrix presented in Figure \ref{Correlogram}. This correlogram is composed of scatter plots with regression lines and confidence intervals (lower-panels), histograms (diagonal), and the results with a 95\% of confidence level of the Spearman rank coefficient between each pair of variables (upper-panels). To maintain consistency with the previous linear regression analysis computed in Section \ref{analysis}, we did not consider the source with the lowest mass for the correlogram.  

\begin{figure}
\centering
\includegraphics[scale=0.27]{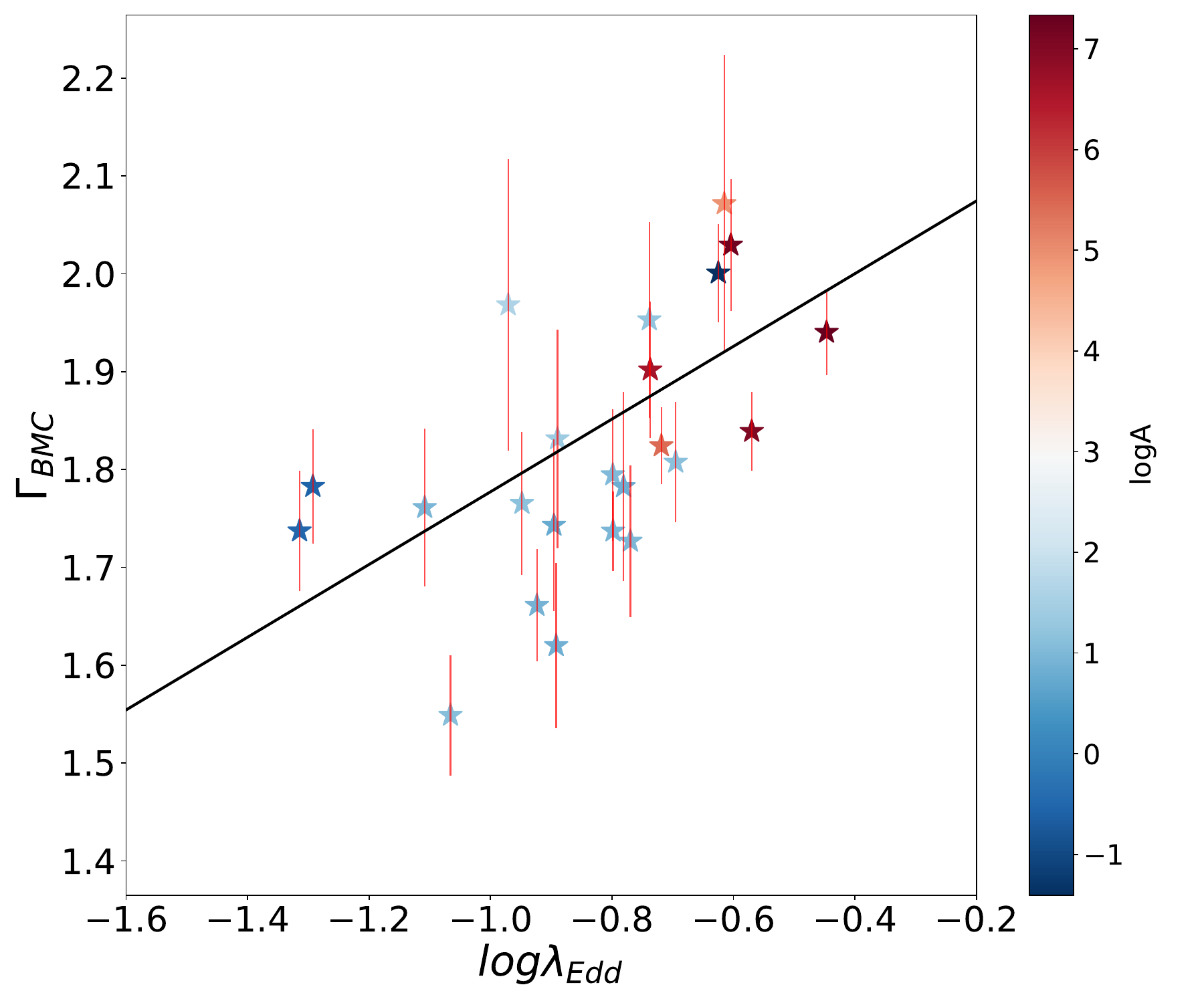}
\includegraphics[scale=0.3]{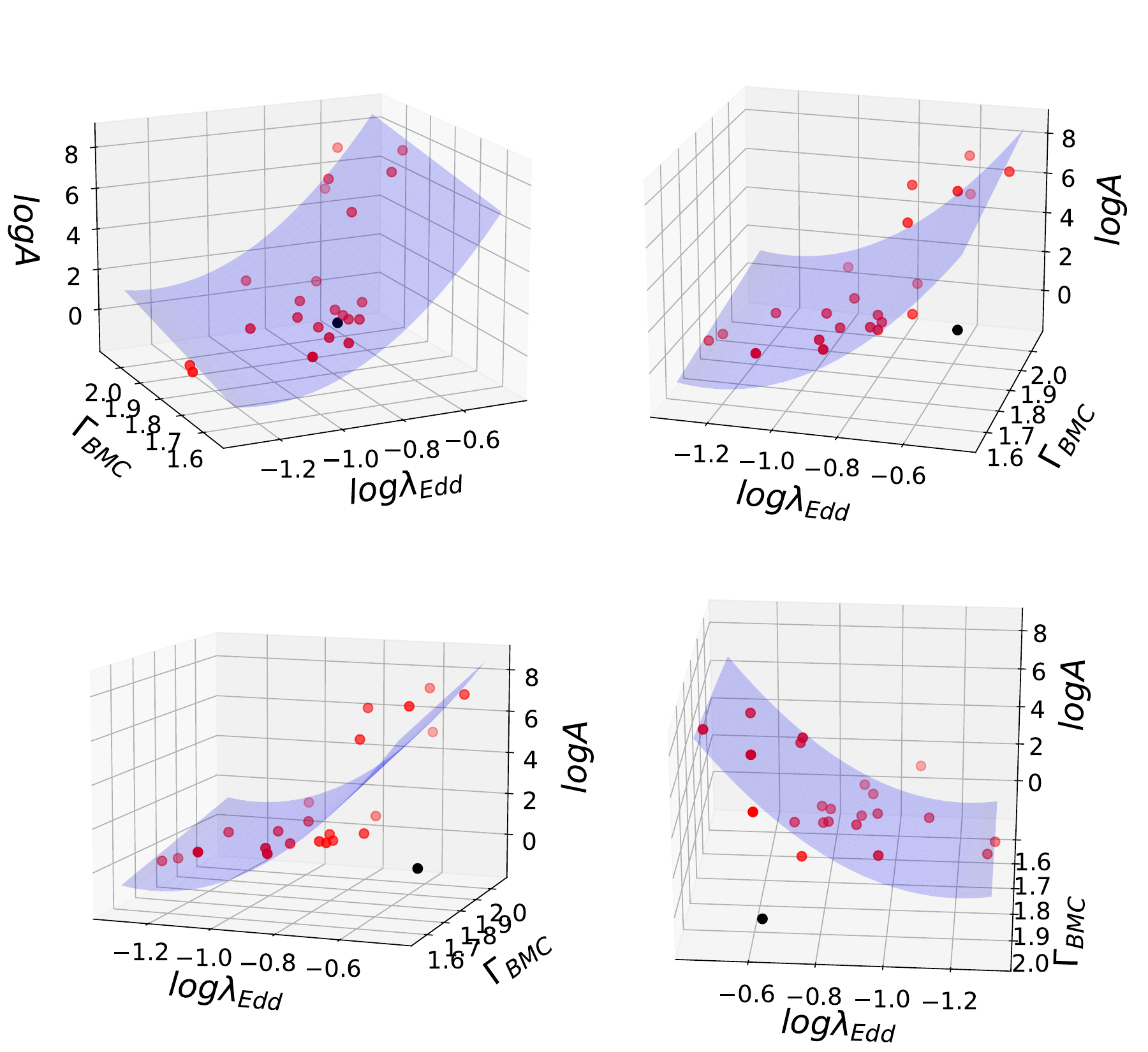}
\caption{$\Gamma_{BMC}$, $\lambda_{Edd}$, and $\log(A)$ distribution of our AGN sample. Upper: 2D view, Lower: 3D view. The black line and the central plane represent the best regression fit of our data.}
\label{3D}
\end{figure}

We found a strong and significant correlations between $\lambda_{Edd} - \Gamma_{BMC}$ ($S = 0.64$, $p_{value} =  0.0014$) and $\lambda_{Edd} - \log(A)$ ($S = 0.56$, $p_{value} =  0.0037$). The relation between those physical properties weighted by the illumination factor is presented in Figure \ref{3D}-upper. The simple linear regression equation for $\lambda_{Edd} - \Gamma$ is expressed as follows:

\begin{equation}
 \label{Gamm_lambda}
    \Gamma_{BMC} = (2.10 \pm 0.09)+ (0.34 \pm 0.11) \log(\lambda_{Edd})
\end{equation}

By including the illumination factor in a third axis, we obtained a 3D plot presented in Figure \ref{3D}-lower. The projected plane that best fits our data is calculated with Equation \ref{logA}.  We obtained a significant relationship with $p_{value} = 3.265 \times 10^{-5}$, which indicates a strong correlation between these three components.  We mark with a black circle the source XID-4, which is the only source that did not follow the $\log(\lambda_{Edd})-\Gamma-\log(A)$ plane. 

\begin{equation}
\begin{split}
\label{logA}
\log(A) = (3.4 \pm 3.2)\Gamma_{BMC} + (30.9 + 10.7)\log(\lambda_{Edd}) + \\ (12.8 + 5.7)\log(\lambda_{Edd}^2) + (12.6 \pm 9.1)
\end{split}
\end{equation}

A similar relation is computed in Equation \ref{Gamma_logA_lambda} to estimate the Eddington ratio as a function of $\Gamma_{BMC}$ and $\log(A)$. Finally, we did not find a clear relation among $\lambda_{Edd}$ and other properties, for instance, the Normal Excess Variance and the redshift. 

\begin{equation}
\label{Gamma_logA_lambda}
\begin{split}
\log(\lambda_{Edd})  = (0.18 \pm 0.05)\log(A) +  (0.11 + 0.28) \Gamma_{BMC} - \\ (0.016 + 0.007)\log(A)^2  - (1.28 \pm 0.49)
\end{split}
\end{equation}

We should be cautious with the interpretation of these results, primarily because the method used to estimate $M_{BH}$ and, consequently, $\lambda_{Edd}$ may introduce bias. On the other hand, this multiple correlation may arise naturally from the connection between the accretion flow and the hot corona. For instance, previous studies have confirmed a strong correlation between the Eddington ratio and the X-ray photon index \citep[e.g.][]{Brightman2013,Sarma2015}, suggesting that the accretion rate represented by $\lambda_{Edd}$ could drive the physical conditions of the hot corona and the accretion disk. In this scenario, higher Eddington ratios are equivalent to higher accretion rates that can lead to a more efficient release of energy near the black hole. This excess energy can result in the production of higher-energy X-ray photons through Comptonization by the hot corona. Furthermore, as AGNs evolve, their accretion rates, the coverage of the hot corona over the effective disk area, and the spectral properties could change, therefore, the $\log(\lambda_{Edd})-\Gamma-\log(A)$ plane could be used to understand the transitional phases in AGN evolution, such as the transition from a high accretion rate Seyfert (NLSy1) to a normal Seyfert.

\section{Summary and Conclusions}
\label{conclusions}
Since X-ray emission is an intrinsic property observed in all Active Galactic Nuclei, X-ray analysis of multi-epoch observations of AGNs is a powerful tool to identify and study AGNs. In this paper, we performed a multi-epoch X-ray spectral analysis on a bright AGN sample with the highest S/N observed in the XMM-UNDF, which is one of the deepest X-ray surveys carried out with the satellite XMM-Newton. A summary of the main correlations presented in this paper is provided in Table \ref{alltables}. The key results of this analysis are listed below:

\begin{table*}
\centering
\caption{Summary of the main correlations.}
\label{alltables}
\begin{tabular}{ccccc} 
\toprule 
Equation & Variables & Formula & $S$ & $p_{value}$\\ \toprule 
\ref{EWdet}  & $EW_{Fe},\, Lx$ &  $\log \left(\frac{EW_{Fe}}{keV} \right)_{det} = (-1.10\pm 0.06) + (-0.23 \pm 0.07)  \log \left(\frac{Lx}{10^{44} \rm{erg}\, s^{-1}}\right) $ & -0.6 & 0.04 \\ \midrule 
\ref{z_NH} & $N_H,\, z$ & $\log(N_H) =  (20.66 \pm 0.16) + (0.31 \pm 0.23) z$ & 0.45 & $> 0.05$ \\  \midrule
\ref{MBH_X} & $\sigma^{2}_{rms},\, M_{BH,X}$ & $\log(\sigma^{2}_{rms}) = (-1.46 \pm 0.05) + (-0.26 \pm 0.05)\log \left (\frac{M_{BH,X}}{10^8\, M_{\sun}} \right)$ & -0.26 & $\approx 0.05$ \\ \midrule
\ref{sigma_lx} & $\sigma^{2}_{rms},\, L_{2-10\, \rm{keV}} $ & $\log(\sigma^{2}_{rms}) = (-1.37 \pm 0.04) + (-0.31 \pm 0.04)\log \left(\frac{L_{2-10\, \rm{keV}}}{10^{44}\, erg\, s^{-1} } \right)$ & -0.4 & $ 0.07$ \\ \midrule     
\ref{LMz} & $L_{bol},\, M_{BH},\, z $& $\log \left(\frac{L_{bol}}{10^{44}\, erg\, s^{-1}} \right) = (0.39 \pm 0.19) + (0.75 \pm 0.17)\log \left( \frac{M_{BH}}{10^8\, M_{\odot}} \right) 
+ (0.57\pm 0.10) z$ & - & $\ll 0.05$ \\ \midrule
\ref{Gamm_lambda} & $\Gamma_{BMC},\, \lambda_{Edd}$ & $\Gamma_{BMC} = (2.10 \pm 0.09)+ (0.34 \pm 0.11) \log(\lambda_{Edd})$ & 0.64 & $0.0014$ \\ \midrule
\ref{Gamma_logA_lambda} & $\lambda_{Edd},\, \log(A),\, \Gamma_{BMC} $ & $\log(\lambda_{Edd})  = (0.18 \pm 0.05)\log(A) +  (0.11 + 0.28) \Gamma_{BMC}$  & - & $10^{-4}$ \\ 
& & $- (0.016 + 0.007)\log(A)^2  - (1.28 \pm 0.49)$& & \\ \bottomrule
\end{tabular}
\end{table*}

\begin{itemize}
 
\item The best model that fits our data is a combination of a simple power-law with a constant Galactic absorption, a neutral intrinsic absorption associated with the host galaxy, and a Fe-K$\alpha$ emission line. We found a mean and standard deviation for the column density and the Fe-K$\alpha$ line equivalent width of $\log(N_H) = 20.92 \pm 0.18\, \rm{cm}^{-2}$ and $ EW_{Fe} = 0.14 \pm 0.11\, \rm{keV}$, respectively. 

\item We found statistically significant anti-correlation between the X-ray luminosity and the Fe-K$\alpha$ equivalent width of our AGN sample, which is consistent with the “Iwasawa-Taniguchi effect” associated with the decreasing of the torus opening angle as a function of $L_x$. It can be well described by $EW_{Fe} \propto  L_{2-10\, \rm{keV}}^{-0.23}$.  

\item For the relation between $N_H$ and $z$, we found a moderate correlation consistent with the reported by  \citet{Iwasawa2020} with the XMM deep survey in the CDFS and \citet{Corral2011} with the XBS. Our results suggest a potential trend in the evolution of the obscured AGN fraction toward higher redshifts. 

\item We obtained a good agreement between the two approaches used to estimate the BH masses. For the X-ray luminosity method, the mean black hole mass was estimated to be $\log(M_{BH,L_x}/M_{\sun}) = 7.59 \pm  0.59$, while the X-ray scaling method yielded a mean black hole mass of  $\log(M_{BH,X}/M_{\sun}) = 7.26 \pm  0.68$. We found a trend of slightly lower masses obtained with the luminosity method, described as $M_{BH,L_x} \sim 0.33\times M_{BH,X}$.

\item The $L_x - \sigma^{2}_{rms}$ and the $M_{BH} - \sigma^{2}_{rms}$ distributions present statistically significant anti-correlations with roughly the same flat slopes with $m = -0.31 \pm 0.4$ and $m = -0.26 \pm 0.05$, respectively. These results support the possibility that the anti-correlation between the luminosity and X-ray variability arises as a consequence of an intrinsic relationship between the BH mass and the X-ray variability.

\item Our AGN sample covers the bolometric luminosity range of $42.5 < \log(L_{bol}) < 46.1$ with moderate dispersion in the Eddington ratio distribution, with a mean of $\lambda_{Edd} = 0.16 \pm 0.07$. Additionally, our analysis reveals strong correlations between $\Gamma_{BMC}$, $\lambda_{Edd}$, and $\log(A)$. Studying this parameter space could offer a novel perspective on the changing stages of AGN evolution.

\end{itemize}

\section*{Acknowledgements}
MEC acknowledges support from CONAHCYT through a postdoctoral fellowship within the program ‘‘Estancias Posdoctorales por México’’. ALL acknowledges support from CONACyT grant CB-2016-286316 and PAPIIT IA-101623. T. Miyaji is supported by UNAM-DGAPA PAPIIT 114423. Y.K. acknowledges support from DGAPA-PAPIIT grant IN102023.

\section*{Data Availability}
The data underlying this article will be shared on reasonable
request to the corresponding author.



\bibliographystyle{mnras}
\bibliography{bibliography} 




\appendix
\section{AGN Sample Spectra}
\label{Spectra}
XMM-Newton  $0.3 - 10\, \rm{keV}$ rest-frame spectra of our sample composed of 23 AGNs detected in the XMM-UNDF with mean X-ray photon counts of $10,000$ cts. Each spectrum is composed by the combination of the 3 cameras with the 11 observations, as was explained in Section \ref{X-ray_spectral_analysis}. The data were modeled with their best fit according to table \ref{statistic1}, i.e. a combination of a simple power-law, constant Galactic absorption of $N_H=3.56 \times 10^{20}\, \rm{cm}^{-2}$, and an iron emission line at 6.4 $\rm{keV}$. While the residuals are presented in terms of the data minus the model weighted by the error.  

y-axis in the upper panel in all plots is in terms of “normalized counts $\rm{s}^{-1}\, \rm{keV}^{-1}\, \rm{cm}^{-2}$”, while the y-axis in the lower panel in all plots is in terms of “(data - model) / error”.

\begin{figure*}
\centering
\includegraphics[scale=0.211]{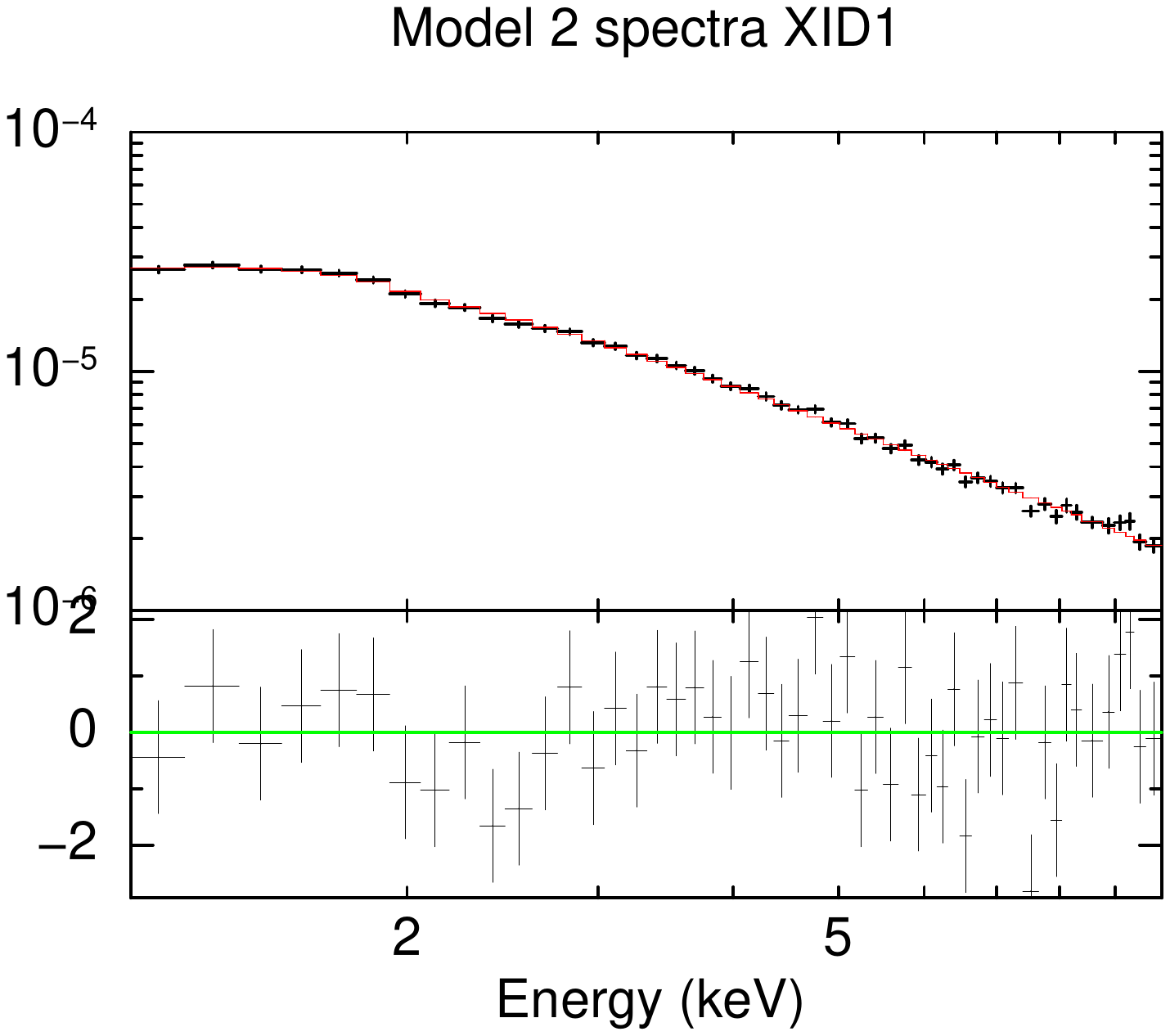}
\includegraphics[scale=0.211]{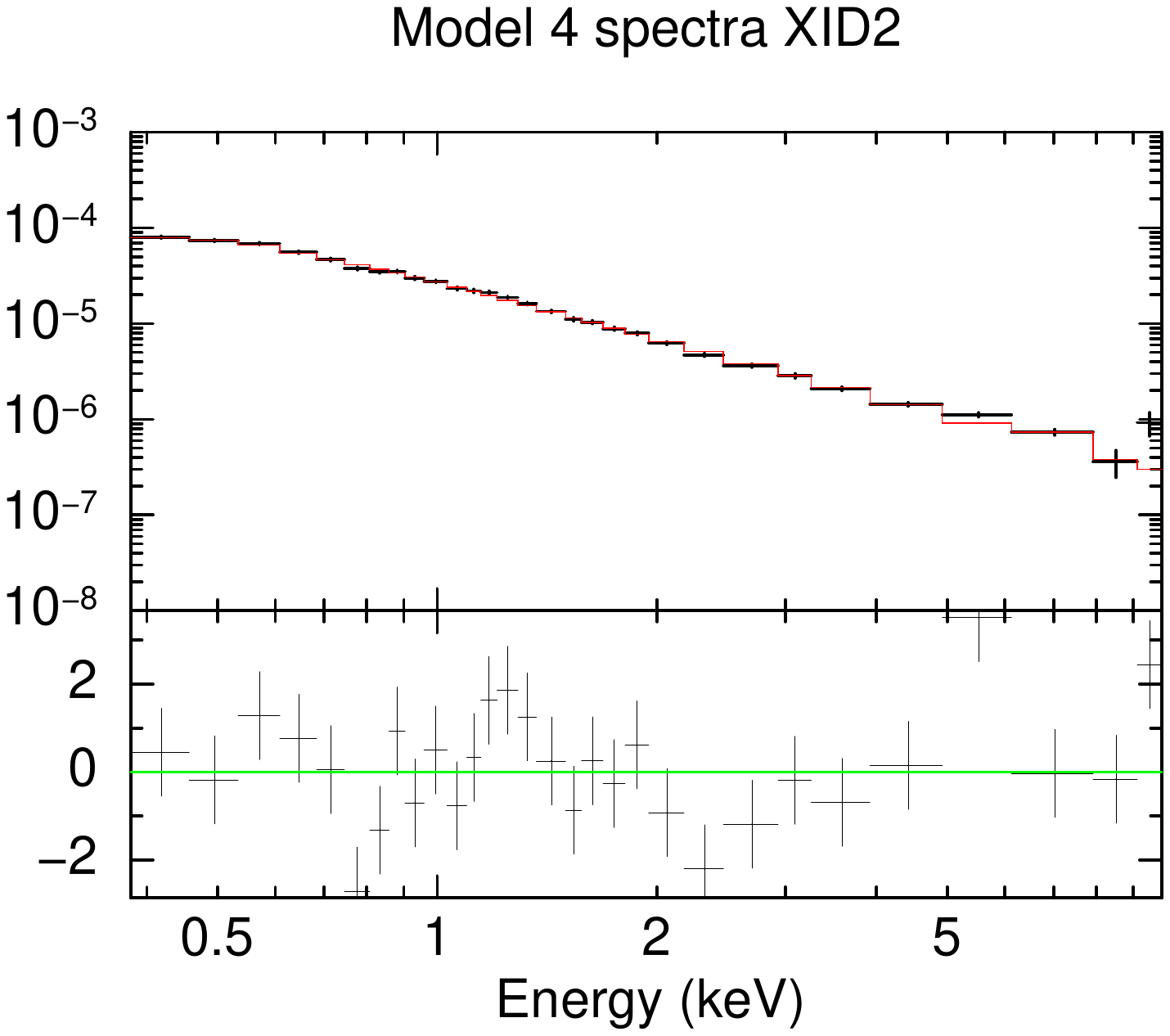} 
\includegraphics[scale=0.211]{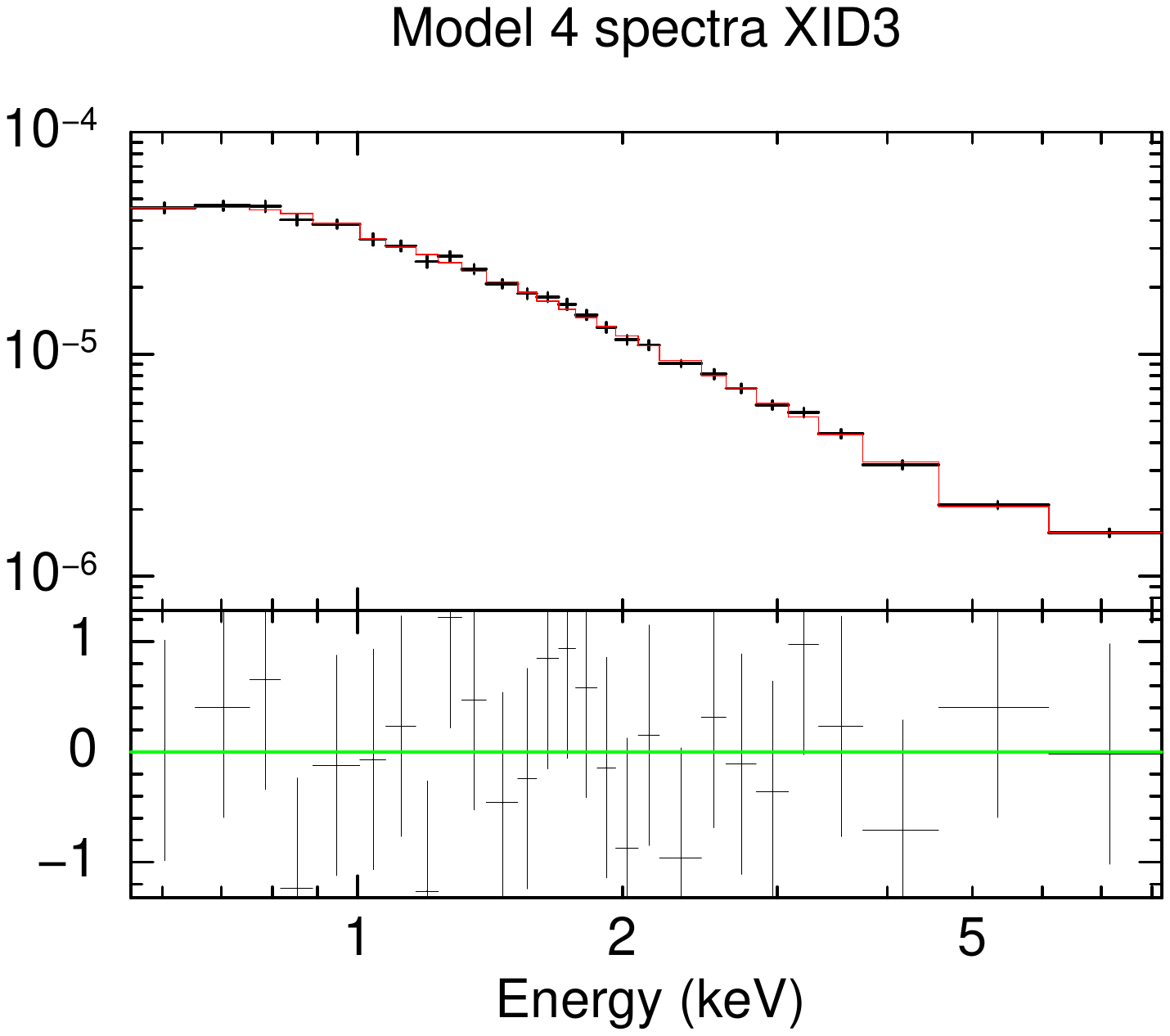} 
\\
\includegraphics[scale=0.211]{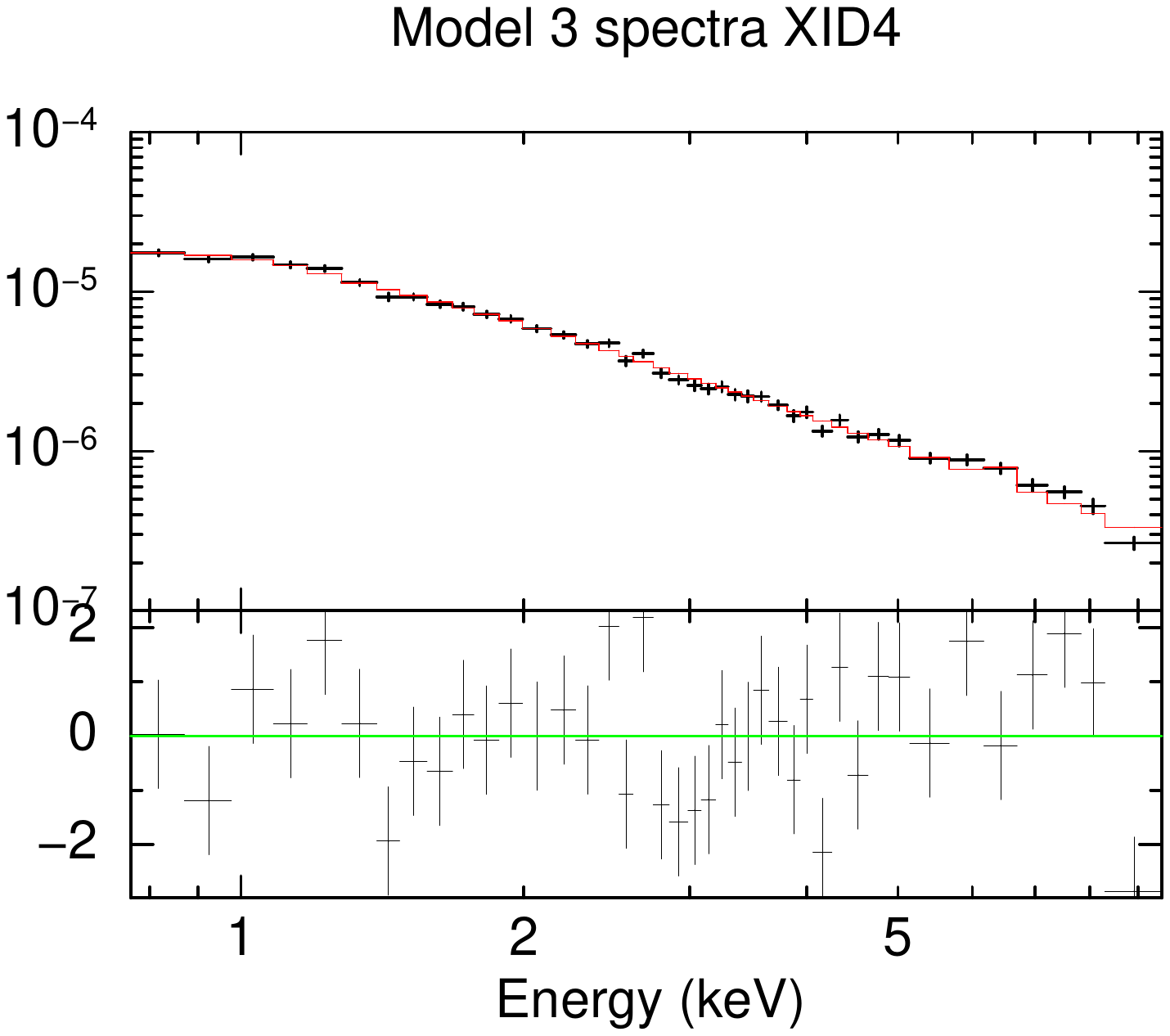}
\includegraphics[scale=0.211]{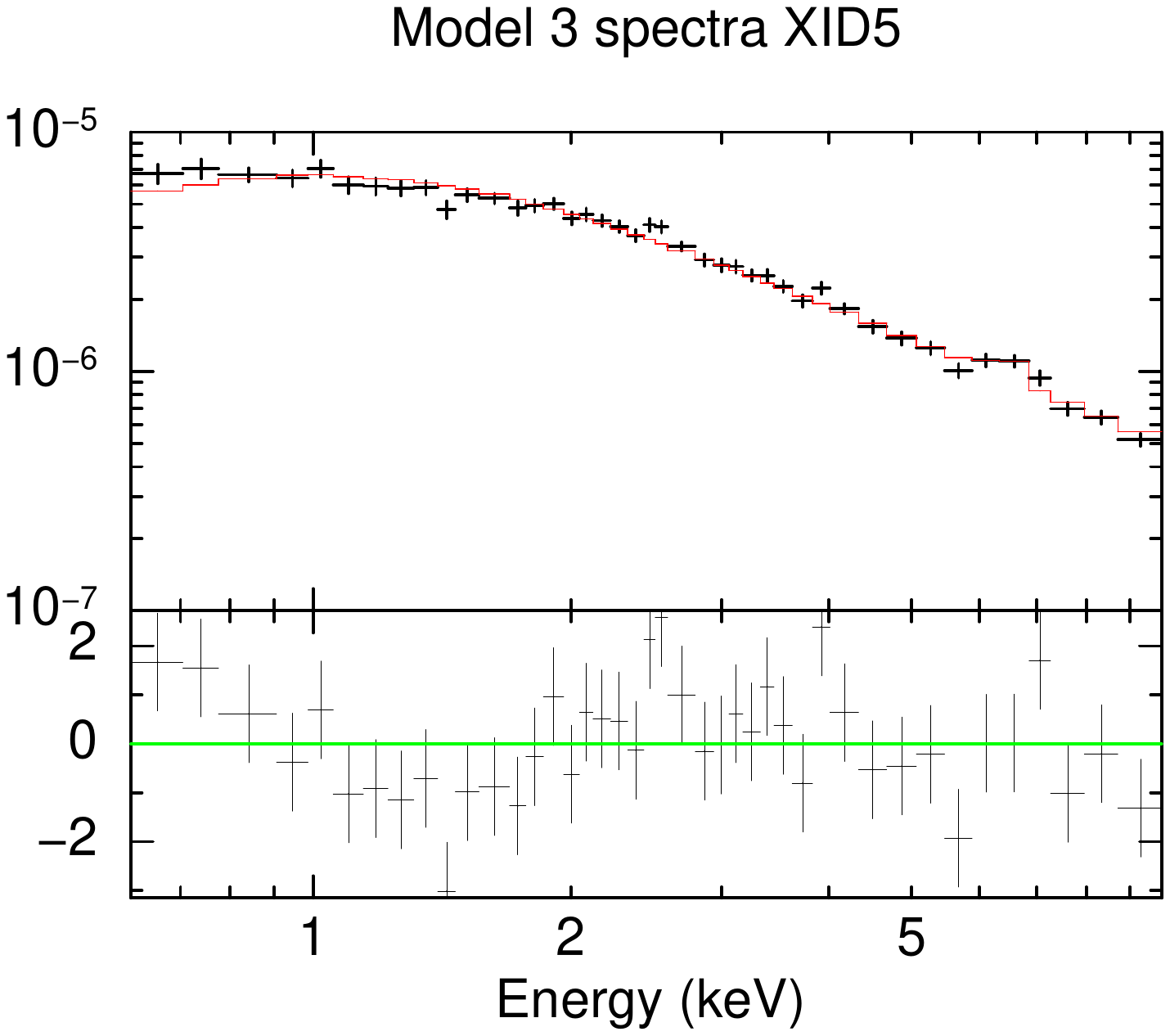} 
\includegraphics[scale=0.211]{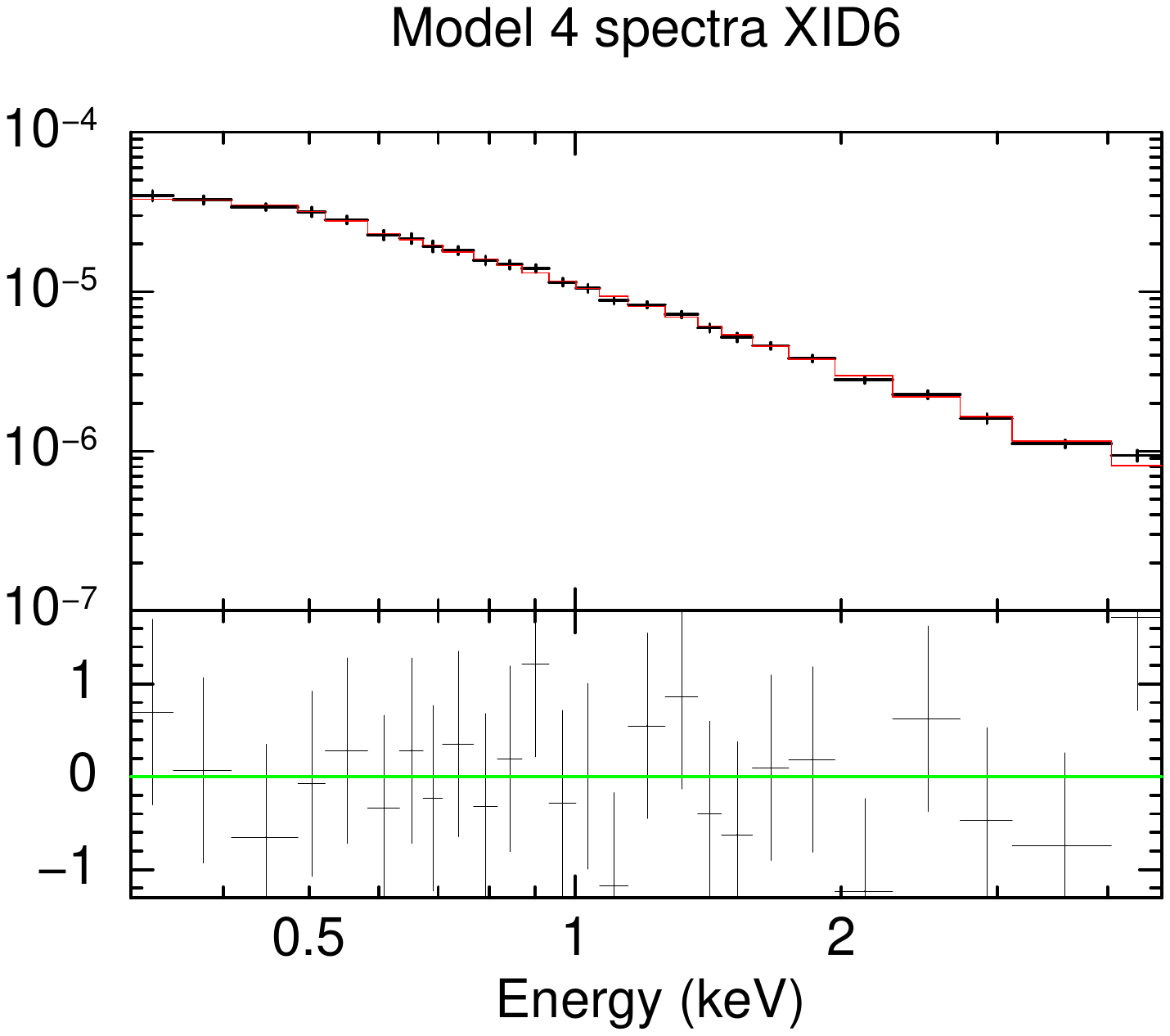}
\\
\includegraphics[scale=0.211]{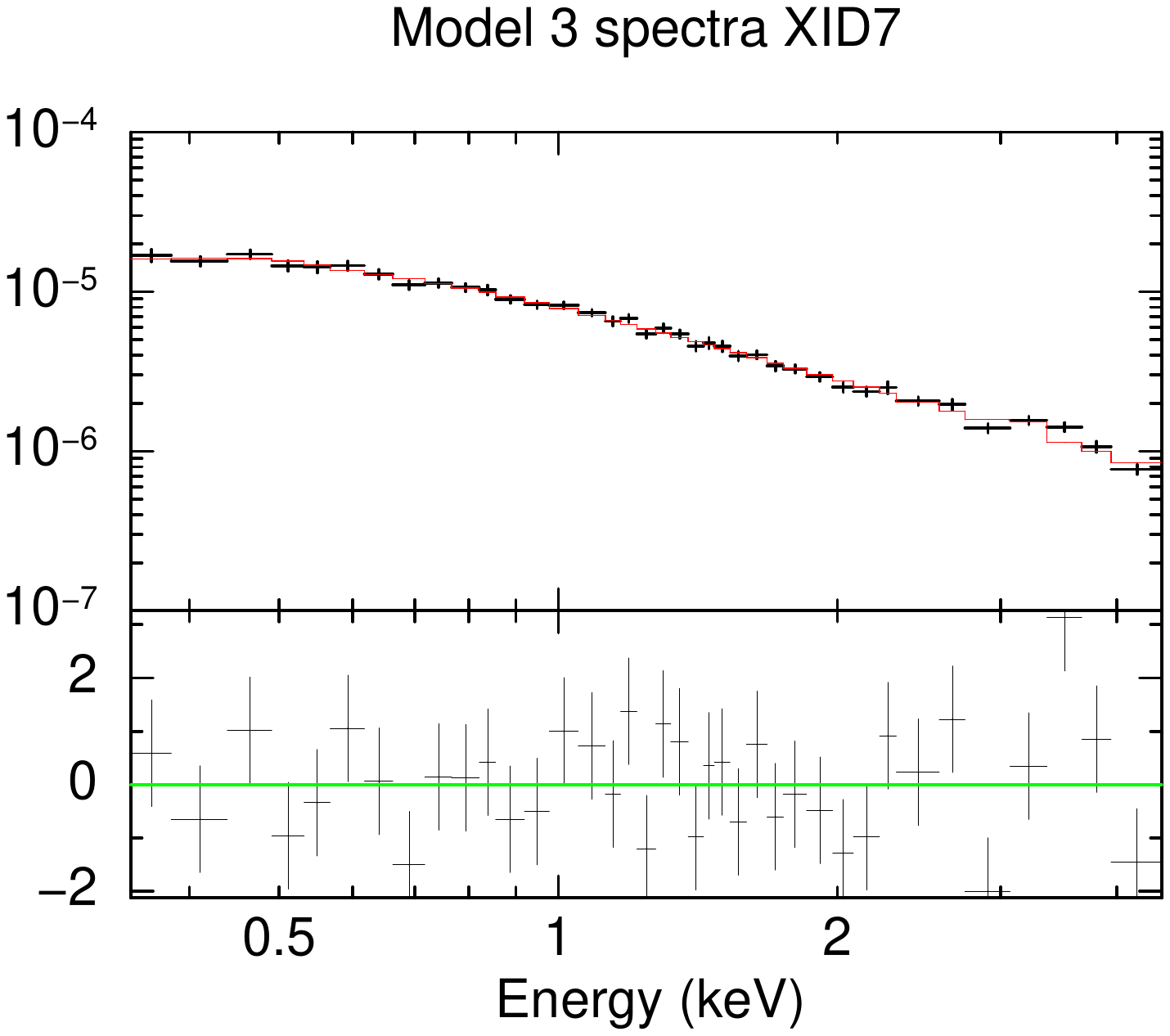}
\includegraphics[scale=0.211]{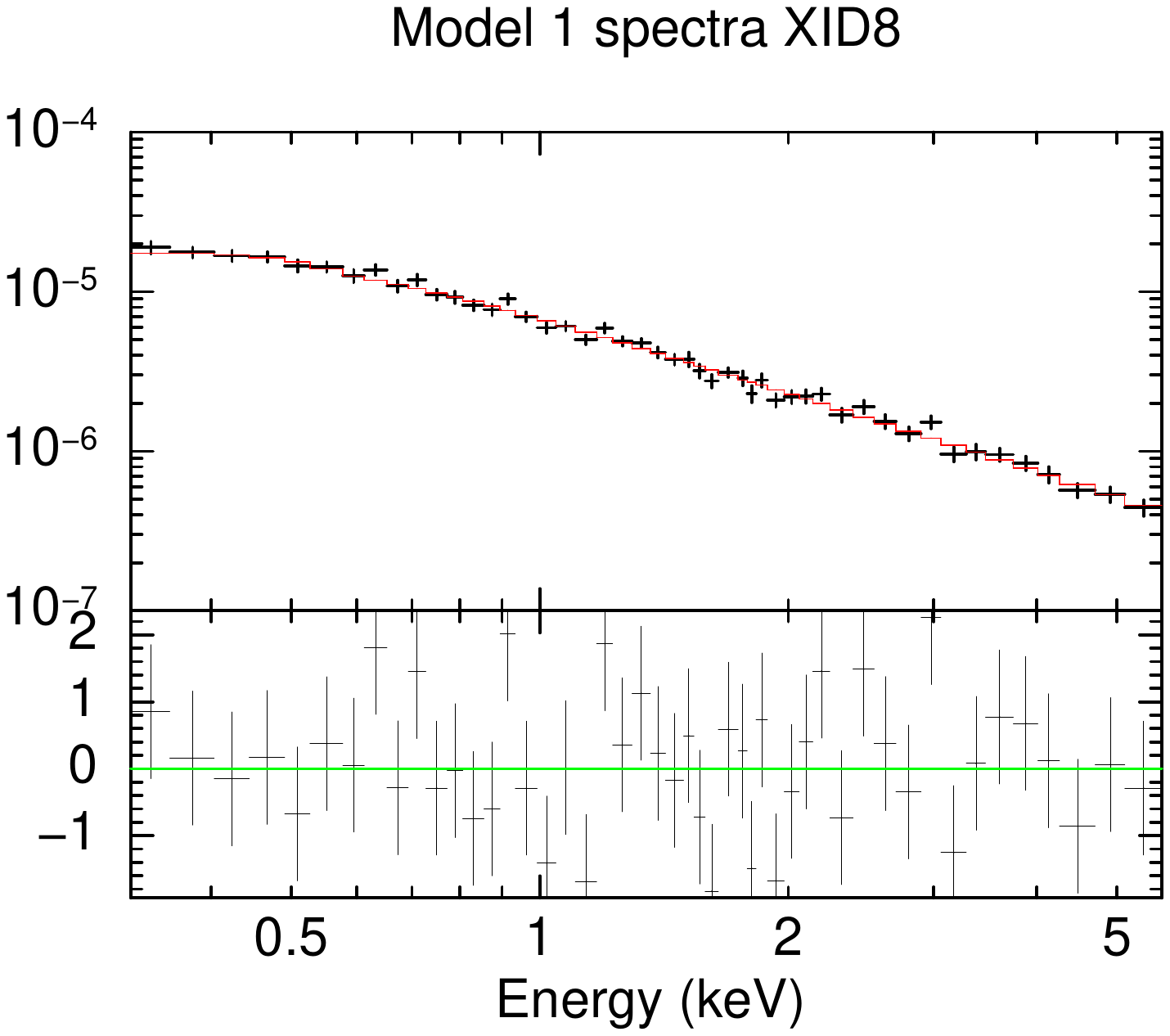} 
\includegraphics[scale=0.211]{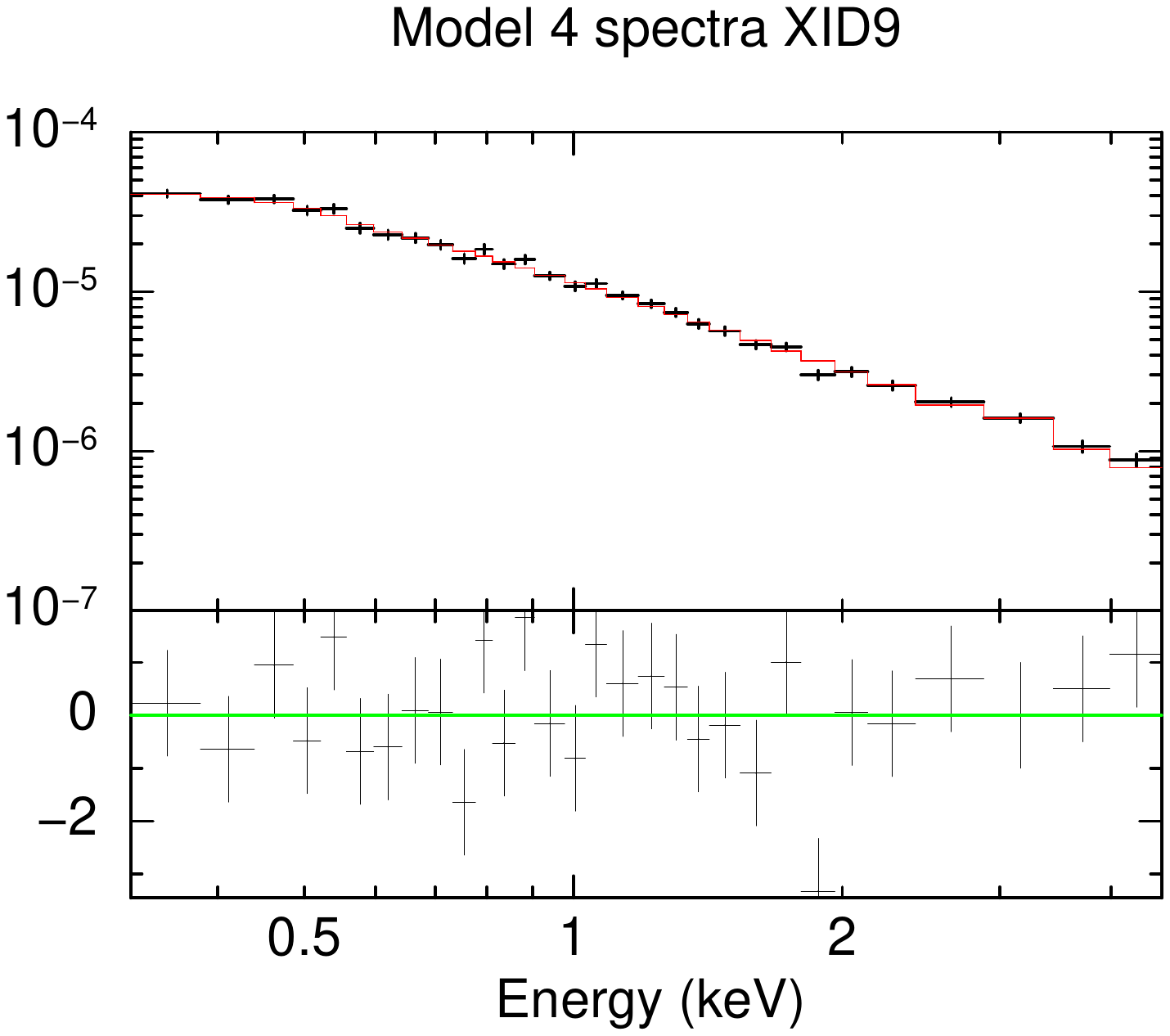} 
\\
\includegraphics[scale=0.211]{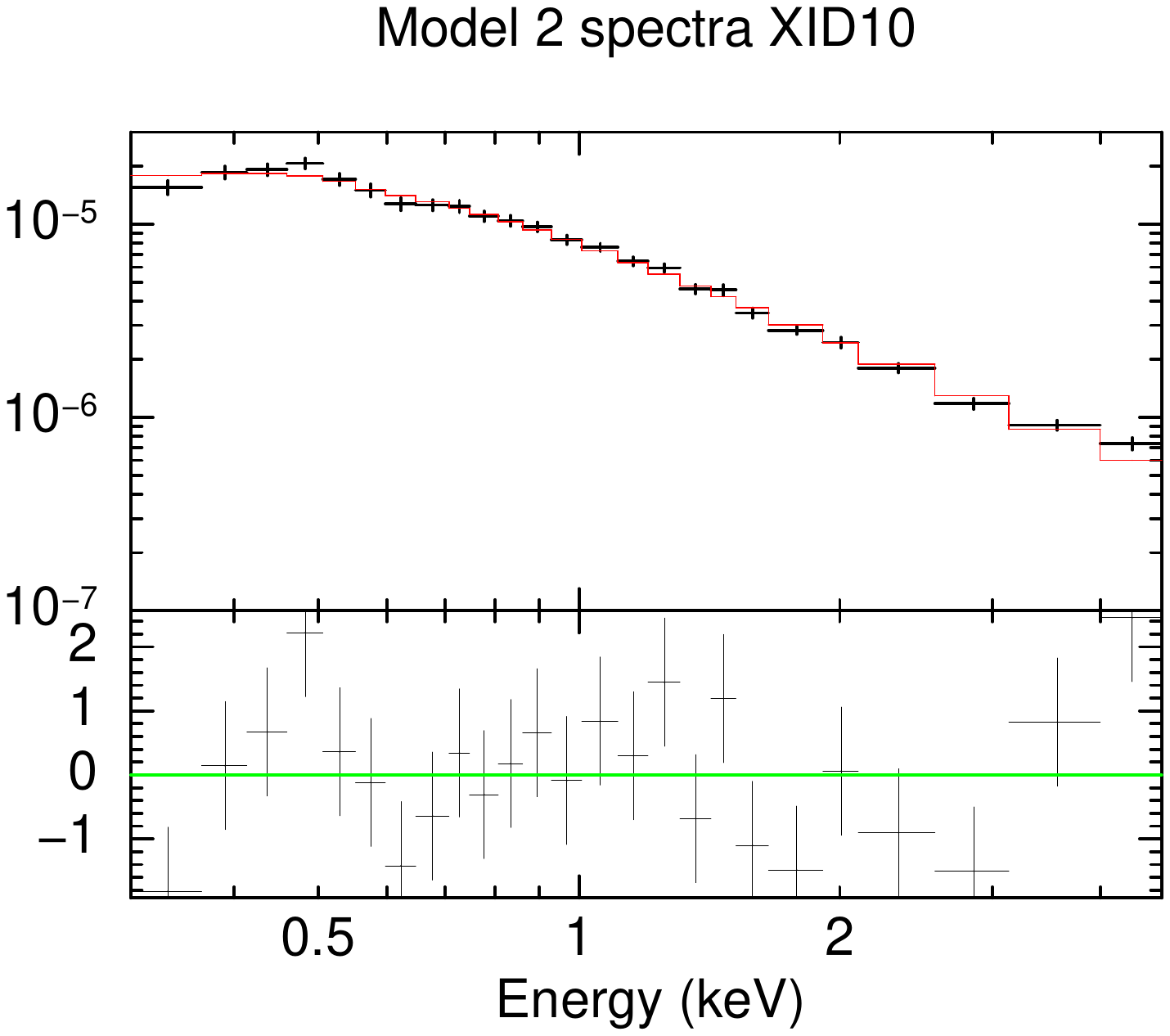}
\includegraphics[scale=0.211]{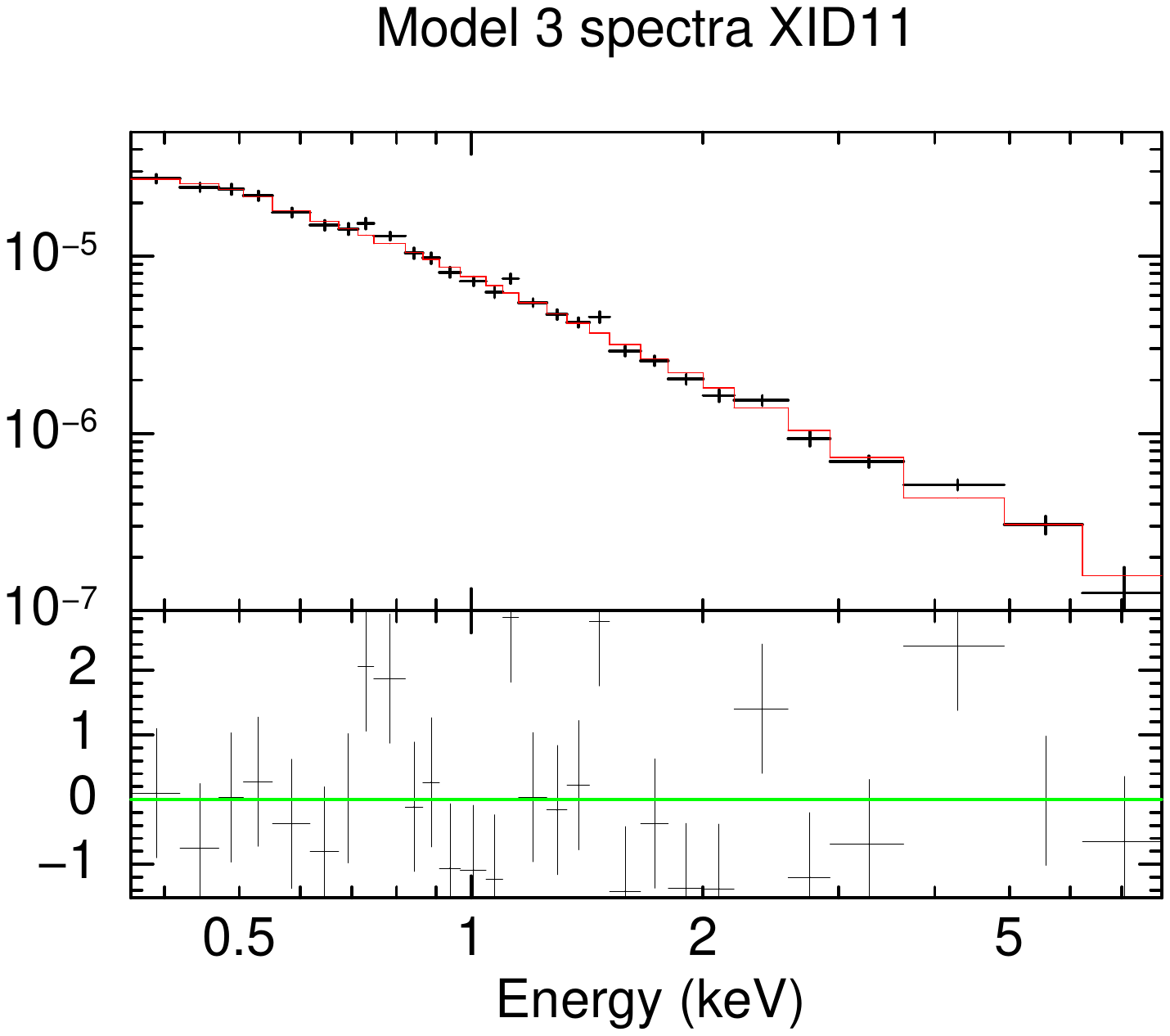} 
\includegraphics[scale=0.211]{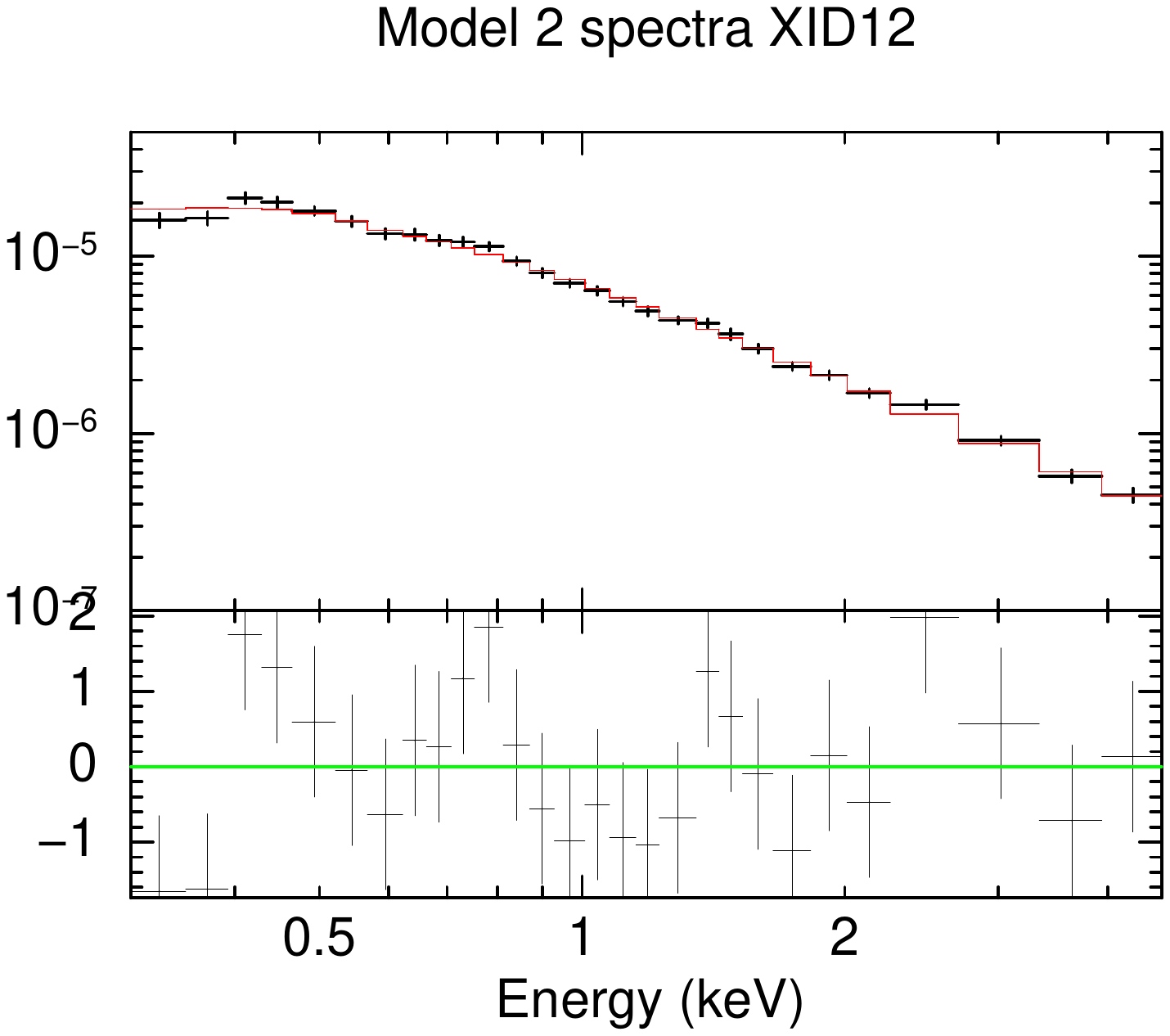} 
\caption{Spectra of the AGN sample and their residuals part-1.}
\end{figure*}

\begin{figure*}
\centering
\includegraphics[scale=0.211]{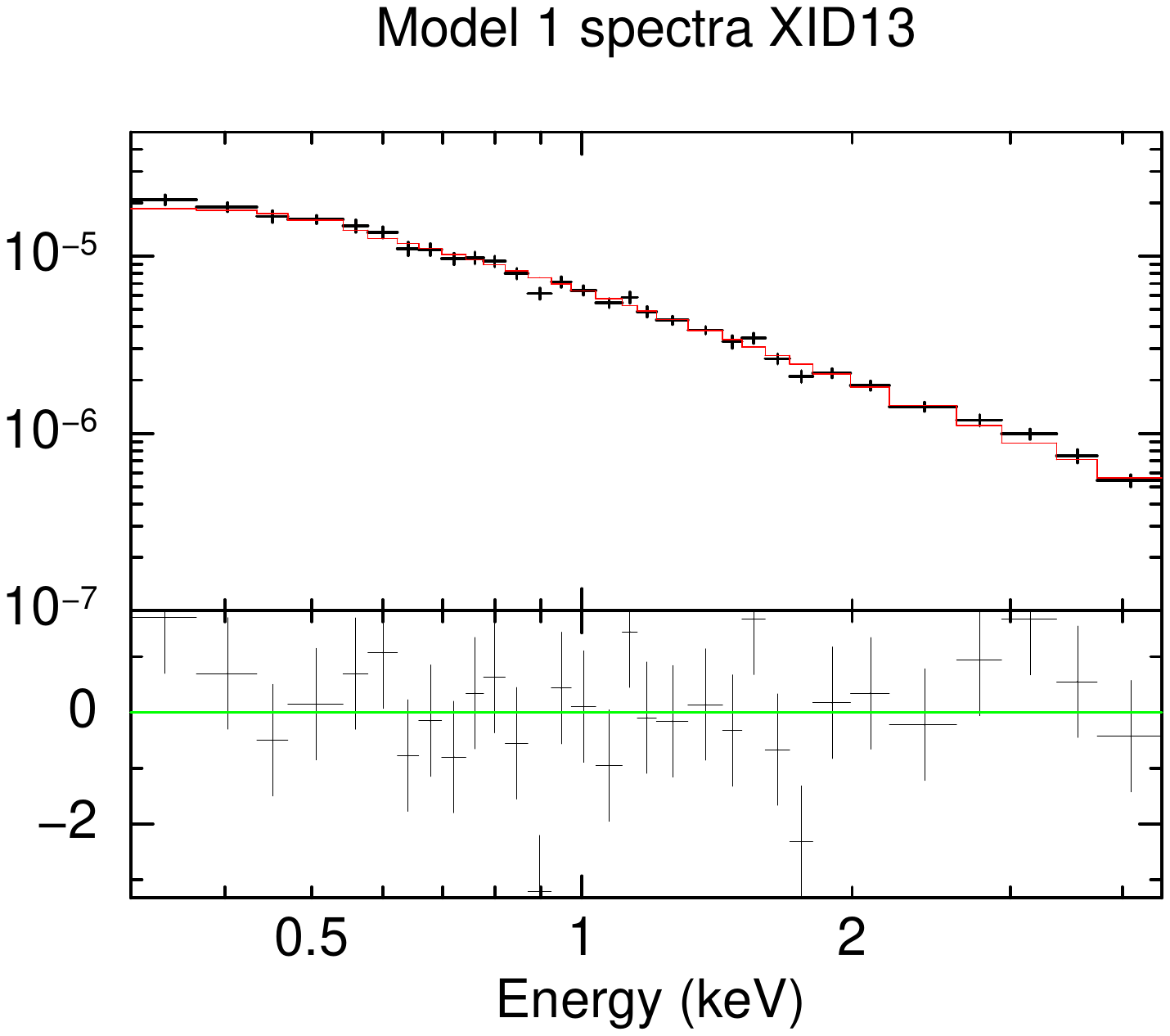}
\includegraphics[scale=0.211]{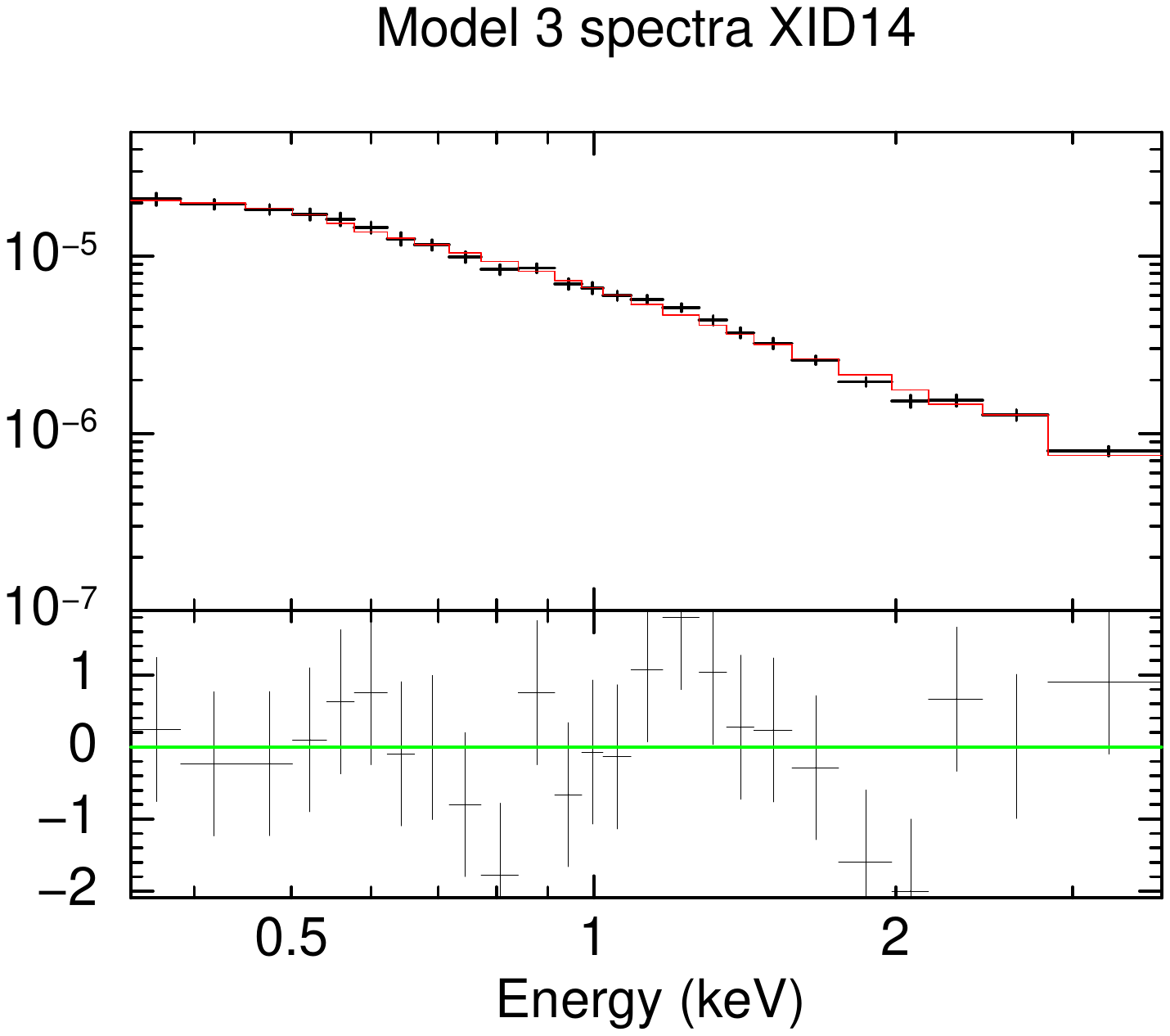} 
\includegraphics[scale=0.211]{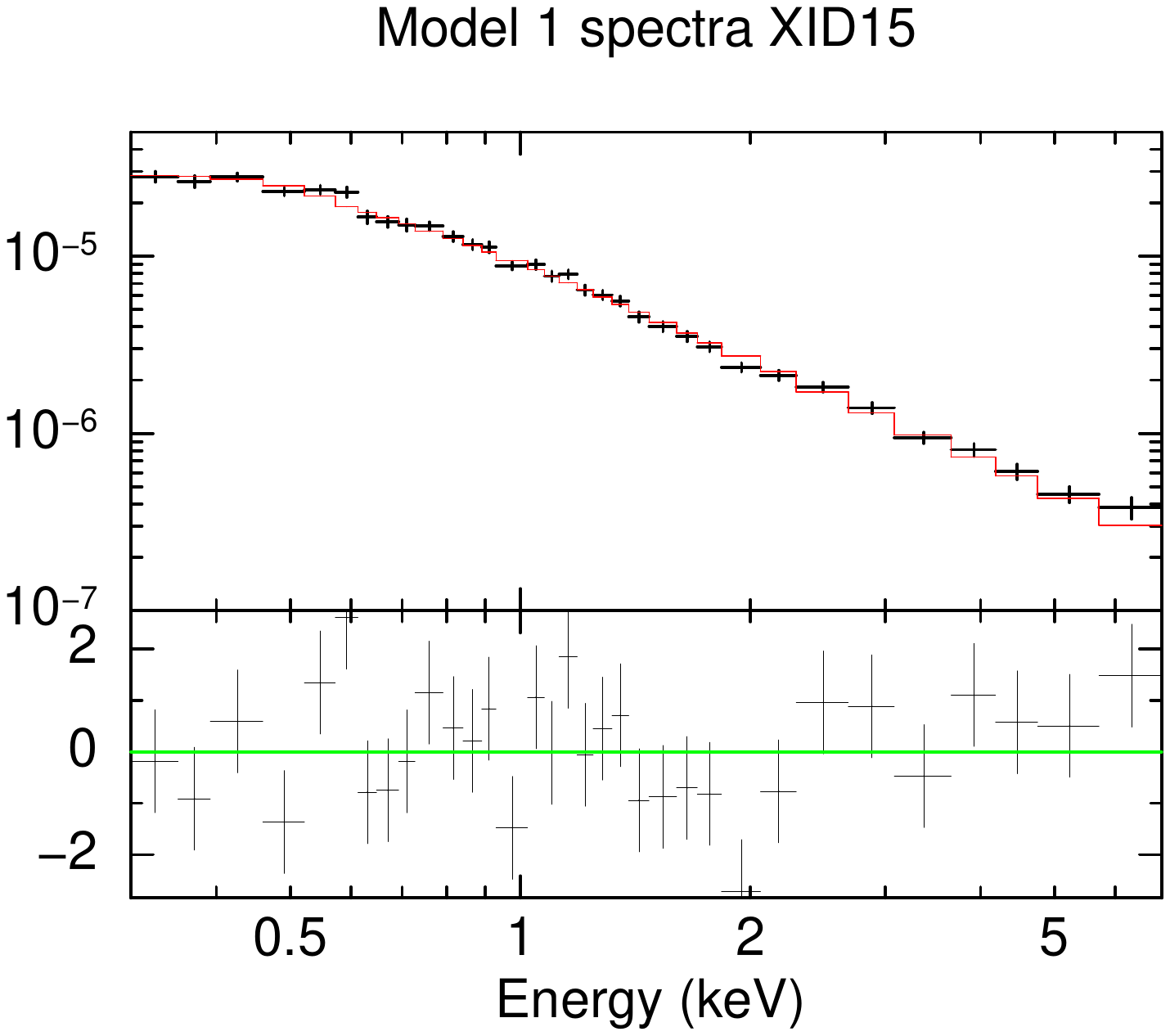} 
\\
\includegraphics[scale=0.211]{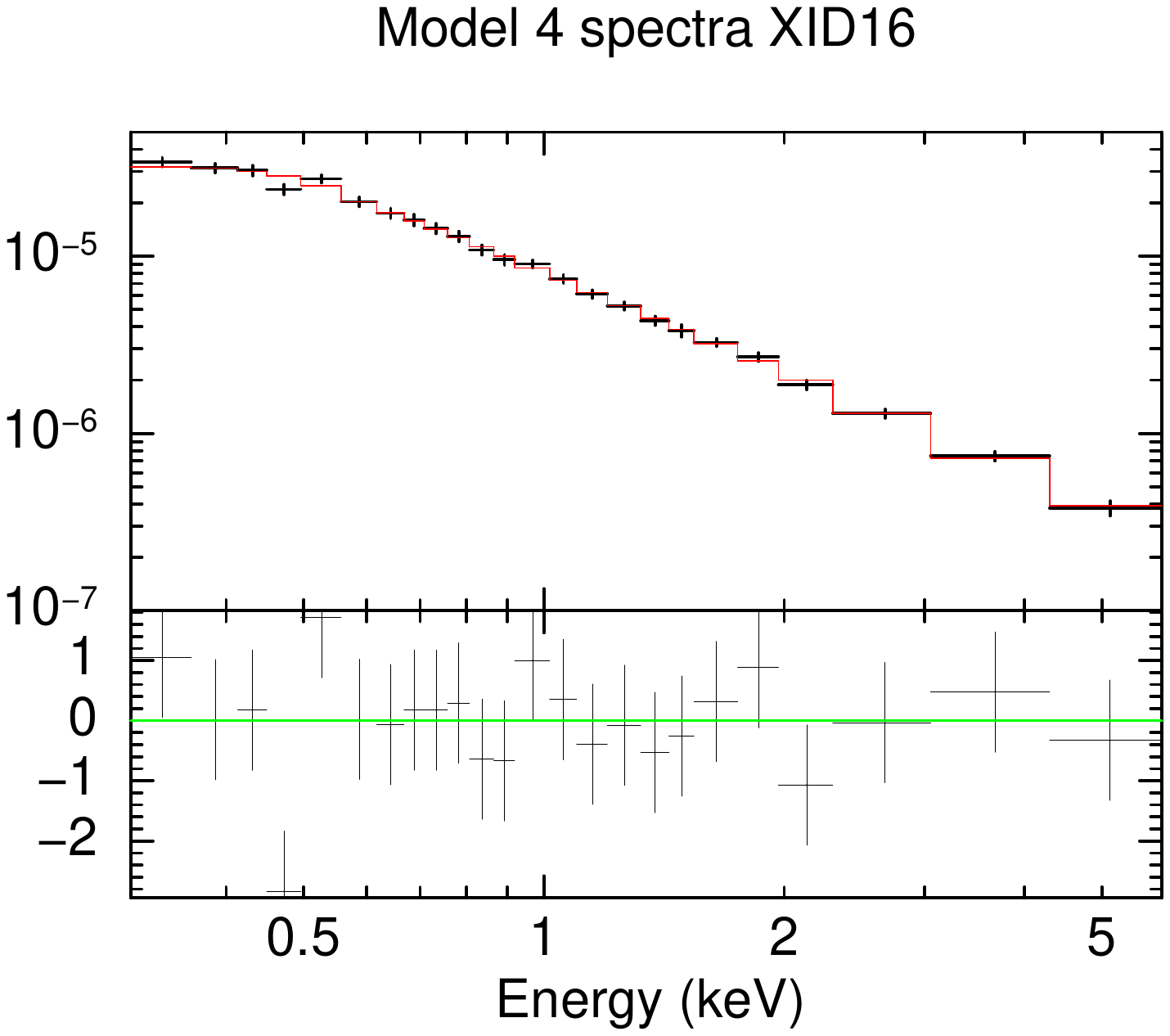}
\includegraphics[scale=0.211]{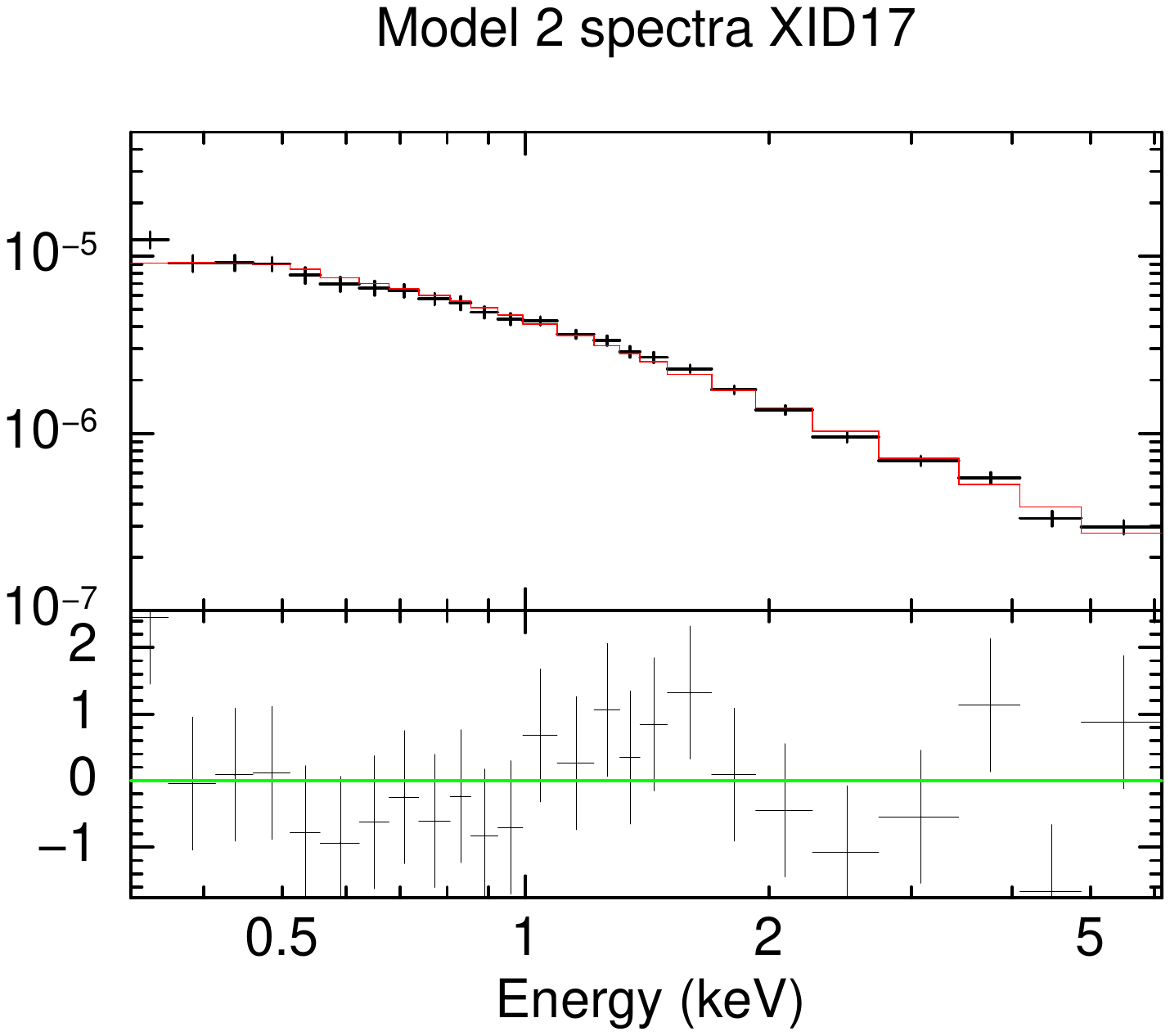} 
\includegraphics[scale=0.211]{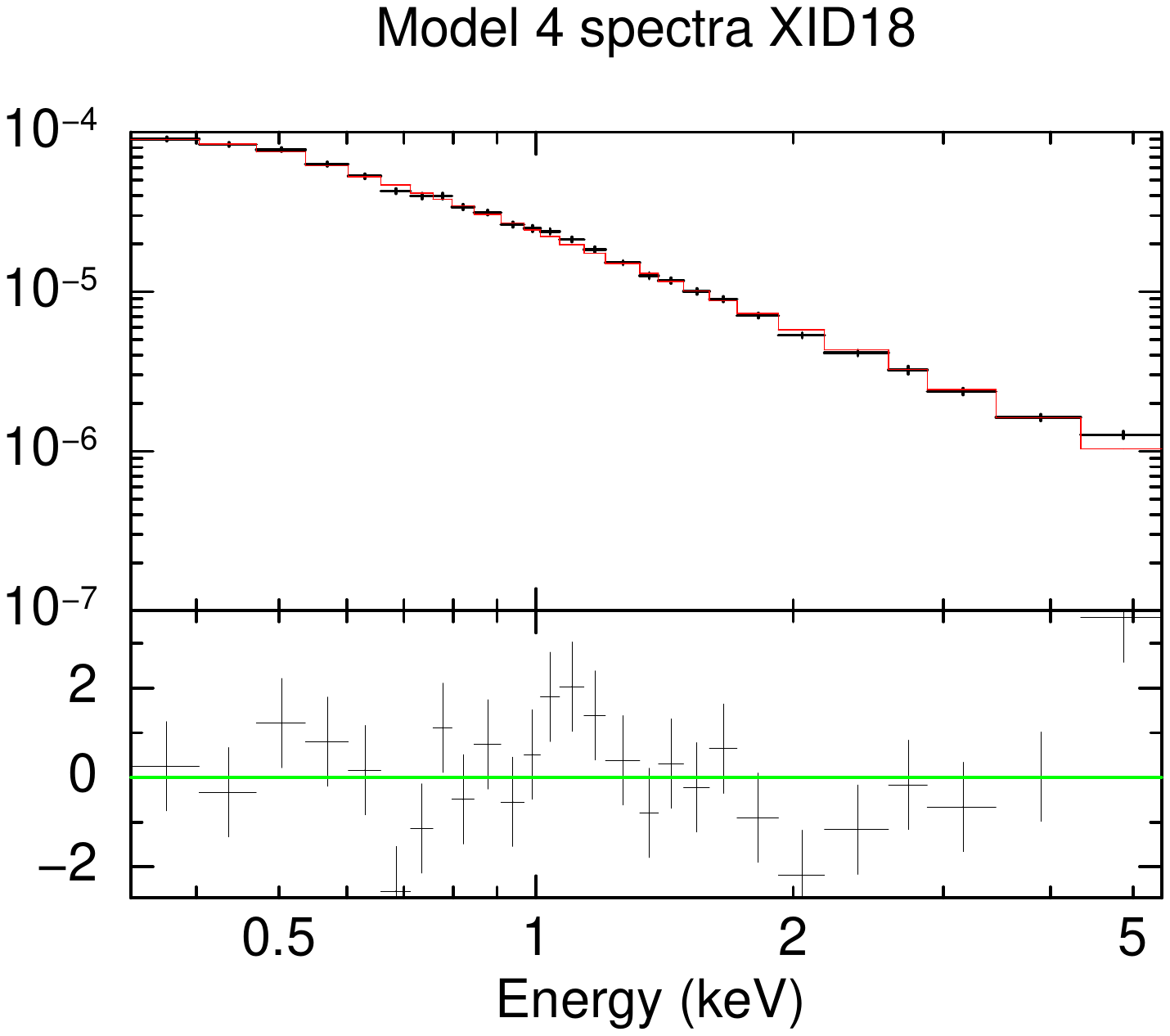} 
\\
\includegraphics[scale=0.211]{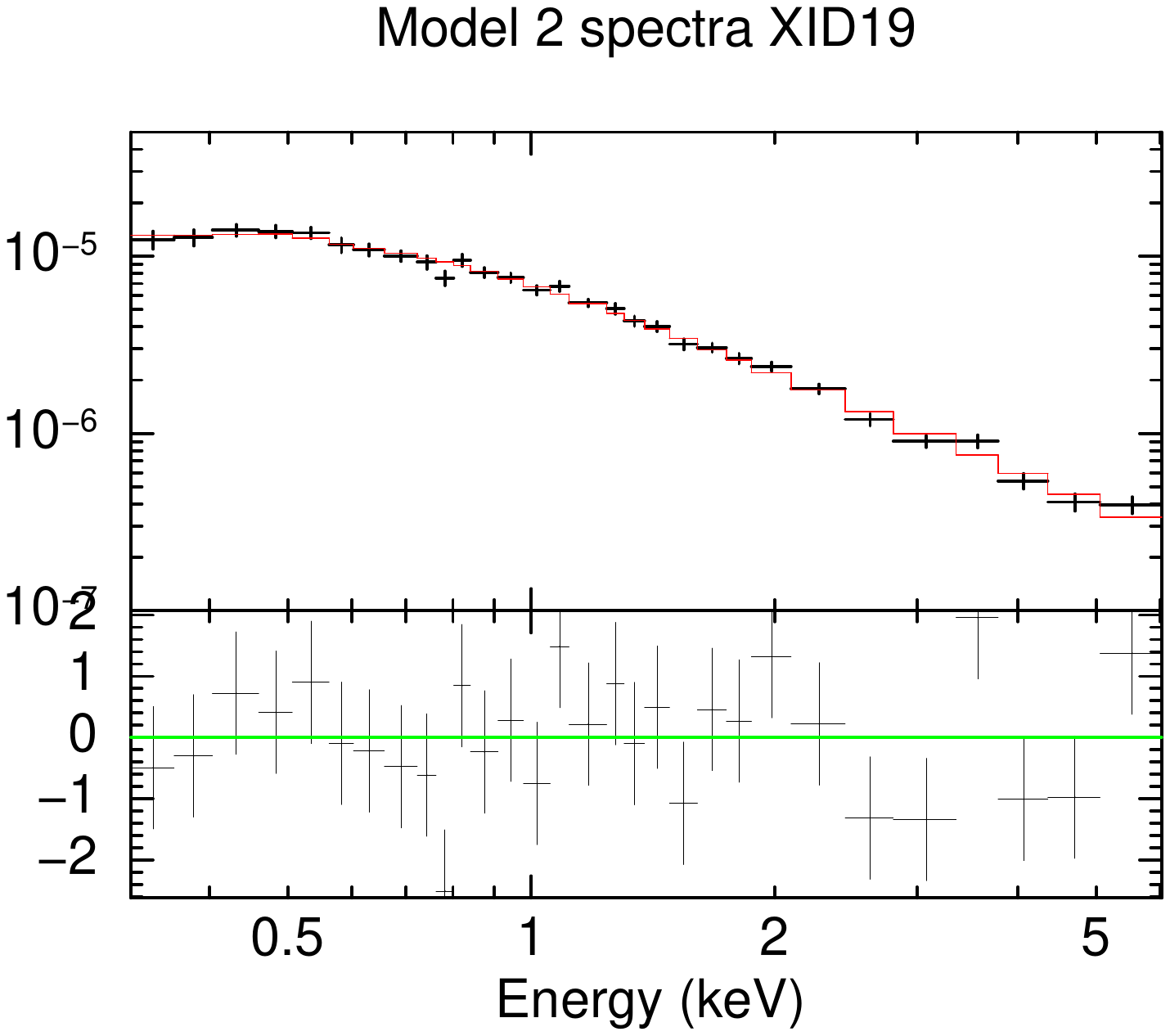}
\includegraphics[scale=0.211]{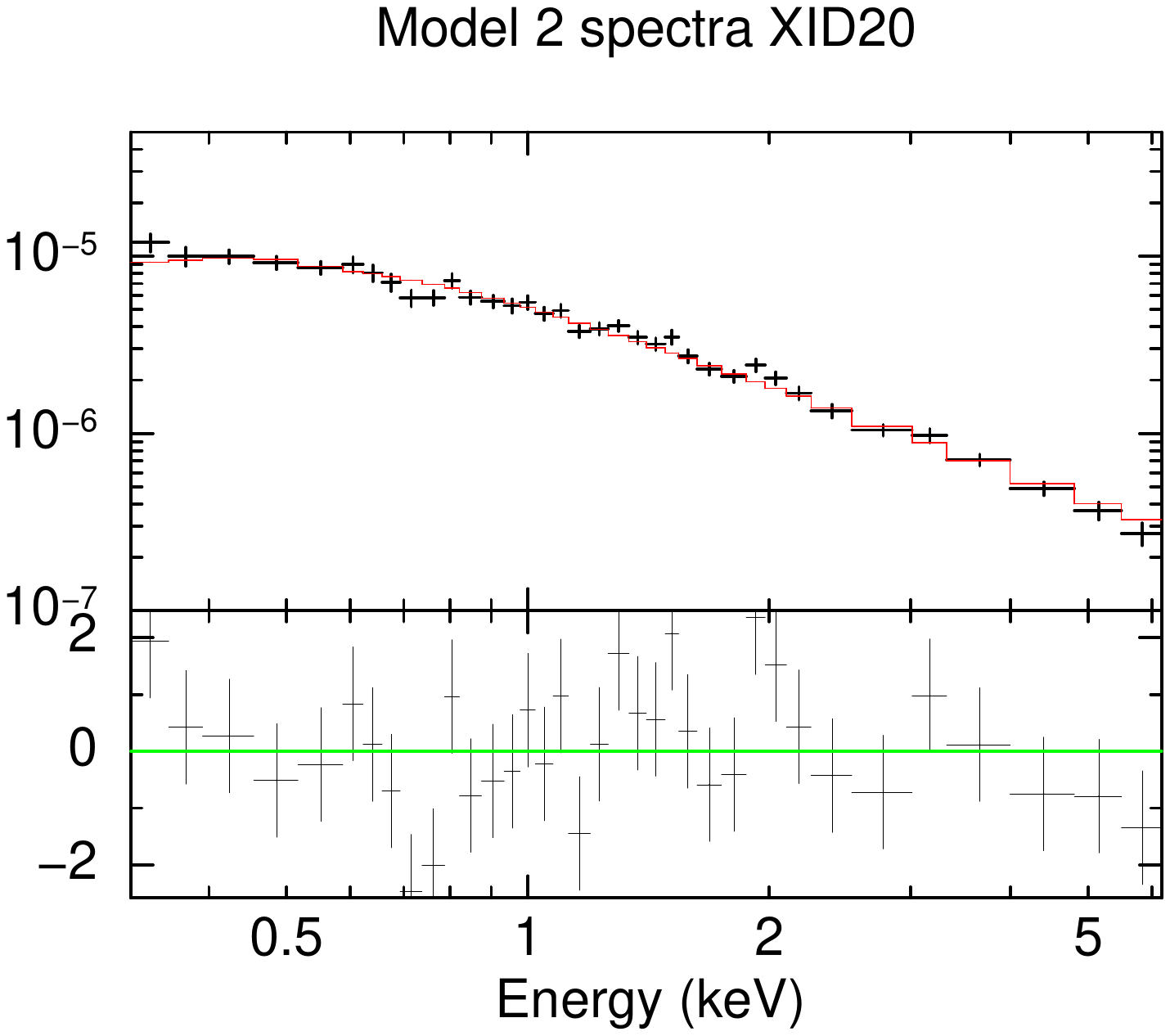} 
\includegraphics[scale=0.211]{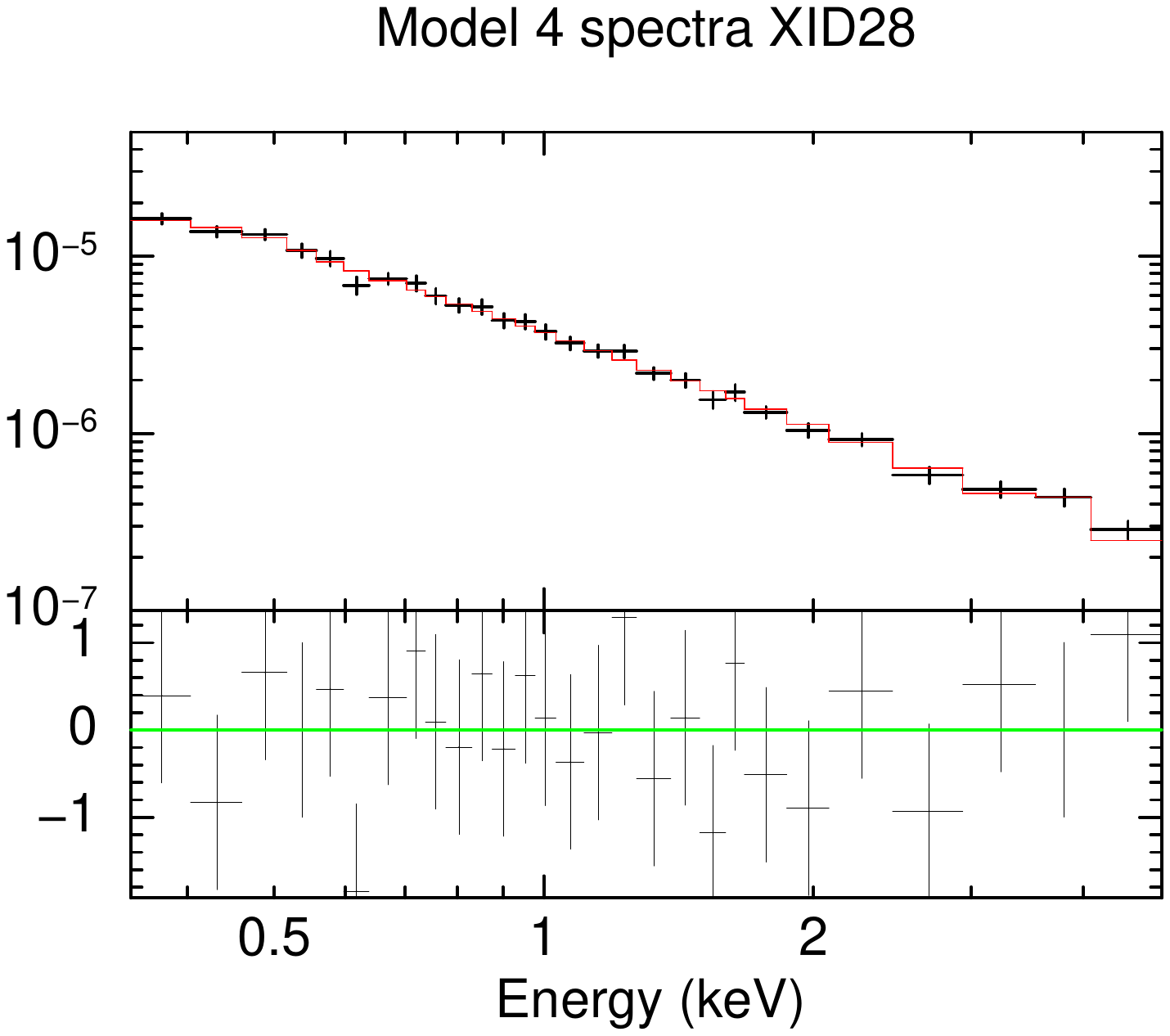}
\\
\includegraphics[scale=0.211]{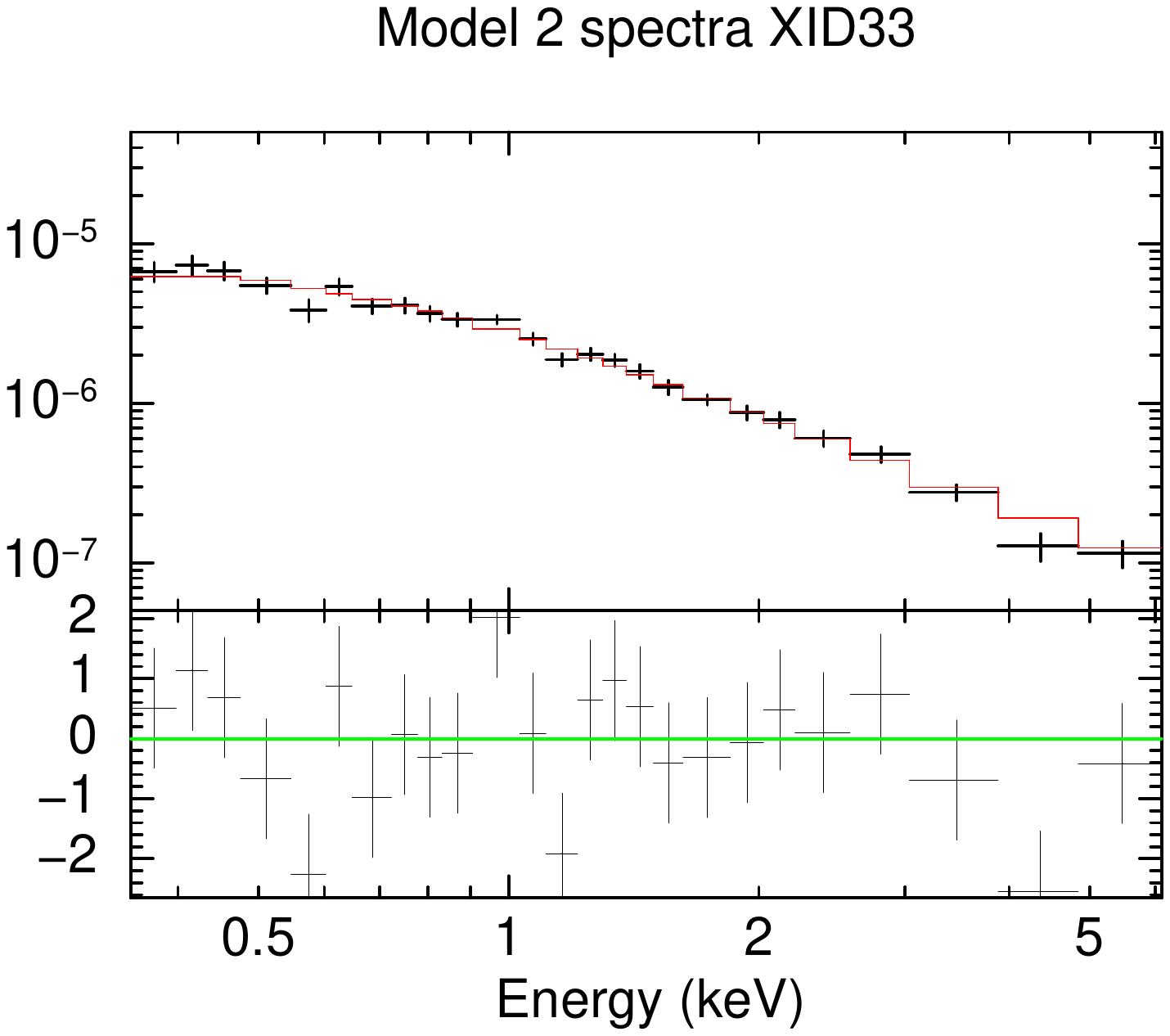} 
\includegraphics[scale=0.211]{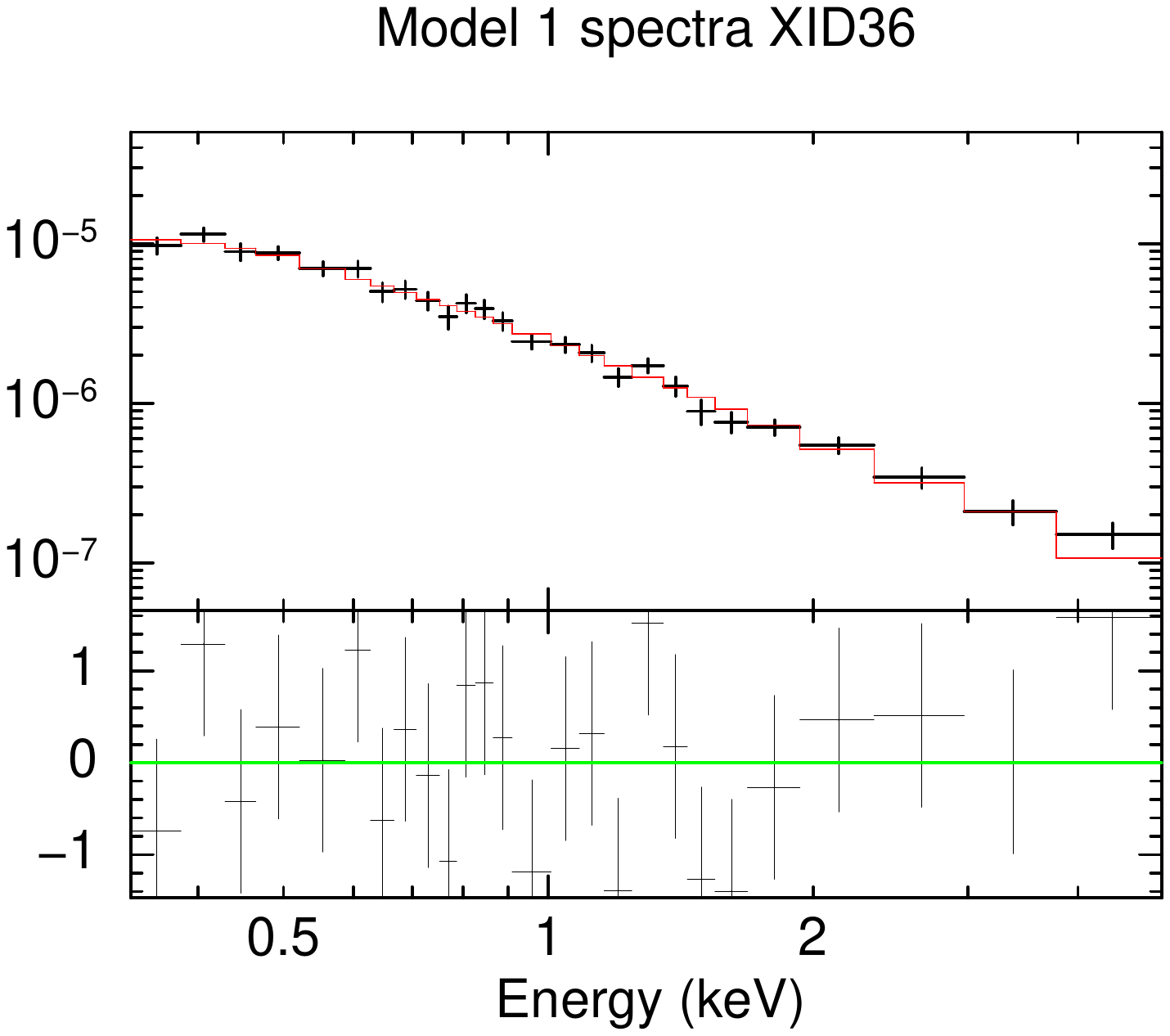} 
\caption{Spectra of the AGN sample and their residuals part-2.}
\end{figure*}


\bsp	
\label{lastpage}
\end{document}